\documentclass[]{imd_techreport}

\usepackage[toc,page,header]{appendix}

\usepackage{minitoc}

\usepackage{amssymb}

\usepackage{subcaption}
\usepackage{xcolor}
\usepackage{hyperref}
\usepackage{xspace}
\usepackage{varwidth}
\usepackage{pifont} 
\usepackage{wrapfig}
\usepackage[ruled,vlined]{algorithm2e}
\usepackage{CJKutf8}
\usepackage{pgfplots}
\pgfplotsset{compat=1.18}
\usepackage{enumitem}
\usepackage{tabularx}
\usepackage{booktabs}
\usepackage{multirow}
\usepackage{makecell}
\usepackage{algorithm}      
\usepackage{algorithmic}
\usepackage{multicol}
\usepackage{etoolbox}

\doparttoc                 
\usepackage[table]{xcolor}   

\title{
ODesign: A World Model for Biomolecular Interaction Design

}

\affiliation{{ODesign Team}\textsuperscript{\hyperref[sec:team]{*}}, \email{odesign@lglab.ac.cn} }

\definecolor{stringcolor}{rgb}{0.25, 0.5, 0.5}
\definecolor{numbercolor}{rgb}{0.0, 0.6, 0.4}
\definecolor{keywordcolor}{rgb}{0.85, 0.18, 0.50}
\lstdefinelanguage{json}{
  basicstyle=\ttfamily\small,
  numbers=none,
  breaklines=true,
  showstringspaces=false,
  morestring=[b]",
  stringstyle=\color{stringcolor},
  literate=
   *{0}{{{\color{numbercolor}0}}}{1}
    {1}{{{\color{numbercolor}1}}}{1}
    {2}{{{\color{numbercolor}2}}}{1}
    {3}{{{\color{numbercolor}3}}}{1}
    {4}{{{\color{numbercolor}4}}}{1}
    {5}{{{\color{numbercolor}5}}}{1}
    {6}{{{\color{numbercolor}6}}}{1}
    {7}{{{\color{numbercolor}7}}}{1}
    {8}{{{\color{numbercolor}8}}}{1}
    {9}{{{\color{numbercolor}9}}}{1}
    {:}{{{\color{black}{:}}}}{1}
    {,}{{{\color{black}{,}}}}{1}
    {_}{{{\char`_}}}{1} 
}

\begin{document}
\useinlineabstract
\maketitle

\newcommand{\figref}[1]{Figure~\ref{#1}}
\newcommand{\todo}[1]{\colorbox{yellow!30}{[{#1}]}}

\newcommand{\outline}[1]{%
  \colorbox{cyan!20}{%
    \begin{varwidth}{\dimexpr\linewidth-2\fboxsep\relax}
    [{\bf Outline}]  #1
    \end{varwidth}%
  }%
}\vspace{-4em}
\begin{graphicalabstract}
  \includegraphics[width=1\textwidth, trim=0 170 420 0, clip]{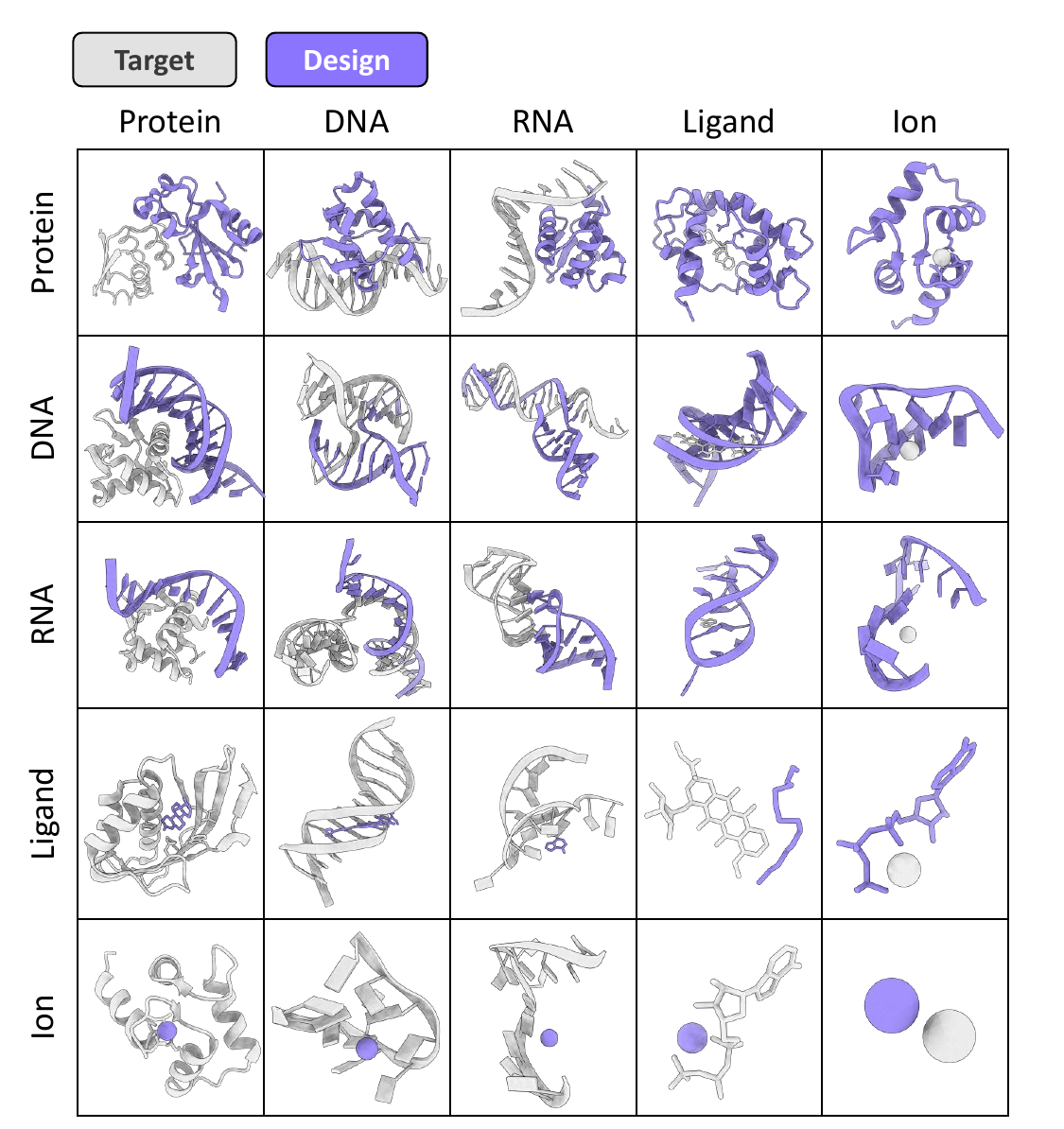}
\end{graphicalabstract}
\begin{abstract}
Biomolecular interactions underpin almost all biological processes, and their rational design is central to programming new biological functions. Generative AI models have emerged as powerful tools for molecular design, yet most remain specialized for individual molecular types and lack fine-grained control over interaction details. Here we present ODesign, an all-atom generative \textit{world model} for all-to-all biomolecular interaction design. ODesign allows scientists to specify epitopes on arbitrary targets and generate diverse classes of binding partners with fine-grained control. Across entity-, token-, and atom-level benchmarks in the protein modality, ODesign demonstrates superior controllability and performance to modality-specific baselines. Extending beyond proteins, it generalizes to nucleic acid and small-molecule design, enabling interaction types such as protein-binding RNA/DNA and RNA/DNA-binding ligands that were previously inaccessible. By unifying multimodal biomolecular interactions within a single generative framework, ODesign moves toward a general-purpose molecular world model capable of programmable design. ODesign is available at \url{https://odesign.lglab.ac.cn}, allowing users to generate binding partners without coding expertise.
\end{abstract}

\section{Introduction}\label{intro}
Life processes are orchestrated by the coordinated actions of multiple molecular modalities, including proteins, nucleic acids, and small molecules---that collectively mediate signal transduction, energy metabolism, and gene regulation, forming the core machinery of biology~\citep{Karp2009,Brini2020_Science}. For decades, scientists have sought to understand, emulate, and ultimately control biological processes through rational molecular design, a pursuit often described as the Holy Grail of structural biology and drug discovery~\citep{Huang2016_Nature,Lippow2007_COB,Mandal2009_EJP}. Bridging this molecular complexity with computational design has long been a central challenge.

Recent advances in artificial intelligence (AI) have transformed molecular design from intuition-driven trial and error into a computational discipline. Breakthroughs such as RFDiffusion for \textit{de novo} protein design~\citep{Watson2023_Nature_RFDiffusion}, ResGen for protein--small-molecule generation~\citep{Zhang2023_NMI_ResGen}, and RhoDesign for RNA design~\citep{Wong2024_NCS_RhoDesign} have each demonstrated remarkable progress within their respective domains. Yet these models cannot perform the cross-modality reasoning required to program complex biological functions. Structural prediction suggests a potential route toward this goal, as it intrinsically learns the shared physicochemical principles governing molecular interactions across modalities. Models such as RosettaFold-All-Atom (RF-AA)~\citep{Krishna2024_Science_RFAA} and AlphaFold3~\citep{Abramson2024_Nature_AF3} jointly represent proteins, nucleic acids, small molecules, and chemical modifications within a unified structure-prediction framework. Leveraging carefully designed architectures and multimodal representations, AlphaFold3 has achieved state-of-the-art performance across nearly all structural prediction benchmarks. While these advances demonstrate the power of cross-modality modeling, their objectives remain confined to predicting existing structures rather than generating new molecules with desired functions. Realizing a generative framework capable of cross-modality biomolecular design remains an open challenge.

Despite the rapid progress of generative foundation models such as Latent-X~\citep{LatentX2025_arXiv} and PXDesign~\citep{Ren2025_bioRxiv_PXDesign}, existing frameworks remain primarily protein-centric, focusing on target--binder generation at the entity level. This formulation simplifies modeling but fails to reflect the multimodal nature of living systems, in which proteins, nucleic acids, metabolites, lipids, and metal ions form dynamic, multiscale networks and regulatory hierarchies that together sustain biological function~\citep{Barabasi2004_NRG,Kitano2002_Science_SystemsBio}. Extending generative modeling beyond this protein-centric view is challenging, as molecules across different modalities follow distinct compositional principles: small molecules consist of atoms or functional groups, proteins of amino acid residues, and nucleic acids of nucleotide chains~\citep{Jorgensen2018_MolInform}. A purely atomic representation struggles to ensure that generated structures conform to the intended molecular modality and often overlooks the chemical and biological priors encoded in higher-order structural units~\citep{Jin2020_ICML_Motif,Qu2024_bioRxiv_PAllAtom}. Achieving truly cross-modality molecular generation therefore requires models that reconcile atomic-level precision with hierarchical organization, enabling conditional control across molecular, motif, and atomic levels to realize programmable functional design~\citep{Joshi2021_Molecules_AI_MolDesign,Zhang2024_JACS_DeepLeadOpt}.

To address these challenges, we developed \textbfr{ODesign}, the first-of-its-kind generative world model for cross-modality and multi-conditional molecular design. Built upon state-of-the-art structure-prediction architectures~\citep{Abramson2024_Nature_AF3}, ODesign abstracts the minimal chemical units of diverse molecular species into a unified token space. This representation enables ``all-to-all'' molecular generation within a single framework. A task-oriented masking mechanism provides conditional control at molecular, motif, and atomic levels, while flexible and rigid conditioning allows the model to either fix target structures or co-generate target--binder interfaces. Across eleven benchmark tasks spanning proteins, small molecules, RNA, and DNA, ODesign consistently outperforms modality-specific models and achieves substantial improvements in design throughput, enabling high-throughput generation across molecular modalities. Collectively, these results demonstrate that ODesign establishes a unified framework for molecular generation across modalities---an important step toward a general-purpose molecular world model.

\section{Architecture and training of ODesign}

\begin{figure}[htbp]
  \centering
  \includegraphics[width=1.\textwidth, trim=20 10 20 10, clip ]{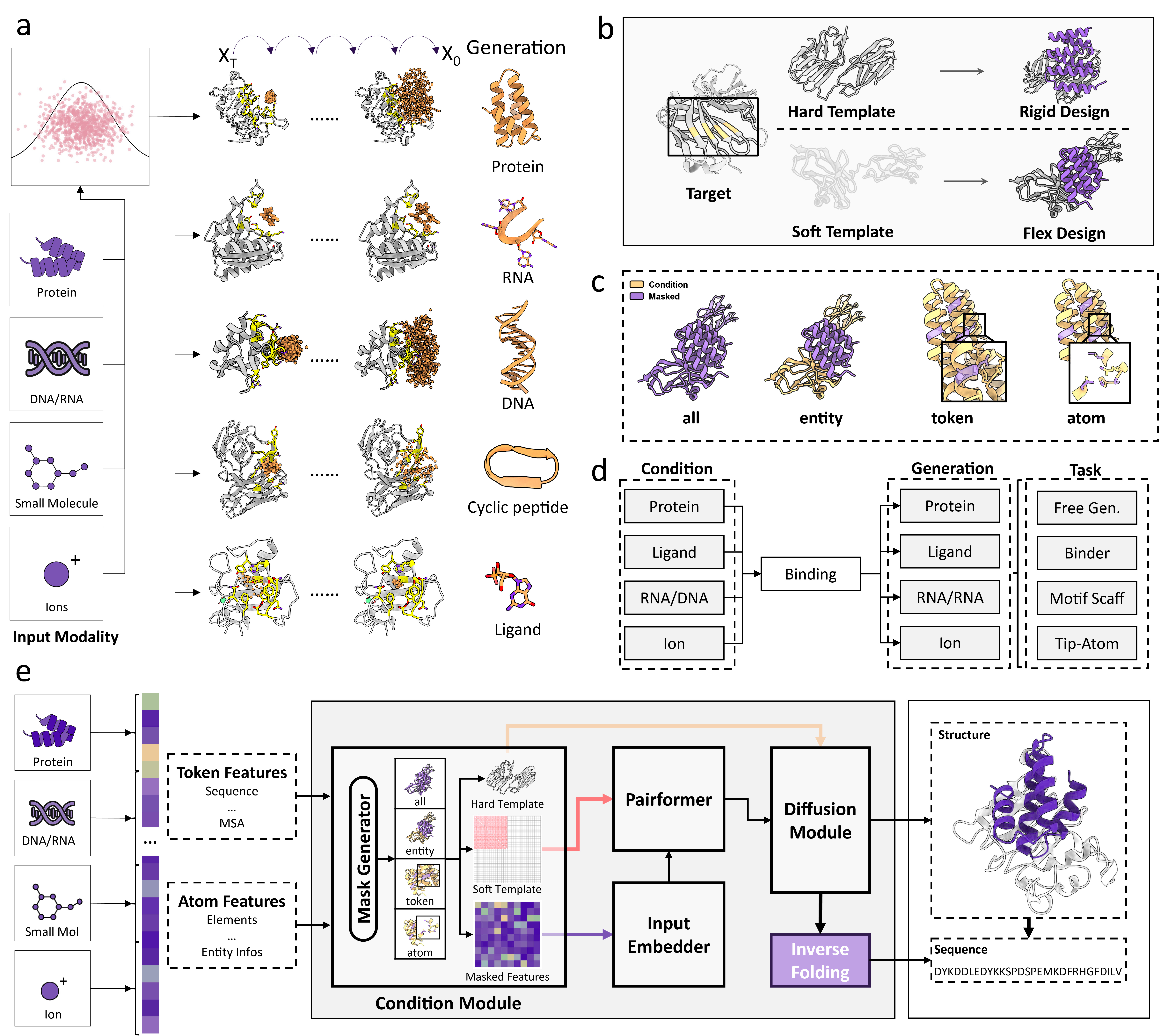}
  \caption{
  Overview of the ODesign framework for all-to-all molecular design. (a) Overview of ODesign. The model takes proteins, DNAs, RNAs, small molecules, and ions as inputs and designs binding partners of specified modalities through a diffusion-based generative process. (b) ODesign supports both rigid and flexible conformations when designing binding partners. (c) Unified masking strategy. Four levels of masking are applied during training, where conditioned regions are shown in yellow and designable regions in blue. (d) Task composition. ODesign supports combinations of molecular modalities—{protein, ligand, RNA, DNA, ion} binding {protein, ligand, RNA, DNA, ion}—and multiple task types including free generation, binder design, motif scaffolding, and tip-atom scaffolding. (e) Architecture of ODesign. Arrows indicate the direction of information flow across model components.
  }
  \label{fig:odesign}
\end{figure}

Molecular design involves two types of outputs: continuous coordinates in three-dimensional space and discrete element types. Accordingly, existing technical routes can be broadly classified into two categories. The first is co-design, which simultaneously optimizes both structure and type; the second is a two-stage strategy, in which a backbone is generated first and the sequence is then designed. Although co-design is theoretically capable of more comprehensively modeling the coevolution between structure and sequence, the two-stage approach always performs better in practical protein design~\citep{Geffner2025_arXiv_Proteina,Campbell2024_arXiv_Multiflow}. Therefore, ODesign adopts the two-stage design paradigm. On this basis, the central challenge of cross-modality molecular generation lies in ensuring that the generated set of atoms corresponds to the correct molecular modality. To address this issue, we abstract the chemical building blocks of different molecule types into a unified set of generative tokens, which are decoded through an all-atom diffusion module to enable multimodal molecular generation.

The primary objective of a structure-prediction model is to infer three-dimensional structures from biological sequences~\citep{Abramson2024_Nature_AF3, team2025intfold}; hence, a high-performance predictor inherently encodes rich knowledge of interatomic interactions. We hypothesized that these learned structural priors could be repurposed to guide molecular design. Therefore, we treat structure prediction as a pretraining task for molecular design and build ODesign upon a state-of-the-art AlphaFold3-like architecture. Within this architecture, ODesign abstracts the minimal generative unit of each chemical modality as a unified token, learns its interactions with the target structure through the Pairformer layer, and decodes the three-dimensional coordinates that satisfy these interaction constraints via an all-atom conditional diffusion module. By combining upsampling and downsampling pathways, the model achieves cross-modality molecular generation at full-atom resolution.

Beyond the generative mechanism, molecular design and structure prediction differ fundamentally in task formulation: the former starts from a known target or motif structure, whereas the latter infers structure from a completely unknown sequence. To enhance the model’s conditional awareness, we introduce three-dimensional structural information into the embedding layer, the Pairformer module, and the diffusion decoder, thereby injecting structural priors throughout the generative process. ODesign also implements two design modes: a rigid-receptor mode that fixes the target structure and a flexible-receptor mode that allows joint generation of ligand binders and receptor conformations. Considering that a single target often contains multiple potential binding epitopes, we further eenable ODesign to condition its generation on specified target epitopes. During training, different epitopes are randomly selected with a certain probability, enabling the model to learn epitope localization and key interaction recognition.

In summary, ODesign consists of five core modules: (1) an embedding layer for initial element representation; (2) a conditional module encoding the target’s 3D structure; (3) a Pairformer module for modeling molecular interactions; (4) a conditional diffusion module for decoding all-atom coordinates; and (5) an inverse folding module for unified type generation. We train ODesign on all monomeric and complex structures from the PDB~\citep{Burley2017_PDB}, adjusting sampling probabilities and masking strategies according to the downstream design tasks.

\section{Protein Designability}

Proteins, owing to their programmability and ease of synthesis, represent the most mature molecular modality in computational design. Protein design has been widely applied to the development of therapeutic molecules~\citep{Carter2022_Cell_Antibodies,Jayaraman2020_EBioMedicine_CART} and synthetic biological systems~\citep{Pinto2022_TrendsBiochemSci_EnzymeEvolution,Eriksen2014_JSB_PathwayEngineering,Drubin2007_GenesDev_DesigningSystems}. Recent advances in structure prediction and deep generative modeling have transformed protein design from a rule-based, intuition-driven process into a data- and computation-driven discipline, enabling solutions to design problems that were previously intractable~\citep{Gloegl2024_Science_TNFR}. From a computational perspective, protein design tasks can be categorized along two dimensions~\citep{Kortemme2024_Cell_Review}: (1) by the scale of structural control—spanning entity-level, motif-level, and atomic-level tasks; and (2) by the type of target molecule—encompassing the design of proteins, small molecules, and nucleic acids.

\begin{figure}
\begin{minipage}{\textwidth}\centering\vspace{-2em}
  \includegraphics[width=0.92\textwidth, trim=00 90 00 15, clip]{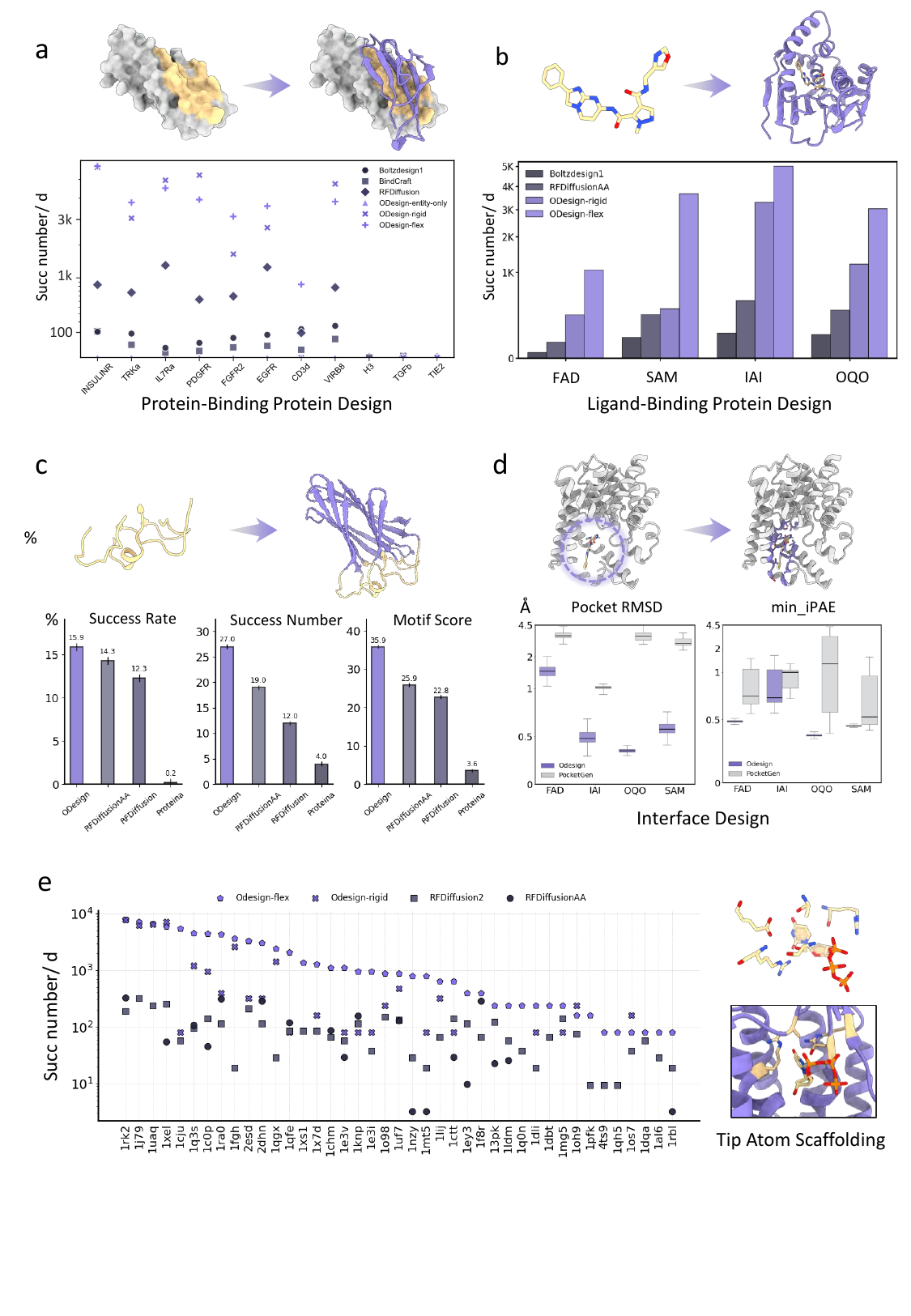}
  \caption{
  Performance of ODesign on protein-centric benchmarks. All panels: designs are shown in purple, targets in pink, and ligands in green (carbon in green, nitrogen in blue, oxygen in red, sulfur in orange). (a) Protein-binding protein design. ODesign was compared with BindCraft, RFDiffusion, and BoltzDesign across ten targets, measured by the number of filter-passing designs generated per GPU per day. (b) Ligand-binding protein design. ODesign was compared with RFDiffusion-AA and BoltzDesign across four targets using the same throughput metric. (c) Motif scaffolding. ODesign was evaluated against RFDiffusion, RFDiffusion-AA, and ProteinA on the MotifBench benchmark. “Success rate” denotes the average across targets; “success number” indicates the count of targets with successful solutions; “MotifScore” is a weighted benchmark score reflecting both difficulty and success number. (d) Interface design. ODesign was compared with PocketGen; distributions of pocket-aligned RMSD and "min\_iPAE" are shown. (e) Tip-atom (atomic motif) scaffolding. ODesign was compared with RFDiffusion-AA and RFDiffusion2 on the AME benchmark, evaluated by the number of filter-passing designs per GPU per day.
  }
  \label{fig:protein_designability}
\end{minipage}
\centering
\end{figure}

Despite this progress, current generative AI frameworks address only subsets of these challenges. For instance, BindCraft~\citep{Pacesa2025_Nature_BindCraft} focuses on \textit{de novo} binder generation; RFDiffusion~\citep{Watson2023_Nature_RFDiffusion} supports both \textit{de novo} and motif-scaffolding tasks; subsequent models, such as RFDiffusion-AA~\citep{Krishna2024_Science_RFAA}, extend designability to ligand-binding protein systems, and RFDiffusion2~\citep{Ahern2025_bioRxiv_RFDiffusion2} enables atomic-level motif scaffolding. However, none of these models have shown generalization across different levels of conditional control. Moreover, AI-based protein design still faces a throughput bottleneck in practice~\citep{Khmelinskaia2024_Nature_FiveTasks,Li2024_JAFC_ThermostableScreening}. For example, for a medium-difficulty target, a single GPU run often yields only a few ready-to-test candidates, which are not even enough to fill a 96-well screening plate. This limitation constrains wet-lab productivity and underscores the need for AI methodologies that can achieve orders-of-magnitude gains in computational efficiency to scale protein design toward practical, large-scale applications.

\subsection{Protein-binding Protein Binder Design}

Protein–protein binder design is one of the most common directions in protein engineering, frequently employed in developing protein inhibitors~\citep{Gainza2020_NMeth_GDLsurfaces}. It is particularly effective for targets with relatively flat, hydrophobic interfaces dominated by nonpolar interactions~\citep{Gainza2020_NMeth_GDLsurfaces,Meenan2010_PNAS_Selectivity}. In this study, eleven benchmark protein targets were selected from previous protein design work~\citep{Cao2022_Nature_TargetOnly}. For each target, 800 candidate binders were generated, and structure refolding was performed using AF3 with the given multiple sequence alignment (MSA) for the target protein. An evaluation criterion~\citep{Zambaldi2024_arXiv_AlphaProteo} was adopted: refolded complexes with iPAE$<$10.85, ipTM$>$0.5, pLDDT$>$0.8, and complex\_RMSD$<$2.5\AA{} were regarded as successful designs, and the number of successful cases obtained within one day on a single H100 GPU was plotted. ODesign was compared with representative existing methods, including the diffusion-based RFDiffusion, the AF2-based hallucination model BindCraft, and the Boltz-based hallucination BoltzDesign~\citep{Cho2025_bioRxiv_Boltzdesign1}.

As shown in Figure~\ref{fig:protein_designability}a, ODesign outperformed all other methods across each target (average (Avg.) 2672), achieving an order-of-magnitude improvement in design throughput compared to the second-best method, RFDiffusion (Avg. 555). Hallucination-based models such as BindCraft and BoltzDesign required prolonged optimization to generate designs that passed internal filters, yielding only a few successful designs within a one-day computation window (BindCraft Avg. 22 while BoltzDesign1 Avg. 61). Among the two ODesign variants, the flexible version (ODesign-Flex) slightly outperformed the rigid version (ODesign-Rigid) in this task, likely because it allows binding poses to adjust during the design process, thereby better accommodating the conformational features of the target interface~\citep{Mandell2009_COB_BackboneFlex}.

\subsection{Ligand-binding Protein Binder Design}

Ligand-binding protein design has broad applications in enzyme engineering and molecular sensing~\citep{Tinberg2013_Nature_LigandBinding}. In contrast to the protein–protein binder design, this task is inherently more geometry-sensitive, as it depends on shaping precise three-dimensional pockets to ensure stable and selective ligand binding~\citep{Yang2017_COSB_LigandDesignReview},requiring more powerful and accurate protein-ligand interaction awareness of the model. RFDiffusion-AA extends the RFDiffusion by introducing the ligand as a conditioning input during the denoising process, whereas BoltzDesign guides a hallucination-based model using the chemical formula of the target ligand as input. Four ligand-binding cases were selected from the  previously-described RFDiffusion-AA study, and 800 protein samples were generated for each target. Since small-molecule targets lack MSA information, Chai-1~\citep{Chai12024_bioRxiv} with ESM embeddings~\citep{Lin2022_bioRxiv_LMStructure} was used as the refolding model. The success criterion was defined as iPAE$<$10 and complex\_pLDDT$>$0.7. As shown in Figure~\ref{fig:protein_designability}b, ODesign outperformed existing methods, achieving 12.1-fold improvement over RFDiffusion-AA and 57.8-fold over BoltzDesign1. Moreover, ODesign-Flex consistently surpassed ODesign-Rigid across all cases, indicating that explicitly modeling ligand flexibility during the design process enhances the identification and formation of binding pockets.
\subsection{Motif Scaffolding}

Beyond \textit{de novo} design for different molecular targets, protein engineering also requires the ability to modify existing proteins, a task formulated as motif scaffolding~\citep{Trippe2022_arXiv_BackboneDiffusion}. The objective of this task is to design protein structures that can accommodate and stabilize a given functional motif. MotifBench~\citep{Zheng2025_arXiv_MotifBench} provides 30 challenging test cases, including orphan peptide fragments and enzyme active sites, representing classical and grand challenges of protein design. Following the MotifBench protocol, ODesign was compared against RFDiffusion-AA, Proteina~\citep{Geffner2025_arXiv_Proteina}, and RFDiffusion. As shown in Figure~\ref{fig:protein_designability}c, ODesign achieved the highest overall success rate and number of successful cases among all competing methods, with an average success rate of 18.4\% compared to 14.3\% for RFDiffusion-AA, and successful designs were obtained in 27 of 30 test cases. On the key MotifScore metric defined in MotifBench, ODesign outperformed the second-best method by 38.6\%. Further case-by-case analyses (Figure~\ref{fig:s4}) demonstrated that ODesign outperformed other models in most test cases, highlighting its strong capability for residue-level controlled generation.

\subsection{Interface Design}

Interface design~\citep{Lucas2020_PLoSCB_SmallMolSites}, also referred to as pocket design, aims to reconstruct a protein’s binding interface while preserving its main backbone structure. This task can be regarded as complementary to motif scaffolding: the former focuses on exploring and sampling diverse binding modes, whereas the latter emphasizes maintaining the stability of an existing motif. The representative method PocketGen~\citep{Zhang2024_NMI_PocketGen} adopts a combined strategy of sequence design and structural optimization to redesign protein–ligand interfaces. For comparison, four ligand-binding targets from the ligand-binding task were selected as benchmarks, and 400 candidate proteins were generated for each case. It should be noted that PocketGen originally employed AlphaFold2~\citep{Jumper2021_Nature_AF2} for structure refolding, which does not account for ligands within the binding pocket during structure prediction. Therefore, Chai-1~\citep{Chai12024_bioRxiv} was used as the refolding model in this study. As shown in Figure~\ref{fig:protein_designability}d, ODesign outperformed PocketGen across all four cases, producing lower mean pocket-aligned RMSD (by 2.24\AA{}) and lower mean min-iPAE (by 0.68), indicating that the pockets designed by ODesign are structurally more stable and better aligned with the intended binding geometry.

\subsection{Atomic Motif Scaffolding}

With the continuous advancement of protein design methods, increasing emphasis has been placed on achieving atomic-level precision. In enzyme design, accurate three-dimensional conformations between catalytic residues and substrates are essential for downstream reactions~\citep{Xu2023_Bioeng_EnzymeReconstruction,Chen2009_PNAS_EnzymeRedesign}. RFDiffusion2~\citep{Ahern2025_bioRxiv_RFDiffusion2} first formalized this problem as the tip-atom task, also known as atomic motif scaffolding. This task is defined as follows: given a ligand and several key residues represented at the atomic level, the model must design a protein that reproduces the spatial geometry of these atomic motifs. For this task, we followed the AME benchmark proposed in RFDiffusion2 to systematically evaluate ODesign. Two methods, RFDiffusion2 and RFDiffusion-AA, are included as the baselines. As shown in Figure~\ref{fig:ligand_designability}e, ODesign generated high-throughput, filter-passing designs across nearly all test cases, achieving an average efficiency improvement of 37.5-fold. We anticipate that such large candidate pools will enable ODesign to maintain high success rates even under subsequent wet-lab-inspired filtering, ultimately allowing single computational runs to yield designs sufficient to fill an entire 96-well plate, laying the groundwork for high-throughput enzyme design.

\section{Nucleic Acid Designability}

\begin{figure}[htbp]
  \centering
  \includegraphics[width=0.95\textwidth, trim=00 180 00 00, clip]{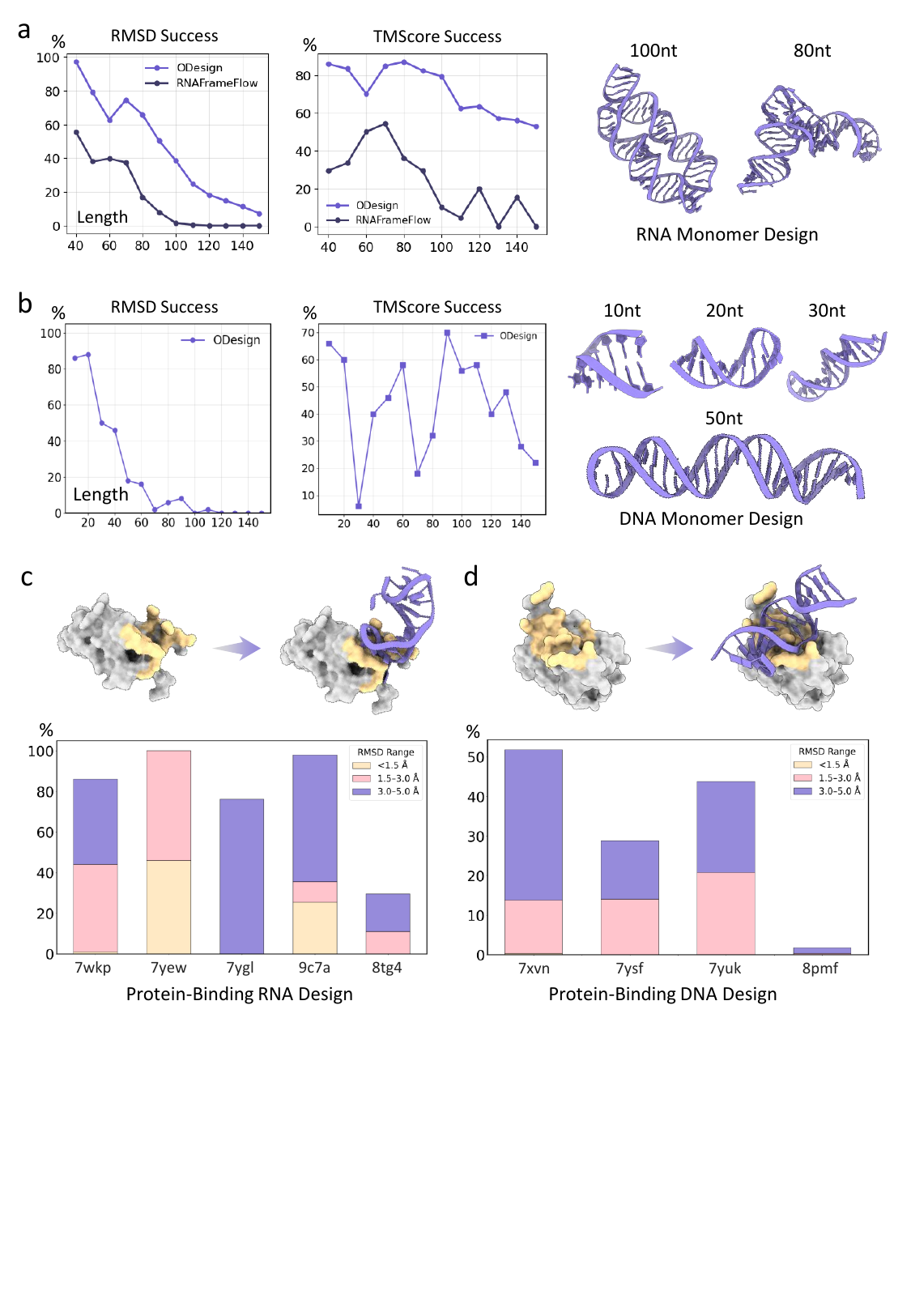}
  \caption{
  Performance of ODesign on nucleic acid-centric benchmarks. All panels: designs are shown in purple, targets in pink, and ligands in green (carbon in green, nitrogen in blue, oxygen in red, sulfur in orange).  (a) RNA monomer design. ODesign was compared with RNAFrameFlow. RMSD (<5.0 Å) and TM-score (>0.45) success rates by nucleic acid length are shown. Generated examples are displayed on the right. (b) DNA monomer design. Same metrics as in (a) are shown for DNA design. (c) Protein-binding RNA design. Distributions of protein-aligned RNA RMSD between designed and refolded complexes are shown in three RMSD intervals. (d) Protein-binding DNA design. The same metric as in (c) is shown for DNA complexes.
  }
  \label{fig:na_designability}
\end{figure}

Nucleic acids (NAs) represent another class of programmable biomolecules capable of performing diverse biological functions. Similar to protein binders, nucleic acids can achieve specific functions through binding to other molecules; such functional nucleic acids are commonly referred to as aptamers~\citep{Tuerk1990_Science_SELEX,Ellington1990_Nature_InVitroSelection}. Because of their low synthesis and modification costs, as well as their high screening and optimization efficiency, aptamers are considered one of the most engineerable modalities in molecular design~\citep{Zhou2017_NatRevDrugDisc_Aptamers,Sun2015_Small_AptamersNanomed}. To date, nucleic acid aptamers have been widely applied to bind proteins, small molecules, other nucleic acids, and even whole cells~\citep{Darmostuk2015_BiotechAdv_SELEXUpdate}. In recent years, several AI-driven nucleic acid design methods have been developed, such as RNAFrameFlow~\citep{Anand2024_arXiv_RNAFlowMatching}, RDesign~\citep{Tan2024_ICLR_RDesign}, and gRNAde~\citep{Joshi2024_Methods_gRNAde}. However, most existing models still focus on generating single nucleic acid molecules and lack the capability to perform more realistic, application-oriented tasks such as target-binding aptamer design.

\subsection{RNA aptamer design}

RNA aptamers are single-stranded RNAs capable of binding to specific targets. They have received broad attention across multiple fields: for example, the RNA aptamer Pegaptanib has been approved for the treatment of neovascular age-related macular degeneration~\citep{Ng2006_NatRevDrugDisc_Pegaptanib}, and fluorescent or radiolabeled RNA aptamers have shown promise in molecular imaging and diagnostics~\citep{Sun2015_Small_AptamersNanomed}. In this study, we first evaluated the performance of ODesign on RNA monomer design tasks. Specifically, RNA molecules ranging from 10 to 150 nucleotides in length were generated and structurally refolded using AF3. RMSD between the refolded and designed structures was then computed, and designs with RMSD$<$5\AA{} or TMScore$>$0.45 were considered successful. Under this criterion, ODesign achieved nearly twice the success rate of the RNA-specific model RNAFrameFlow (Figure~\ref{fig:na_designability}a).

To further assess ODesign’s capability for conditional generation, we performed RNA aptamer design against protein targets. Because no standardized RNA design benchmark currently exists, we curated ten protein–RNA complexes from the Protein Data Bank (PDB) that were released after January 13, 2023, and used the protein structures as design conditions. For each target, 400 RNA aptamers were generated and refolded using AF3. The RMSD of the RNA, aligned to the corresponding protein partner, was used as the success metric. As shown in Figure~\ref{fig:na_designability}c, ODesign successfully generated a range of RNA aptamers for unseen targets in this zero-shot scenario, achieving a 77.95\% success rate under a protein-aligned RMSD threshold of 5\AA{}.

\subsection{DNA aptamer design}

Compared with RNA, computational design of DNA aptamers is far less common. This is partly because naturally occurring functional nucleic acids are almost exclusively RNA, which historically shaped the early conceptual framework of “functional nucleic acids”~\citep{Ellington1990_Nature_InVitroSelection}. In addition, most early aptamer design studies focused on RNA systems, resulting in a research and application path dependency~\citep{Dunn2017_NatRevChem_AptamerDiscovery}. Nevertheless, DNA aptamers possess distinct advantages in \textit{in vitro} stability and resistance to nuclease degradation, making them highly attractive for diagnostic and biosensing applications~\citep{Huizenga1995_Biochemistry_ADNA,Breaker1997_COChBiol_DNAAptamers}. Therefore, we extended ODesign to the DNA modality, broadening the design space of aptamers.

ODesign is one of the first AI models capable of rational DNA design. Since our focus was on aptamer design, all generated DNAs were single-stranded. Following the same protocol as the RNA monomer design task, we examined design success rates as a function of DNA length (Figure~\ref{fig:na_designability}b). Notably, the designability of DNA aptamers declined more rapidly than that of RNA aptamers, likely because most DNAs in the training data were double-stranded, making single-stranded sequences less represented and thus harder to model as the sequence length increased. For targeted DNA aptamer design, we followed the same approach used for RNA: four protein–DNA complexes released after January 13, 2023, with DNA lengths ranging from 5 to 100 nucleotides, were selected as benchmarks. For each target, 400 DNA aptamers were generated and refolded using AF3, and protein-aligned DNA RMSD was used as the success criterion (Figure~\ref{fig:na_designability}d). Figure~\ref{fig:na_designability}d also illustrates an aptamer example designed by ODesign. The designed DNA aptamer tend to localize around electrostatically complementary regions on the protein surface, including Lys113, Arg114, Lys131, His136, indicating the physical plausibility of the designs.

\section{Small Molecule Designability}

Small molecules represent one of the most challenging modalities in multimodal molecular design~\citep{Zhang2025_ChemRev_GNNReview}. In contrast to the linear sequences of proteins and nucleic acids, small molecules possess far more complex topologies, including ring systems, branched chains, chiral centers, and diverse functional groups~\citep{Bano2023_Elsevier_DrugDiscoveryChallenges}. This structural irregularity and diversity impose greater demands on the model’s learning capacity. From an application perspective, small molecules remain the dominant class of therapeutics due to their favorable drug-like properties—high oral bioavailability, structural stability, and controllable metabolic half-lives~\citep{Lipinski2004_DDT_Tech_Ro5}. Statistically, more than 75\% of FDA-approved drugs are small molecules~\citep{Zheng2024_STTT_TargetedApprovals}. Although AI-based small-molecule design methods have achieved meaningful progress, most of these models are trained specifically for small-molecule data~\citep{GomezBombarelli2018_ACSCS_VAE}. Moreover, the scarcity of experimental binding data makes these models use synthetic or physics-derived datasets for model training~\citep{Francoeur2020_JCIM_CrossDocked3D}.

\begin{figure}[htbp]
  \centering
  \includegraphics[width=0.95\textwidth, trim=00 220 00 00, clip]{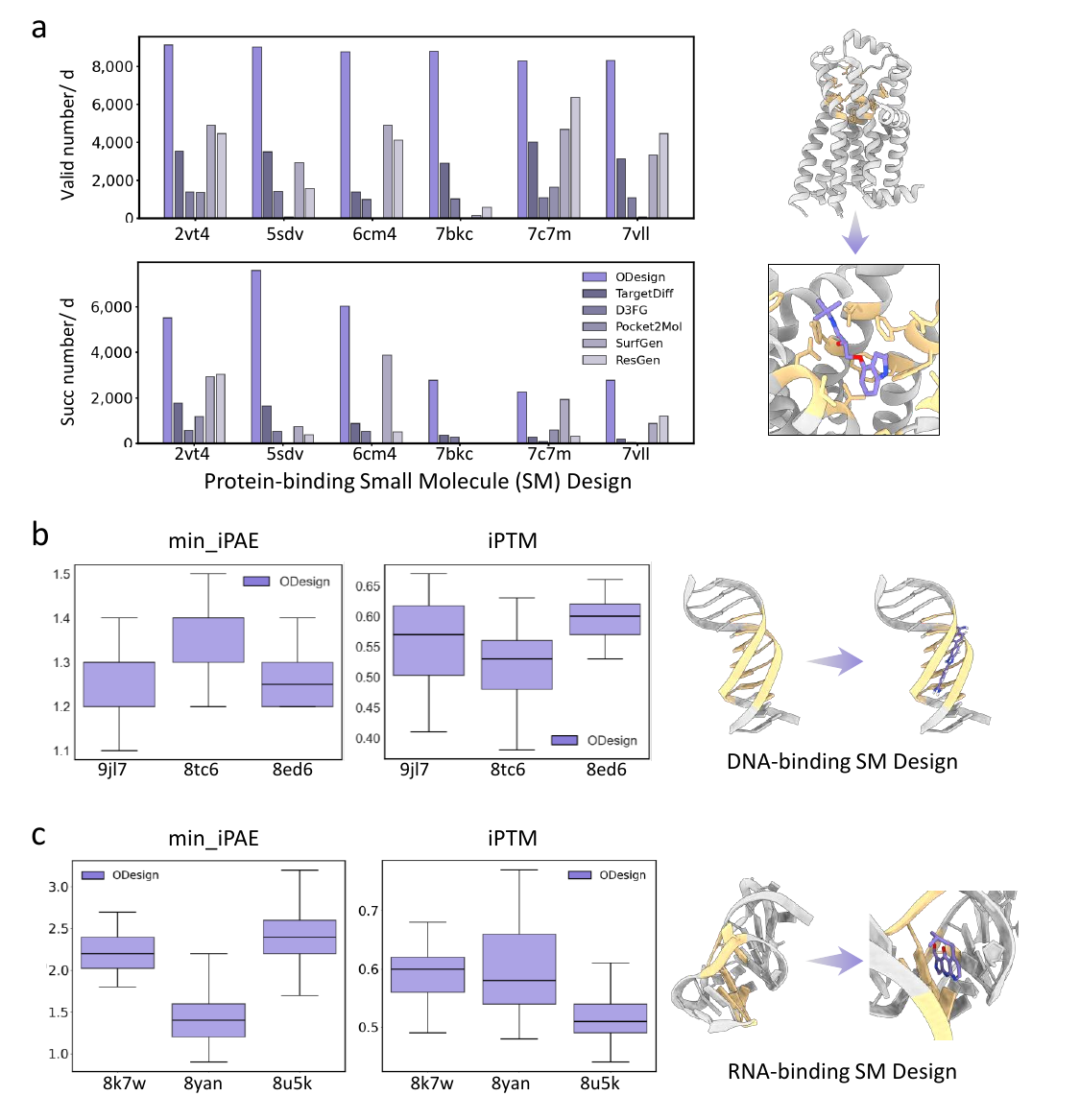}
  \caption{
  Performance of ODesign on ligand-centric benchmarks. All panels: designs are shown in purple, targets in pink, and ligands in green (carbon in green, nitrogen in blue, oxygen in red, sulfur in orange). (a) Protein-binding small-molecule design. ODesign was compared with TargetDiff, D3FG, Pocket2Mol, SurfGen, and ResGen in terms of the number of valid and successful designs generated per GPU per day. (b) DNA-binding small-molecule design. Distributions of min\_iPAE and iPTM across three representative cases are shown. (c) RNA-binding small-molecule design. The same metrics as in (b) are shown.
  }
  \label{fig:ligand_designability}
\end{figure}

\subsection{Protein-binding Small Molecule Design}

Proteins are the primary targets in modern drug discovery, with approximately 90\% of approved small-molecule drugs acting on protein targets~\citep{Santos2017_NatRevDrugDisc_DrugTargets}. Because the protein–small-molecule system is the most mature and extensively studied, most AI-based small-molecule design methods have been developed for protein targets, including diffusion-based approaches~\citep{Hoogeboom2022_ICML_E3Diff} and autoregressive approaches~\citep{Luo2021_NeurIPS_SBDD3DGen}. In this study, representative methods from these two categories, including TargetDiff~\citep{guan20233d_targetdiff}, D3FG~\citep{Lin2024_NeurIPS_FGDiff}, ResGen~\citep{Zhang2023_NMI_ResGen}, SurfGen~\citep{Zhang2023_NatCompSci_TopSurf}, and Pkt2Mol~\citep{Peng2022_ICML_Pocket2Mol} were selected as benchmark methods. For benchmarking, two therapeutic targets from previous studies, DRD2~\citep{Wang2018_Nature_D2DR} and ADRB1~\citep{Nicoulina2010_GenTesting_ADRB1}, along with four ligand-binding protein cases from the protein design task, were used. For each target, following CBGBench~\citep{lin2024cbgbench} and previous evaluation protocol~\citep{Lin2022_arXiv_DiffBP,Luo2021_NeurIPS_SBDD3DGen}, 100 candidate molecules were generated, and two metrics were computed: (1) validity, whether the generated molecules could be recognized by RDKit~\citep{Landrum2013_RDKit} and successfully docked; and (2) successful designs, which include both docking energy (Vina Dock score) and a drug-likeness constraint of QED$>$0.25~\citep{Clark2000_DDT_DrugLikeness}. As shown in Figure~\ref{fig:ligand_designability}a, ODesign outperformed all other methods across all six targets, producing the highest number of valid and successful molecules per day. For example, it exceeded the diffusion-based TargetDiff by 3.9-fold in valid molecule count and 9.0-fold in successful molecule count, and outperformed the autoregressive model SurfGen by 14.0-fold and 48.4-fold, respectively, demonstrating clear advantages in both chemical validity and binding throughput.

\subsection{DNA-binding Small Molecule Design}

Small-molecule drugs targeting DNA occupy a unique niche in therapeutic development~\citep{Hurley2002_NatRevCancer_DNA}. For double-stranded DNA targets within the cell nucleus, small molecules remain among the most feasible intervention strategies owing to their compact size and high cell permeability. Such mechanisms are commonly exploited by antibiotics and anticancer agents~\citep{Sheng2013_MedResRev_DNA_Targeting}. However, structure-based small-molecule design for DNA faces severe data sparsity. For example, in the PDBBind dataset, nucleic acid–small-molecule complexes number only 234 entries, which is less than 1\% of protein–small-molecule complexes. No specialized models are available for this task~\citep{Liu2015_Bioinformatics_PDBbind}. In this context, ODesign is among the first models capable of designing small molecules that target DNA structures. To benchmark ODesign, DNA–small-molecule complexes released after January 13, 2023, were collected from the PDB and AF3 was used as the refolding model. As shown in Figure~\ref{fig:ligand_designability}b, ODesign-generated ligands formed high-confidence interfaces with their DNA targets, with mean min-iPAE of 1.29 and mean iPTM of 0.55, supporting the physical plausibility of the designs. A representative case involves a DNA duplex and its small-molecule binder DB1476. In the native 8ED6 complex, DB1476 binds within the minor groove of DNA through aromatic stacking and guanidinium hydrogen bonds~\citep{Ogbonna2023_ACS_BMC_Au_ATselectivity}. ODesign-generated ligands reproduced the same binding mode while introducing substituent and geometric refinements that better fit the groove contours, further validating the model’s design capability.

\subsection{RNA-binding Small Molecule Design}

Unlike DNA, RNA typically exists as a single strand that can fold into diverse, highly flexible secondary and tertiary structures, providing numerous potential binding pockets for small molecules~\citep{Disney2019_JACS_TargetingRNA}. Clinically, several small-molecule drugs targeting RNA have been successful, for instance, antibiotics that act on ribosomal function and the splicing modulator risdiplam~\citep{Sivaramakrishnan2017_NatComm_SMN2}. Nevertheless, RNA–small-molecule design remains highly challenging for two main reasons: (1) high-resolution RNA–ligand complex structures are extremely limited~\citep{Warner2018_NatRevDrugDisc_RNAprinciples}, and (2) RNA’s structural flexibility and conformational heterogeneity make accurate modeling difficult~\citep{Costales2020_JMedChem_TargetingRNA}. In this context, ODesign is the first model capable of generating small molecules targeting RNA structures. To evaluate the model’s capability, three RNA–small-molecule complexes released after January 13, 2023 were selected, and AF3 was used as the refolding model. As illustrated in Figure~\ref{fig:ligand_designability}c, ODesign generated valid ligands across multiple RNA targets, achieving a mean min-iPAE of 2.0 and a mean iPTM of 0.57. The riboswitch design case is illustrated in Figure~\ref{fig:ligand_designability}c. In this case, the RNA target folds into a complex single-stranded pocket where the native ligand binds through aromatic insertion and stacking interactions. ODesign-generated ligands preserved this binding mode while exhibiting strong adaptability to the highly flexible RNA conformations and forming additional interfacial interactions, demonstrating the model’s ability to address flexible RNA targets.

\section{Conclusions}
ODesign can be viewed as a world model for the molecular universe: a unified foundation model that learns cross-modality state representations and supports controllable generation under multiple biological conditions. Through multimodal and multitask training, ODesign integrates information across proteins, nucleic acids, and small molecules, enabling internal simulation of design interventions and improving few-shot and zero-shot performance, such as protein-binding nucleic acid design and nucleic-acid-binding small-molecule design. Across eleven benchmark tasks, it consistently outperforms modality-specific models and achieves up to two to four orders of magnitude higher design throughput. With ODesign, researchers can specify epitopes on arbitrary targets and design molecules across modalities in a fully controllable manner. Although ODesign demonstrates superior performance across multiple molecular design tasks, certain limitations remain. Its performance still declines under some out-of-distribution conditions; for instance, in DNA aptamer design the model’s designability decreases more rapidly with sequence length than in RNA aptamer design. It also lacks fine-grained semantic control over properties such as protein secondary-structure composition or small-molecule aromaticity. Looking ahead, ODesign can be further advanced toward a self-evolving molecular world model, capable of autonomously improving through in silico evolution guided by its own simulation results. By integrating physics-informed constraints and drug-likeness priors, and coupling with LLM-based reasoning agents to perform imagination rollouts and closed-loop design–evaluation–update cycles, ODesign could evolve from a powerful cross-modal generator into an autonomous, reasoning-capable molecular world model, laying the foundation for the next generation of intelligent molecular design.

\section*{Acknowledgements}
We thank David Baker and Gaurav Bhardwaj for his valuable advice and inspiration that greatly influenced the development of this project. We thank Chunbin Gu, Zijun Gao, Duo An, Hao Wu, Xinheng He, Shizhuo Cheng, Meihui Song, and Kezhi Fu for their valuable discussions and insightful feedback. 

\section*{Author Contributions Statement}
\begin{itemize}
    \item \textbfr{Conceptualization and Project Leadership:} \\Odin Zhang conceived the research idea, led the overall project, and, together with Xujun Zhang, designed the initial model architecture. Shuangjia Zheng contributed to the conceptual idea, provided foundational resources and strategic guidance throughout the research. Tingjun Hou advised on small-molecule modeling, while Siqi Sun guided the nucleic acid design. Pheng Ann Heng advised on the machine learning architecture. Chang-yu Hsieh and Xiaoyu Chen helped refine the scientific concepts and identify potential applications. Lei Bai provided critical training infrastructure, computational resources, and advice on machine learning algorithms.
    \item \textbfr{Machine Learning Development:}\\
Odin Zhang, Xujun Zhang, and Haitao Lin developed the main architecture. Cheng Tan led the inverse folding module; Qinghan Wang oversaw the protein design tasks; Yuanle Mo supervised small-molecule design; and Qiantai Feng established the ODesign benchmarking framework. Yuntao Yu led the nucleic acid design tasks, while Zichang Jin and Ziyi You contributed to the early-stage model development. Yijie Zhang further optimized the inverse folding module. Yunpeng Xia, Yuyang Tao, Xiaoyu Chen, Weibo Zhao, Kejun Ying, Runze Ma, Chenqing Hua, Bo Qiang, and Jiaqi Wang participated in the development of ODesign v1. Peicong Lin managed the large-scale training and integration of the universal model.
    \item \textbfr{Experimental and Applied Design:}
Gang Du led the in silico design efforts. Shicheng Chen, Peichen Pan and Jian Zhang contributed to real-world design campaigns and experimental validations.
\end{itemize}

\section*{Authors and Affiliations} \label{sec:team}
Odin Zhang$^{1,3,5,\star,\dag}$, 
Xujun Zhang$^{1,3,\star}$, 
Haitao Lin$^{1,5,\star}$, 
Cheng Tan$^{5,6,\star}$, 
Qinghan Wang$^{1,3,\star}$, 
Yuanle Mo$^{1,5,\star}$, 
Qiantai Feng$^{4,6,\star}$, 
Gang Du$^{1,3}$, 
Yuntao Yu$^{1,3}$, 
Zichang Jin$^{1,3}$, 
Ziyi You$^{1,3}$, 
Peicong Lin$^{1}$, 
Yijie Zhang$^{7}$, 
Yuyang Tao$^{1}$, 
Shicheng Chen$^{3}$, 
Jack Xiaoyu Chen$^{8}$, 
Chenqing Hua$^{7}$, 
Weibo Zhao$^{5}$, 
Runze Ma$^{1,2}$, 
Yunpeng Xia$^{1}$, 
Kejun Ying$^{9}$, 
Jun Li$^{1}$, 
Yundian Zeng$^{3}$, 
Lijun Lang$^{5}$, 
Peichen Pan$^{3}$, 
Hanqun Cao$^{5}$, 
Zihao Song$^{10}$, 
Bo Qiang$^{10}$, 
Jiaqi Wang$^{10}$, 
Pengfei Ji$^{11}$, 
Lei Bai$^{6}$, 
Jian Zhang$^{12}$, 
Chang-yu Hsieh$^{3}$, 
Pheng Ann Heng$^{5,\dag}$,
Siqi Sun$^{6,\dag}$, 
Tingjun Hou$^{3,\dag}$, 
Shuangjia Zheng$^{2,1,\dag}$.

\textsuperscript{1}Lingang Laboratory, Shanghai, China\\
\textsuperscript{2}Global Institute of Future Technology, Shanghai Jiao Tong University, Shanghai, China\\
\textsuperscript{3}College of Pharmaceutical Sciences, Zhejiang University, Zhejiang, China\\
\textsuperscript{4}Research Institute of Intelligent Complex Systems, Fudan University, Shanghai, China\\
\textsuperscript{5}Department of Computer Science and Engineering, The Chinese University of Hong Kong, Hong Kong, China\\
\textsuperscript{6}Shanghai Artificial Intelligence Laboratory, Shanghai, China\\
\textsuperscript{7}McGill University, Montreal, Canada\\
\textsuperscript{8}Institute for Medical Engineering Science and Department of Biological Engineering, Massachusetts Institute of Technology, Cambridge, MA, USA\\
\textsuperscript{9}T. H. Chan School of Public Health, Harvard University, Boston, MA, USA\\
\textsuperscript{10}Department of Biochemistry, University of Washington, Seattle, WA, USA\\
\textsuperscript{11}Institute of Catalysis, Department of Chemistry, Zhejiang University, Hangzhou, China\\
\textsuperscript{12}School of Medicine, Shanghai Jiao Tong University, Shanghai, China

\section*{Competing Interests Statement}
There are no conflicts to declare. 

\bibliographystyle{icml2025}
\bibliography{odesign_ref}

\begin{thebibliography}{111}
\providecommand{\natexlab}[1]{#1}
\providecommand{\url}[1]{\texttt{#1}}
\expandafter\ifx\csname urlstyle\endcsname\relax
  \providecommand{\doi}[1]{doi: #1}\else
  \providecommand{\doi}{doi: \begingroup \urlstyle{rm}\Url}\fi

\bibitem[Abramson et~al.(2024)Abramson, Adler, Dunger, et~al.]{Abramson2024_Nature_AF3}
Abramson, J., Adler, J., Dunger, J., et~al.
\newblock Accurate structure prediction of biomolecular interactions with {AlphaFold 3}.
\newblock \emph{Nature}, 630:\penalty0 493--500, 2024.

\bibitem[Ahdritz et~al.(2024)Ahdritz, Bouatta, Floristean, Kadyan, Xia, Gerecke, O’Donnell, Berenberg, Fisk, Zanichelli, et~al.]{ahdritz2024openfold}
Ahdritz, G., Bouatta, N., Floristean, C., Kadyan, S., Xia, Q., Gerecke, W., O’Donnell, T.~J., Berenberg, D., Fisk, I., Zanichelli, N., et~al.
\newblock Openfold: Retraining alphafold2 yields new insights into its learning mechanisms and capacity for generalization.
\newblock \emph{Nature methods}, 21\penalty0 (8):\penalty0 1514--1524, 2024.

\bibitem[Ahern et~al.(2025)Ahern, Yim, Tischer, et~al.]{Ahern2025_bioRxiv_RFDiffusion2}
Ahern, W., Yim, J., Tischer, D., et~al.
\newblock Atom level enzyme active site scaffolding using rfdiffusion2.
\newblock \emph{bioRxiv}, pp.\  2025.04.09.648075, 2025.

\bibitem[Anand et~al.(2024)Anand, Joshi, Morehead, et~al.]{Anand2024_arXiv_RNAFlowMatching}
Anand, R., Joshi, C., Morehead, A., et~al.
\newblock Flow matching for de novo 3d rna backbone design.
\newblock \emph{arXiv preprint arXiv:2406.13839}, 2024.

\bibitem[Anand et~al.(2025)Anand, Joshi, Morehead, Jamasb, Harris, Mathis, Didi, Ying, Hooi, and Li{\`o}]{anand2025rna_rnaframeflow}
Anand, R., Joshi, C.~K., Morehead, A., Jamasb, A.~R., Harris, C., Mathis, S.~V., Didi, K., Ying, R., Hooi, B., and Li{\`o}, P.
\newblock Rna-frameflow: Flow matching for de novo 3d rna backbone design.
\newblock \emph{ArXiv}, pp.\  arXiv--2406, 2025.

\bibitem[Bano et~al.(2023)Bano, Butt, and Mohsan]{Bano2023_Elsevier_DrugDiscoveryChallenges}
Bano, I., Butt, U.~D., and Mohsan, S. A.~H.
\newblock New challenges in drug discovery.
\newblock In \emph{Novel Platforms for Drug Delivery Applications}, pp.\  619--643. Elsevier, 2023.

\bibitem[Barab{\'a}si \& Oltvai(2004)Barab{\'a}si and Oltvai]{Barabasi2004_NRG}
Barab{\'a}si, A.-L. and Oltvai, Z.~N.
\newblock Network biology: understanding the cell's functional organization.
\newblock \emph{Nature Reviews Genetics}, 5\penalty0 (2):\penalty0 101--113, 2004.

\bibitem[Breaker(1997)]{Breaker1997_COChBiol_DNAAptamers}
Breaker, R.~R.
\newblock Dna aptamers and dna enzymes.
\newblock \emph{Current Opinion in Chemical Biology}, 1\penalty0 (1):\penalty0 26--31, 1997.

\bibitem[Brini et~al.(2020)Brini, Simmerling, and Dill]{Brini2020_Science}
Brini, E., Simmerling, C., and Dill, K.~A.
\newblock Protein storytelling through physics.
\newblock \emph{Science}, 370\penalty0 (6520):\penalty0 eaaz3041, 2020.

\bibitem[Burley et~al.(2017)Burley, Berman, Kleywegt, et~al.]{Burley2017_PDB}
Burley, S.~K., Berman, H.~M., Kleywegt, G.~J., et~al.
\newblock Protein data bank (pdb): the single global macromolecular structure archive.
\newblock In \emph{Protein Crystallography}, volume 1607, pp.\  627--641. 2017.

\bibitem[Butcher et~al.(2025)Butcher, Krishna, Mitra, Brent, Li, Corley, Kim, Funk, Mathis, Salike, et~al.]{butcher2025novo_rfdiffusion3}
Butcher, J. K.~V., Krishna, R., Mitra, R., Brent, R.~I., Li, Y., Corley, N., Kim, P., Funk, J., Mathis, S.~V., Salike, S., et~al.
\newblock De novo design of all-atom biomolecular interactions with rfdiffusion3.
\newblock \emph{bioRxiv}, 2025.

\bibitem[Buttenschoen et~al.(2024)Buttenschoen, Morris, and Deane]{buttenschoen2024posebusters}
Buttenschoen, M., Morris, G.~M., and Deane, C.~M.
\newblock Posebusters: Ai-based docking methods fail to generate physically valid poses or generalise to novel sequences.
\newblock \emph{Chemical Science}, 15\penalty0 (9):\penalty0 3130--3139, 2024.

\bibitem[Campbell et~al.(2024)Campbell, Yim, Barzilay, et~al.]{Campbell2024_arXiv_Multiflow}
Campbell, A., Yim, J., Barzilay, R., et~al.
\newblock Generative flows on discrete state-spaces: Enabling multimodal flows with applications to protein co-design.
\newblock \emph{arXiv preprint arXiv:2402.04997}, 2024.

\bibitem[Cao et~al.(2022)Cao, Coventry, Goreshnik, et~al.]{Cao2022_Nature_TargetOnly}
Cao, L., Coventry, B., Goreshnik, I., et~al.
\newblock Design of protein-binding proteins from the target structure alone.
\newblock \emph{Nature}, 605\penalty0 (7910):\penalty0 551--560, 2022.

\bibitem[Carter \& Rajpal(2022)Carter and Rajpal]{Carter2022_Cell_Antibodies}
Carter, P.~J. and Rajpal, A.
\newblock Designing antibodies as therapeutics.
\newblock \emph{Cell}, 185\penalty0 (15):\penalty0 2789--2805, 2022.

\bibitem[Chen et~al.(2009)Chen, Georgiev, Anderson, et~al.]{Chen2009_PNAS_EnzymeRedesign}
Chen, C.-Y., Georgiev, I., Anderson, A.~C., et~al.
\newblock Computational structure-based redesign of enzyme activity.
\newblock \emph{Proceedings of the National Academy of Sciences}, 106\penalty0 (10):\penalty0 3764--3769, 2009.

\bibitem[Cho et~al.(2025)Cho, Pacesa, Zhang, et~al.]{Cho2025_bioRxiv_Boltzdesign1}
Cho, Y., Pacesa, M., Zhang, Z., et~al.
\newblock Boltzdesign1: Inverting all-atom structure prediction model for generalized biomolecular binder design.
\newblock \emph{bioRxiv}, pp.\  2025.04.06.647261, 2025.

\bibitem[Clark \& Pickett(2000)Clark and Pickett]{Clark2000_DDT_DrugLikeness}
Clark, D.~E. and Pickett, S.~D.
\newblock Computational methods for the prediction of `drug-likeness'.
\newblock \emph{Drug Discovery Today}, 5\penalty0 (2):\penalty0 49--58, 2000.

\bibitem[Costales et~al.(2020)Costales, Childs-Disney, Haniff, et~al.]{Costales2020_JMedChem_TargetingRNA}
Costales, M.~G., Childs-Disney, J.~L., Haniff, H.~S., et~al.
\newblock How we think about targeting rna with small molecules.
\newblock \emph{Journal of Medicinal Chemistry}, 63\penalty0 (17):\penalty0 8880--8900, 2020.

\bibitem[Darmostuk et~al.(2015)Darmostuk, Rimpelova, Gbelcova, et~al.]{Darmostuk2015_BiotechAdv_SELEXUpdate}
Darmostuk, M., Rimpelova, S., Gbelcova, H., et~al.
\newblock Current approaches in selex: An update to aptamer selection technology.
\newblock \emph{Biotechnology Advances}, 33\penalty0 (6):\penalty0 1141--1161, 2015.

\bibitem[Dauparas et~al.(2022)Dauparas, Anishchenko, Bennett, Bai, Ragotte, Milles, Wicky, Courbet, de~Haas, Bethel, et~al.]{dauparas2022robust_proteinmpnn}
Dauparas, J., Anishchenko, I., Bennett, N., Bai, H., Ragotte, R.~J., Milles, L.~F., Wicky, B.~I., Courbet, A., de~Haas, R.~J., Bethel, N., et~al.
\newblock Robust deep learning--based protein sequence design using proteinmpnn.
\newblock \emph{Science}, 378\penalty0 (6615):\penalty0 49--56, 2022.

\bibitem[Dauparas et~al.(2025)Dauparas, Lee, Pecoraro, An, Anishchenko, Glasscock, and Baker]{dauparas2025atomic_ligandmpnn}
Dauparas, J., Lee, G.~R., Pecoraro, R., An, L., Anishchenko, I., Glasscock, C., and Baker, D.
\newblock Atomic context-conditioned protein sequence design using ligandmpnn.
\newblock \emph{Nature Methods}, pp.\  1--7, 2025.

\bibitem[Disney(2019)]{Disney2019_JACS_TargetingRNA}
Disney, M.~D.
\newblock Targeting rna with small molecules to capture opportunities at the intersection of chemistry, biology, and medicine.
\newblock \emph{Journal of the American Chemical Society}, 141\penalty0 (17):\penalty0 6776--6790, 2019.

\bibitem[Drubin et~al.(2007)Drubin, Way, and Silver]{Drubin2007_GenesDev_DesigningSystems}
Drubin, D.~A., Way, J.~C., and Silver, P.~A.
\newblock Designing biological systems.
\newblock \emph{Genes \& Development}, 21\penalty0 (3):\penalty0 242--254, 2007.

\bibitem[Dunn et~al.(2017)Dunn, Jimenez, and Chaput]{Dunn2017_NatRevChem_AptamerDiscovery}
Dunn, M.~R., Jimenez, R.~M., and Chaput, J.~C.
\newblock Analysis of aptamer discovery and technology.
\newblock \emph{Nature Reviews Chemistry}, 1\penalty0 (10):\penalty0 0076, 2017.

\bibitem[Eberhardt et~al.(2021)Eberhardt, Santos-Martins, Tillack, and Forli]{eberhardt2021autodock_vina}
Eberhardt, J., Santos-Martins, D., Tillack, A.~F., and Forli, S.
\newblock Autodock vina 1.2. 0: new docking methods, expanded force field, and python bindings.
\newblock \emph{Journal of chemical information and modeling}, 61\penalty0 (8):\penalty0 3891--3898, 2021.

\bibitem[Ellington \& Szostak(1990)Ellington and Szostak]{Ellington1990_Nature_InVitroSelection}
Ellington, A.~D. and Szostak, J.~W.
\newblock In vitro selection of rna molecules that bind specific ligands.
\newblock \emph{Nature}, 346\penalty0 (6287):\penalty0 818--822, 1990.

\bibitem[Eriksen et~al.(2014)Eriksen, Lian, and Zhao]{Eriksen2014_JSB_PathwayEngineering}
Eriksen, D.~T., Lian, J., and Zhao, H.
\newblock Protein design for pathway engineering.
\newblock \emph{Journal of Structural Biology}, 185\penalty0 (2):\penalty0 234--242, 2014.

\bibitem[Francoeur et~al.(2020)Francoeur, Masuda, Sunseri, et~al.]{Francoeur2020_JCIM_CrossDocked3D}
Francoeur, P.~G., Masuda, T., Sunseri, J., et~al.
\newblock Three-dimensional convolutional neural networks and a cross-docked data set for structure-based drug design.
\newblock \emph{Journal of Chemical Information and Modeling}, 60\penalty0 (9):\penalty0 4200--4215, 2020.

\bibitem[Gainza et~al.(2020)Gainza, S{\"v}errisson, Monti, et~al.]{Gainza2020_NMeth_GDLsurfaces}
Gainza, P., S{\"v}errisson, F., Monti, F., et~al.
\newblock Deciphering interaction fingerprints from protein molecular surfaces using geometric deep learning.
\newblock \emph{Nature Methods}, 17\penalty0 (2):\penalty0 184--192, 2020.

\bibitem[Geffner et~al.(2025)Geffner, Didi, Zhang, et~al.]{Geffner2025_arXiv_Proteina}
Geffner, T., Didi, K., Zhang, Z., et~al.
\newblock Proteina: Scaling flow-based protein structure generative models.
\newblock \emph{arXiv preprint arXiv:2503.00710}, 2025.

\bibitem[Gl{\"o}gl et~al.(2024)Gl{\"o}gl, Krishnakumar, Ragotte, et~al.]{Gloegl2024_Science_TNFR}
Gl{\"o}gl, M., Krishnakumar, A., Ragotte, R.~J., et~al.
\newblock Target-conditioned diffusion generates potent tnfr superfamily antagonists and agonists.
\newblock \emph{Science}, 386\penalty0 (6726):\penalty0 1154--1161, 2024.

\bibitem[G{\'o}mez-Bombarelli et~al.(2018)G{\'o}mez-Bombarelli, Wei, Duvenaud, et~al.]{GomezBombarelli2018_ACSCS_VAE}
G{\'o}mez-Bombarelli, R., Wei, J.~N., Duvenaud, D., et~al.
\newblock Automatic chemical design using a data-driven continuous representation of molecules.
\newblock \emph{ACS Central Science}, 4\penalty0 (2):\penalty0 268--276, 2018.

\bibitem[Guan et~al.(2023)Guan, Qian, Peng, Su, Peng, and Ma]{guan20233d_targetdiff}
Guan, J., Qian, W.~W., Peng, X., Su, Y., Peng, J., and Ma, J.
\newblock 3d equivariant diffusion for target-aware molecule generation and affinity prediction.
\newblock \emph{arXiv preprint arXiv:2303.03543}, 2023.

\bibitem[Hayes et~al.(2025)Hayes, Rao, Akin, Sofroniew, Oktay, Lin, Verkuil, Tran, Deaton, Wiggert, et~al.]{hayes2025simulating_esm3}
Hayes, T., Rao, R., Akin, H., Sofroniew, N.~J., Oktay, D., Lin, Z., Verkuil, R., Tran, V.~Q., Deaton, J., Wiggert, M., et~al.
\newblock Simulating 500 million years of evolution with a language model.
\newblock \emph{Science}, 387\penalty0 (6736):\penalty0 850--858, 2025.

\bibitem[Ho et~al.(2020)Ho, Jain, and Abbeel]{ho2020denoising_ddpm}
Ho, J., Jain, A., and Abbeel, P.
\newblock Denoising diffusion probabilistic models.
\newblock \emph{Advances in neural information processing systems}, 33:\penalty0 6840--6851, 2020.

\bibitem[Hoogeboom et~al.(2022)Hoogeboom, Satorras, Vignac, et~al.]{Hoogeboom2022_ICML_E3Diff}
Hoogeboom, E., Satorras, V.~G., Vignac, C., et~al.
\newblock Equivariant diffusion for molecule generation in 3d.
\newblock In \emph{Proceedings of the International Conference on Machine Learning}. PMLR, 2022.

\bibitem[Huang et~al.(2016)Huang, Boyken, and Baker]{Huang2016_Nature}
Huang, P.-S., Boyken, S.~E., and Baker, D.
\newblock The coming of age of de novo protein design.
\newblock \emph{Nature}, 537\penalty0 (7620):\penalty0 320--327, 2016.

\bibitem[Huizenga \& Szostak(1995)Huizenga and Szostak]{Huizenga1995_Biochemistry_ADNA}
Huizenga, D.~E. and Szostak, J.~W.
\newblock A dna aptamer that binds adenosine and atp.
\newblock \emph{Biochemistry}, 34\penalty0 (2):\penalty0 656--665, 1995.

\bibitem[Hurley(2002)]{Hurley2002_NatRevCancer_DNA}
Hurley, L.~H.
\newblock Dna and its associated processes as targets for cancer therapy.
\newblock \emph{Nature Reviews Cancer}, 2\penalty0 (3):\penalty0 188--200, 2002.

\bibitem[Ingraham et~al.(2023)Ingraham, Baranov, Costello, Barber, Wang, Ismail, Frappier, Lord, Ng-Thow-Hing, Van~Vlack, et~al.]{ingraham2023illuminating_chroma}
Ingraham, J.~B., Baranov, M., Costello, Z., Barber, K.~W., Wang, W., Ismail, A., Frappier, V., Lord, D.~M., Ng-Thow-Hing, C., Van~Vlack, E.~R., et~al.
\newblock Illuminating protein space with a programmable generative model.
\newblock \emph{Nature}, 623\penalty0 (7989):\penalty0 1070--1078, 2023.

\bibitem[Jayaraman et~al.(2020)Jayaraman, Mellody, Hou, et~al.]{Jayaraman2020_EBioMedicine_CART}
Jayaraman, J., Mellody, M.~P., Hou, A.~J., et~al.
\newblock Car-t design: Elements and their synergistic function.
\newblock \emph{EBioMedicine}, 58, 2020.

\bibitem[Jin et~al.(2020)Jin, Barzilay, and Jaakkola]{Jin2020_ICML_Motif}
Jin, W., Barzilay, R., and Jaakkola, T.
\newblock Hierarchical generation of molecular graphs using structural motifs.
\newblock In \emph{Proceedings of the International Conference on Machine Learning}. PMLR, 2020.

\bibitem[J{\o}rgensen et~al.(2018)J{\o}rgensen, Schmidt, and Winther]{Jorgensen2018_MolInform}
J{\o}rgensen, P.~B., Schmidt, M.~N., and Winther, O.
\newblock Deep generative models for molecular science.
\newblock \emph{Molecular Informatics}, 37\penalty0 (1-2):\penalty0 1700133, 2018.

\bibitem[Joshi \& Li{\`o}(2024)Joshi and Li{\`o}]{Joshi2024_Methods_gRNAde}
Joshi, C.~K. and Li{\`o}, P.
\newblock grnade: a geometric deep learning pipeline for 3d rna inverse design.
\newblock In \emph{RNA Design: Methods and Protocols}, pp.\  121--135. Springer, 2024.

\bibitem[Joshi \& Kumar(2021)Joshi and Kumar]{Joshi2021_Molecules_AI_MolDesign}
Joshi, R.~P. and Kumar, N.
\newblock Artificial intelligence for autonomous molecular design: A perspective.
\newblock \emph{Molecules}, 26\penalty0 (22):\penalty0 6761, 2021.

\bibitem[Jumper et~al.(2021)Jumper, Evans, Pritzel, et~al.]{Jumper2021_Nature_AF2}
Jumper, J., Evans, R., Pritzel, A., et~al.
\newblock Highly accurate protein structure prediction with alphafold.
\newblock \emph{Nature}, 596\penalty0 (7873):\penalty0 583--589, 2021.

\bibitem[Karp(2009)]{Karp2009}
Karp, G.
\newblock \emph{Cell and Molecular Biology: Concepts and Experiments}.
\newblock John Wiley \& Sons, 2009.

\bibitem[Karras et~al.(2022)Karras, Aittala, Aila, and Laine]{karras2022elucidating_edm}
Karras, T., Aittala, M., Aila, T., and Laine, S.
\newblock Elucidating the design space of diffusion-based generative models.
\newblock \emph{Advances in neural information processing systems}, 35:\penalty0 26565--26577, 2022.

\bibitem[Khmelinskaia(2024)]{Khmelinskaia2024_Nature_FiveTasks}
Khmelinskaia, A.
\newblock Five tasks that still challenge protein designers.
\newblock \emph{Nature}, 635:\penalty0 7, 2024.

\bibitem[Kitano(2002)]{Kitano2002_Science_SystemsBio}
Kitano, H.
\newblock Systems biology: a brief overview.
\newblock \emph{Science}, 295\penalty0 (5560):\penalty0 1662--1664, 2002.

\bibitem[Kortemme(2024)]{Kortemme2024_Cell_Review}
Kortemme, T.
\newblock De novo protein design---from new structures to programmable functions.
\newblock \emph{Cell}, 187\penalty0 (3):\penalty0 526--544, 2024.

\bibitem[Krishna et~al.(2024)Krishna, Wang, Ahern, et~al.]{Krishna2024_Science_RFAA}
Krishna, R., Wang, J., Ahern, W., et~al.
\newblock Generalized biomolecular modeling and design with {RoseTTAFold All-Atom}.
\newblock \emph{Science}, 384\penalty0 (6693):\penalty0 eadl2528, 2024.

\bibitem[Landrum(2013)]{Landrum2013_RDKit}
Landrum, G.
\newblock Rdkit: A software suite for cheminformatics, computational chemistry, and predictive modeling, 2013.
\newblock Available at \url{http://www.rdkit.org}.

\bibitem[Li et~al.(2024)Li, Liu, Bai, et~al.]{Li2024_JAFC_ThermostableScreening}
Li, L., Liu, X., Bai, Y., et~al.
\newblock High-throughput screening techniques for the selection of thermostable enzymes.
\newblock \emph{Journal of Agricultural and Food Chemistry}, 72\penalty0 (8):\penalty0 3833--3845, 2024.

\bibitem[Lin et~al.(2022{\natexlab{a}})Lin, Huang, Liu, et~al.]{Lin2022_arXiv_DiffBP}
Lin, H., Huang, Y., Liu, M., et~al.
\newblock Diffbp: Generative diffusion of 3d molecules for target protein binding.
\newblock \emph{arXiv preprint arXiv:2211.11214}, 2022{\natexlab{a}}.

\bibitem[Lin et~al.(2024{\natexlab{a}})Lin, Huang, Zhang, et~al.]{Lin2024_NeurIPS_FGDiff}
Lin, H., Huang, Y., Zhang, O., et~al.
\newblock Functional-group-based diffusion for pocket-specific molecule generation and elaboration.
\newblock \emph{Advances in Neural Information Processing Systems}, 36:\penalty0 34603--34626, 2024{\natexlab{a}}.

\bibitem[Lin et~al.(2025)Lin, Zhao, Zhang, Huang, Wu, Tan, Liu, Gao, and Li]{lin2024cbgbench}
Lin, H., Zhao, G., Zhang, O., Huang, Y., Wu, L., Tan, C., Liu, Z., Gao, Z., and Li, S.~Z.
\newblock {CBGB}ench: Fill in the blank of protein-molecule complex binding graph.
\newblock In \emph{The Thirteenth International Conference on Learning Representations}, 2025.
\newblock URL \url{https://openreview.net/forum?id=mOpNrrV2zH}.

\bibitem[Lin et~al.(2024{\natexlab{b}})Lin, Lee, Zhang, and AlQuraishi]{lin2024out_genie2}
Lin, Y., Lee, M., Zhang, Z., and AlQuraishi, M.
\newblock Out of many, one: Designing and scaffolding proteins at the scale of the structural universe with genie 2.
\newblock \emph{arXiv preprint arXiv:2405.15489}, 2024{\natexlab{b}}.

\bibitem[Lin et~al.(2022{\natexlab{b}})Lin, Akin, Rao, et~al.]{Lin2022_bioRxiv_LMStructure}
Lin, Z., Akin, H., Rao, R., et~al.
\newblock Language models of protein sequences at the scale of evolution enable accurate structure prediction.
\newblock \emph{bioRxiv}, pp.\  2022.05., 2022{\natexlab{b}}.

\bibitem[Lin et~al.(2023)Lin, Akin, Rao, Hie, Zhu, Lu, Smetanin, Verkuil, Kabeli, Shmueli, et~al.]{lin2023evolutionary_esmfold}
Lin, Z., Akin, H., Rao, R., Hie, B., Zhu, Z., Lu, W., Smetanin, N., Verkuil, R., Kabeli, O., Shmueli, Y., et~al.
\newblock Evolutionary-scale prediction of atomic-level protein structure with a language model.
\newblock \emph{Science}, 379\penalty0 (6637):\penalty0 1123--1130, 2023.

\bibitem[Lipinski(2004)]{Lipinski2004_DDT_Tech_Ro5}
Lipinski, C.~A.
\newblock Lead-and drug-like compounds: the rule-of-five revolution.
\newblock \emph{Drug Discovery Today: Technologies}, 1\penalty0 (4):\penalty0 337--341, 2004.

\bibitem[Lippow \& Tidor(2007)Lippow and Tidor]{Lippow2007_COB}
Lippow, S.~M. and Tidor, B.
\newblock Progress in computational protein design.
\newblock \emph{Current Opinion in Biotechnology}, 18\penalty0 (4):\penalty0 305--311, 2007.

\bibitem[Liu et~al.(2015)Liu, Li, Han, et~al.]{Liu2015_Bioinformatics_PDBbind}
Liu, Z., Li, Y., Han, L., et~al.
\newblock Pdb-wide collection of binding data: current status of the pdbbind database.
\newblock \emph{Bioinformatics}, 31\penalty0 (3):\penalty0 405--412, 2015.

\bibitem[Lucas \& Kortemme(2020)Lucas and Kortemme]{Lucas2020_PLoSCB_SmallMolSites}
Lucas, J.~E. and Kortemme, T.
\newblock New computational protein design methods for de novo small molecule binding sites.
\newblock \emph{PLoS Computational Biology}, 16\penalty0 (10):\penalty0 e1008178, 2020.

\bibitem[Luo et~al.(2021)Luo, Guan, Ma, et~al.]{Luo2021_NeurIPS_SBDD3DGen}
Luo, S., Guan, J., Ma, J., et~al.
\newblock A 3d generative model for structure-based drug design.
\newblock \emph{Advances in Neural Information Processing Systems}, 34:\penalty0 6229--6239, 2021.

\bibitem[Mandal et~al.(2009)Mandal, Moudgil, and Mandal]{Mandal2009_EJP}
Mandal, S., Moudgil, M.~N., and Mandal, S.~K.
\newblock Rational drug design.
\newblock \emph{European Journal of Pharmacology}, 625\penalty0 (1--3):\penalty0 90--100, 2009.

\bibitem[Mandell \& Kortemme(2009)Mandell and Kortemme]{Mandell2009_COB_BackboneFlex}
Mandell, D.~J. and Kortemme, T.
\newblock Backbone flexibility in computational protein design.
\newblock \emph{Current Opinion in Biotechnology}, 20\penalty0 (4):\penalty0 420--428, 2009.

\bibitem[Meenan et~al.(2010)Meenan, Sharma, Fleishman, et~al.]{Meenan2010_PNAS_Selectivity}
Meenan, N.~A., Sharma, A., Fleishman, S.~J., et~al.
\newblock The structural and energetic basis for high selectivity in a high-affinity protein-protein interaction.
\newblock \emph{Proceedings of the National Academy of Sciences}, 107\penalty0 (22):\penalty0 10080--10085, 2010.

\bibitem[Ng et~al.(2006)Ng, Shima, Calias, et~al.]{Ng2006_NatRevDrugDisc_Pegaptanib}
Ng, E.~W., Shima, D.~T., Calias, P., et~al.
\newblock Pegaptanib, a targeted anti-vegf aptamer for ocular vascular disease.
\newblock \emph{Nature Reviews Drug Discovery}, 5\penalty0 (2):\penalty0 123--132, 2006.

\bibitem[Nicoulina et~al.(2010)Nicoulina, Shulman, Shesternya, et~al.]{Nicoulina2010_GenTesting_ADRB1}
Nicoulina, S., Shulman, V., Shesternya, P., et~al.
\newblock Association of adrb1 gene polymorphism with atrial fibrillation.
\newblock \emph{Genetic Testing and Molecular Biomarkers}, 14\penalty0 (2):\penalty0 249--253, 2010.

\bibitem[Ogbonna et~al.(2023)Ogbonna, Paul, Farahat, et~al.]{Ogbonna2023_ACS_BMC_Au_ATselectivity}
Ogbonna, E.~N., Paul, A., Farahat, A.~A., et~al.
\newblock X-ray structure characterization of the selective recognition of at base pair sequences.
\newblock \emph{ACS Bio \& Med Chem Au}, 3\penalty0 (4):\penalty0 335--348, 2023.

\bibitem[Pacesa et~al.(2025)Pacesa, Nickel, Schellhaas, et~al.]{Pacesa2025_Nature_BindCraft}
Pacesa, M., Nickel, L., Schellhaas, C., et~al.
\newblock One-shot design of functional protein binders with bindcraft.
\newblock \emph{Nature}, pp.\  1--10, 2025.

\bibitem[Peng et~al.(2022)Peng, Luo, Guan, et~al.]{Peng2022_ICML_Pocket2Mol}
Peng, X., Luo, S., Guan, J., et~al.
\newblock Pocket2mol: Efficient molecular sampling based on 3d protein pockets.
\newblock In \emph{Proceedings of the International Conference on Machine Learning}. PMLR, 2022.

\bibitem[Pinto et~al.(2022)Pinto, Corbella, Demkiv, et~al.]{Pinto2022_TrendsBiochemSci_EnzymeEvolution}
Pinto, G.~P., Corbella, M., Demkiv, A.~O., et~al.
\newblock Exploiting enzyme evolution for computational protein design.
\newblock \emph{Trends in Biochemical Sciences}, 47\penalty0 (5):\penalty0 375--389, 2022.

\bibitem[Qu et~al.(2024{\natexlab{a}})Qu, Guan, Ma, et~al.]{Qu2024_bioRxiv_PAllAtom}
Qu, W., Guan, J., Ma, R., et~al.
\newblock P (all-atom) is unlocking new path for protein design.
\newblock \emph{bioRxiv}, pp.\  2024.08.16.608235, 2024{\natexlab{a}}.

\bibitem[Qu et~al.(2024{\natexlab{b}})Qu, Qiu, Song, Gong, Han, Zheng, Zhou, and Ma]{qu2024molcraft}
Qu, Y., Qiu, K., Song, Y., Gong, J., Han, J., Zheng, M., Zhou, H., and Ma, W.-Y.
\newblock Molcraft: structure-based drug design in continuous parameter space.
\newblock \emph{arXiv preprint arXiv:2404.12141}, 2024{\natexlab{b}}.

\bibitem[Ren et~al.(2025)Ren, Sun, Guan, et~al.]{Ren2025_bioRxiv_PXDesign}
Ren, M., Sun, J., Guan, J., et~al.
\newblock Pxdesign: Fast, modular, and accurate de novo design of protein binders.
\newblock \emph{bioRxiv}, pp.\  2025.08.15.670450, 2025.

\bibitem[Santos et~al.(2017)Santos, Ursu, Gaulton, et~al.]{Santos2017_NatRevDrugDisc_DrugTargets}
Santos, R., Ursu, O., Gaulton, A., et~al.
\newblock A comprehensive map of molecular drug targets.
\newblock \emph{Nature Reviews Drug Discovery}, 16\penalty0 (1):\penalty0 19--34, 2017.

\bibitem[Sheng et~al.(2013)Sheng, Gan, and Huang]{Sheng2013_MedResRev_DNA_Targeting}
Sheng, J., Gan, J., and Huang, Z.
\newblock Structure-based dna-targeting strategies with small molecule ligands for drug discovery.
\newblock \emph{Medicinal Research Reviews}, 33\penalty0 (5):\penalty0 1119--1173, 2013.

\bibitem[Sivaramakrishnan et~al.(2017)Sivaramakrishnan, McCarthy, Campagne, et~al.]{Sivaramakrishnan2017_NatComm_SMN2}
Sivaramakrishnan, M., McCarthy, K.~D., Campagne, S., et~al.
\newblock Binding to smn2 pre-mrna-protein complex elicits specificity for small molecule splicing modifiers.
\newblock \emph{Nature Communications}, 8\penalty0 (1):\penalty0 1476, 2017.

\bibitem[Steinegger \& S{\"o}ding(2017)Steinegger and S{\"o}ding]{steinegger2017mmseqs2}
Steinegger, M. and S{\"o}ding, J.
\newblock Mmseqs2 enables sensitive protein sequence searching for the analysis of massive data sets.
\newblock \emph{Nature biotechnology}, 35\penalty0 (11):\penalty0 1026--1028, 2017.

\bibitem[Sun \& Zu(2015)Sun and Zu]{Sun2015_Small_AptamersNanomed}
Sun, H. and Zu, Y.
\newblock Aptamers and their applications in nanomedicine.
\newblock \emph{Small}, 11\penalty0 (20):\penalty0 2352--2364, 2015.

\bibitem[Suzek et~al.(2007)Suzek, Huang, McGarvey, Mazumder, and Wu]{suzek2007uniref}
Suzek, B.~E., Huang, H., McGarvey, P., Mazumder, R., and Wu, C.~H.
\newblock Uniref: comprehensive and non-redundant uniprot reference clusters.
\newblock \emph{Bioinformatics}, 23\penalty0 (10):\penalty0 1282--1288, 2007.

\bibitem[Tan et~al.(2024)Tan, Zhang, Gao, et~al.]{Tan2024_ICLR_RDesign}
Tan, C., Zhang, Y., Gao, Z., et~al.
\newblock Rdesign: Hierarchical data-efficient representation learning for tertiary structure-based rna design.
\newblock In \emph{Proceedings of the Twelfth International Conference on Learning Representations (ICLR)}, 2024.

\bibitem[Team et~al.(2024)Team, Boitreau{d}, Dent, et~al.]{Chai12024_bioRxiv}
Team, C.-., Boitreau{d}, J., Dent, J., et~al.
\newblock Chai-1: Decoding the molecular interactions of life.
\newblock \emph{bioRxiv}, pp.\  2024.10.10.615955, 2024.

\bibitem[Team et~al.(2025{\natexlab{a}})Team, Bridgland, Crabb{\'e}, et~al.]{LatentX2025_arXiv}
Team, L.-X., Bridgland, A., Crabb{\'e}, J., et~al.
\newblock Latent-x: An atom-level frontier model for de novo protein binder design.
\newblock \emph{arXiv preprint arXiv:2507.19375}, 2025{\natexlab{a}}.

\bibitem[Team et~al.(2025{\natexlab{b}})Team, Qiao, Bai, Yan, Liu, Xi, Zhang, and Sun]{team2025intfold}
Team, T.~I., Qiao, L., Bai, W., Yan, H., Liu, G., Xi, N., Zhang, X., and Sun, S.
\newblock Intfold: A controllable foundation model for general and specialized biomolecular structure prediction.
\newblock \emph{arXiv preprint arXiv:2507.02025}, 2025{\natexlab{b}}.

\bibitem[Tinberg et~al.(2013)Tinberg, Khare, Dou, et~al.]{Tinberg2013_Nature_LigandBinding}
Tinberg, C.~E., Khare, S.~D., Dou, J., et~al.
\newblock Computational design of ligand-binding proteins with high affinity and selectivity.
\newblock \emph{Nature}, 501\penalty0 (7466):\penalty0 212--216, 2013.

\bibitem[Trippe et~al.(2022)Trippe, Yim, Tischer, et~al.]{Trippe2022_arXiv_BackboneDiffusion}
Trippe, B.~L., Yim, J., Tischer, D., et~al.
\newblock Diffusion probabilistic modeling of protein backbones in 3d for the motif-scaffolding problem.
\newblock \emph{arXiv preprint arXiv:2206.04119}, 2022.

\bibitem[Tuerk \& Gold(1990)Tuerk and Gold]{Tuerk1990_Science_SELEX}
Tuerk, C. and Gold, L.
\newblock Systematic evolution of ligands by exponential enrichment: Rna ligands to bacteriophage t4 dna polymerase.
\newblock \emph{Science}, 249\penalty0 (4968):\penalty0 505--510, 1990.

\bibitem[Van~Kempen et~al.(2024)Van~Kempen, Kim, Tumescheit, Mirdita, Lee, Gilchrist, S{\"o}ding, and Steinegger]{van2024fast_foldseek}
Van~Kempen, M., Kim, S.~S., Tumescheit, C., Mirdita, M., Lee, J., Gilchrist, C.~L., S{\"o}ding, J., and Steinegger, M.
\newblock Fast and accurate protein structure search with foldseek.
\newblock \emph{Nature biotechnology}, 42\penalty0 (2):\penalty0 243--246, 2024.

\bibitem[Wang et~al.(2018)Wang, Che, Levitz, et~al.]{Wang2018_Nature_D2DR}
Wang, S., Che, T., Levitz, J., et~al.
\newblock Structure of the {D2} dopamine receptor bound to the atypical antipsychotic drug risperidone.
\newblock \emph{Nature}, 555\penalty0 (7695):\penalty0 269--273, 2018.

\bibitem[Warner et~al.(2018)Warner, Hajdin, and Weeks]{Warner2018_NatRevDrugDisc_RNAprinciples}
Warner, K.~D., Hajdin, C.~E., and Weeks, K.~M.
\newblock Principles for targeting rna with drug-like small molecules.
\newblock \emph{Nature Reviews Drug Discovery}, 17\penalty0 (8):\penalty0 547--586, 2018.

\bibitem[Watson et~al.(2023)Watson, Juergens, Bennett, et~al.]{Watson2023_Nature_RFDiffusion}
Watson, J.~L., Juergens, D., Bennett, N.~R., et~al.
\newblock De novo design of protein structure and function with rfdiffusion.
\newblock \emph{Nature}, 620\penalty0 (7976):\penalty0 1089--1100, 2023.

\bibitem[Wohlwend et~al.(2025)Wohlwend, Corso, Passaro, Getz, Reveiz, Leidal, Swiderski, Atkinson, Portnoi, Chinn, et~al.]{wohlwend2025boltz}
Wohlwend, J., Corso, G., Passaro, S., Getz, N., Reveiz, M., Leidal, K., Swiderski, W., Atkinson, L., Portnoi, T., Chinn, I., et~al.
\newblock Boltz-1 democratizing biomolecular interaction modeling.
\newblock \emph{BioRxiv}, pp.\  2024--11, 2025.

\bibitem[Wong et~al.(2024)Wong, He, Krishnan, et~al.]{Wong2024_NCS_RhoDesign}
Wong, F., He, D., Krishnan, A., et~al.
\newblock Deep generative design of rna aptamers using structural predictions.
\newblock \emph{Nature Computational Science}, 4\penalty0 (11):\penalty0 829--839, 2024.

\bibitem[Xu et~al.(2023)Xu, Zhou, Xu, et~al.]{Xu2023_Bioeng_EnzymeReconstruction}
Xu, S., Zhou, L., Xu, Y., et~al.
\newblock Recent advances in structure-based enzyme engineering for functional reconstruction.
\newblock \emph{Biotechnology and Bioengineering}, 120\penalty0 (12):\penalty0 3427--3445, 2023.

\bibitem[Xu et~al.(2025)Xu, Feng, Qiao, Wu, Shen, Cheng, Zheng, and Sun]{xu2025foldbench}
Xu, S., Feng, Q., Qiao, L., Wu, H., Shen, T., Cheng, Y., Zheng, S., and Sun, S.
\newblock Foldbench: An all-atom benchmark for biomolecular structure prediction.
\newblock \emph{bioRxiv}, pp.\  2025--05, 2025.

\bibitem[Yang \& Lai(2017)Yang and Lai]{Yang2017_COSB_LigandDesignReview}
Yang, W. and Lai, L.
\newblock Computational design of ligand-binding proteins.
\newblock \emph{Current Opinion in Structural Biology}, 45:\penalty0 67--73, 2017.

\bibitem[Yim et~al.(2023{\natexlab{a}})Yim, Campbell, Foong, Gastegger, Jim{\'e}nez-Luna, Lewis, Satorras, Veeling, Barzilay, Jaakkola, et~al.]{yim2023fast_frameflow}
Yim, J., Campbell, A., Foong, A.~Y., Gastegger, M., Jim{\'e}nez-Luna, J., Lewis, S., Satorras, V.~G., Veeling, B.~S., Barzilay, R., Jaakkola, T., et~al.
\newblock Fast protein backbone generation with se (3) flow matching.
\newblock \emph{arXiv preprint arXiv:2310.05297}, 2023{\natexlab{a}}.

\bibitem[Yim et~al.(2023{\natexlab{b}})Yim, Trippe, De~Bortoli, Mathieu, Doucet, Barzilay, and Jaakkola]{yim2023se_framediff}
Yim, J., Trippe, B.~L., De~Bortoli, V., Mathieu, E., Doucet, A., Barzilay, R., and Jaakkola, T.
\newblock Se (3) diffusion model with application to protein backbone generation.
\newblock \emph{arXiv preprint arXiv:2302.02277}, 2023{\natexlab{b}}.

\bibitem[Zambaldi et~al.(2024)Zambaldi, La, Chu, et~al.]{Zambaldi2024_arXiv_AlphaProteo}
Zambaldi, V., La, D., Chu, A.~E., et~al.
\newblock De novo design of high-affinity protein binders with alphaproteo.
\newblock \emph{arXiv preprint arXiv:2409.08022}, 2024.

\bibitem[Zhang et~al.(2023{\natexlab{a}})Zhang, Wang, Weng, et~al.]{Zhang2023_NatCompSci_TopSurf}
Zhang, O., Wang, T., Weng, G., et~al.
\newblock Learning on topological surface and geometric structure for 3d molecular generation.
\newblock \emph{Nature Computational Science}, 3:\penalty0 849--859, 2023{\natexlab{a}}.

\bibitem[Zhang et~al.(2023{\natexlab{b}})Zhang, Zhang, Jin, et~al.]{Zhang2023_NMI_ResGen}
Zhang, O., Zhang, J., Jin, J., et~al.
\newblock Resgen is a pocket-aware 3d molecular generation model based on parallel multiscale modelling.
\newblock \emph{Nature Machine Intelligence}, 5\penalty0 (9):\penalty0 1020--1030, 2023{\natexlab{b}}.

\bibitem[Zhang et~al.(2024{\natexlab{a}})Zhang, Lin, Zhang, et~al.]{Zhang2024_JACS_DeepLeadOpt}
Zhang, O., Lin, H., Zhang, H., et~al.
\newblock Deep lead optimization: leveraging generative ai for structural modification.
\newblock \emph{Journal of the American Chemical Society}, 146\penalty0 (46):\penalty0 31357--31370, 2024{\natexlab{a}}.

\bibitem[Zhang et~al.(2025)Zhang, Lin, Zhang, et~al.]{Zhang2025_ChemRev_GNNReview}
Zhang, O., Lin, H., Zhang, X., et~al.
\newblock Graph neural networks in modern ai-aided drug discovery.
\newblock \emph{Chemical Reviews}, 2025.

\bibitem[Zhang et~al.(2024{\natexlab{b}})Zhang, Shen, Liu, et~al.]{Zhang2024_NMI_PocketGen}
Zhang, Z., Shen, W.~X., Liu, Q., et~al.
\newblock Efficient generation of protein pockets with pocketgen.
\newblock \emph{Nature Machine Intelligence}, 6\penalty0 (11):\penalty0 1382--1395, 2024{\natexlab{b}}.

\bibitem[Zheng et~al.(2024)Zheng, Wang, and Sun]{Zheng2024_STTT_TargetedApprovals}
Zheng, L., Wang, W., and Sun, Q.
\newblock Targeted drug approvals in 2023: breakthroughs by the fda and nmpa.
\newblock \emph{Signal Transduction and Targeted Therapy}, 9\penalty0 (1):\penalty0 46, 2024.

\bibitem[Zheng et~al.(2025)Zheng, Zhang, Didi, et~al.]{Zheng2025_arXiv_MotifBench}
Zheng, Z., Zhang, B., Didi, K., et~al.
\newblock Motifbench: A standardized protein design benchmark for motif-scaffolding problems.
\newblock \emph{arXiv preprint arXiv:2502.12479}, 2025.

\bibitem[Zhou \& Rossi(2017)Zhou and Rossi]{Zhou2017_NatRevDrugDisc_Aptamers}
Zhou, J. and Rossi, J.
\newblock Aptamers as targeted therapeutics: current potential and challenges.
\newblock \emph{Nature Reviews Drug Discovery}, 16\penalty0 (3):\penalty0 181--202, 2017.

\end{thebibliography}

\newpage
\appendix
\onecolumn
\renewcommand{\thefigure}{S\arabic{figure}}
\renewcommand{\thetable}{S\arabic{table}}
\renewcommand{\thealgorithm}{S\arabic{algorithm}}
\setcounter{figure}{0}
\setcounter{table}{0}
\setcounter{algorithm}{0}

\renewcommand \thepart{}
\renewcommand\partname{} 
\renewcommand{\mtcskip}{\vskip 8pt}
\part{\textbf{\Large \textcolor{darkblue}{Supplementary Information}} }\vspace{-1em}

{\color{darkblue}\rule{\textwidth}{0.5pt}}

{\setlength{\parskip}{8pt}   
 \setlength{\parindent}{0pt} 
 \parttoc
}

\section{ODesign Architecture and Training Regimen}
\label{app:archtrain}

\subsection{Data Pipeline}
\subsubsection{Dataset}
To train a cross-modality molecular design foundation model, comprehensive training across all molecular modalities is essential. The training dataset for ODesign comprises two primary sources: experimentally determined crystal structures from the RCSB Protein Data Bank (Weighted PDB) resolved by X-ray crystallography, and distilled datasets generated through structure prediction using AlphaFold2~\cite{Jumper2021_Nature_AF2} and OpenFold~\cite{ahdritz2024openfold} models.

For crystal structures from the PDB database, we implemented a rigorous quality control pipeline:
\begin{enumerate}
    \item \textbfr{Time split}: Only structures released before September 30, 2021, were retained for the training set to prevent data leakage;
    \item \textbfr{Sequence clustering}: Monomeric chains were clustered based on sequence similarity, and molecular interfaces were systematically clustered according to chain type;
    \item \textbfr{Resolution filtering}: Structures with resolution exceeding 9\AA~were excluded to ensure structural accuracy;
    \item \textbfr{Water \& hydrogen cleanup}: Hydrogen atoms and water molecules were systematically removed, focusing on the heavy atoms;
    \item \textbfr{Completeness filtering}: Chains without resolved atoms or those composed entirely of unknown amino acids were excluded;
    \item \textbfr{Structural plausibility validation}: Chains with adjacent C$\alpha$ atoms separated by more than 10\AA~were removed as discontinuous;
    \item \textbfr{Chain number restriction}: For complex systems containing more than 20 chains, core interface tokens were randomly selected based on interface interaction strength, and only the 20 closest chains were retained;
    \item \textbfr{Molecular type classification}: Biological entities were rigorously categorized by modality type, distinguishing polysaccharides and common cofactors from drug-like small molecules.
\end{enumerate}

For model evaluation, we employed the Low Homology Recent PDB Set~\cite{Abramson2024_Nature_AF3} and PoseBusters~V2~\cite{buttenschoen2024posebusters} as validation sets, in contrast to AlphaFold3~\cite{Abramson2024_Nature_AF3}, which utilized these datasets for testing. The Low Homology Recent PDB Set comprises protein and nucleic acid complex crystal structures released between May~1,~2022, and January~12,~2023, with less than 40\% sequence similarity to the training set, ensuring fair and challenging evaluation. PoseBusters~V2 encompasses 308 high-quality protein–ligand complex structures released after September~30,~2019, providing a reliable benchmark for assessing protein–ligand binding pose prediction accuracy. To mitigate potential data leakage arising from temporal overlap between PoseBusters~V2 complexes and training set structures, all 308 targets were explicitly excluded from the training data.

\subsubsection{Input Features}

ODesign's input features comprise both token-level and atom-level representations, as detailed in Table~\ref{tab:input_features}. Notably, ODesign extends the feature set of AlphaFold3~\cite{Abramson2024_Nature_AF3} by introducing two additional features: \texttt{is\_hotspot}, which indicates whether a token corresponds to a hotspot residue at the binding interface, and \texttt{is\_masked}, which specifies whether a token is marked as a \textit{unified generative token}. 

\subsubsection{MaskGenerator}

To enable cross-modality and controllable molecular design, we introduce a masking training strategy in the ODesign generator. Masking operations are hierarchically categorized into four granularity levels: \textbfr{All}, \textbfr{Entity}, \textbfr{Token}, and \textbfr{Atom}, corresponding to control from complete molecular structures down to individual atomic elements. 

The \textbfr{All} strategy (Algorithm~\ref{alg:all_mask_strategy}) corresponds to free generation tasks. Under this strategy, all standard tokens (proteins, nucleic acids, and small-molecule ligands as defined through Algorithm~\ref{alg:get_standard_token}) are masked, removing side-chain atoms and retaining only backbone atoms, requiring the model to generate complete molecular structures from scratch.

The \textbfr{Entity} strategy (Algorithm~\ref{alg:entity_mask_strategy}) applies to tasks involving the generation of new entities given partial entities (i.e., binder design). The system randomly selects 1 to $N-1$ complete entities for masking (where $N$ is the total number of entities in the structure). All standard tokens within masked entities are processed, with only backbone atoms retained.

The \textbfr{Token} strategy (Algorithm~\ref{alg:token_mask_strategy}) enables coarse-grained control over local structures and is applicable to motif scaffolding tasks. The system randomly selects $M$ contiguous token segments for masking ($M\in[1,4]$). Specifically, for a token sequence of total length $L$, the system partitions it into $M$ consecutive segments. The average length of each segment is calculated by dividing the remaining length by the number of remaining segments, after which both the starting position and mask length are randomly determined within each segment.

The \textbfr{Atom} strategy (Algorithm~\ref{alg:atom_mask_strategy}) provides atomic-level structural control and is designed for atomic motif scaffolding tasks. While all standard tokens are initially masked, the system additionally preserves condition atoms for a randomly selected 20\% of tokens to form functional atomic motifs. The condition atom selection protocol (Algorithm~\ref{alg:get_sym_condition_atoms}) operates as follows: with 80\% probability, the atom furthest from the backbone oxygen is selected as the seed atom; with 20\% probability, the seed atom is randomly chosen. Subsequently, the number of chemical bonds $n$ is sampled from a geometric distribution ($p=0.5$), and all atoms within $n$ chemical bonds from the seed atom are retained.

To prevent data leakage and ensure symmetry generation, the mask generator implements a symmetry-aware masking mechanism (Algorithm~\ref{alg:get_symmetric_mask}). Atomic identifiers are constructed through \texttt{entity\_mol\_id} and \texttt{mol\_atom\_index}, enabling the grouping of symmetry atoms. When an atom is masked, all its symmetry atoms are simultaneously masked, preventing data leakage.

For multi-ligand complexes, the mask generator controls the masking scope based on predefined reference chain indices. When reference chains are specified, the mask generator only performs masking operations on tokens within these chains and removes other ligand molecules after masking completion, ensuring that the final structure retains only one target ligand.

To achieve unified generation of minimal chemical units, we define a \textit{unified generative token} \texttt{"-"} at the token granularity level. To control the design target modality, we append a modality-specific suffix to the \textit{unified generative token}: \texttt{-P} for proteins, \texttt{-N} for nucleic acids, and \texttt{-L} for ligands. During structural information processing, masked tokens in proteins and nucleic acids have their side-chain atoms removed, retaining only backbone atoms and specific condition atoms, with corresponding updates to token-atom index mappings. Reference conformations (\texttt{ref\_pos}) employ representative residues according to modality: glycine (GLY) for proteins, deoxycytidine (DC) for DNA, cytidine (C) for RNA, and zero coordinates for small molecules. Multiple sequence alignment (MSA) feature zeroing: for protein entities containing masked tokens, all MSA-related features are reset to corresponding masked values (Table~\ref{tab:feature_masking}). Bond connectivity clearing: all bond connections of masked tokens are set to 0 in the \texttt{token\_bonds} matrix. Ligand feature anonymization: atomic element types and atom names of masked ligands are replaced with \texttt{"-"}, and atomic charges are set to 0. Nucleic acid backbone standardization: N1 and N9 in the backbone atoms of masked nucleic acid tokens are uniformly renamed to N to prevent sequence information leakage. This systematic masking strategy effectively conceals specific chemical information in masked regions while maintaining molecular structural integrity.

Building upon these four granularity levels of masking, we further implemented a probabilistic sampling strategy that varies across modalities and masking methods, as detailed in Table~\ref{tab:masking_sampling}. At the modality level, sampling probabilities were assigned as 0.4 for proteins, 0.3 for nucleic acids, and 0.3 for ligands. Within each modality, masking strategies were sampled with the following probability distribution: 0.1 for All-level masking (free backbone generation task), 0.1 for Entity-level masking (binder design), 0.7 for Token-level masking (motif scaffolding), and 0.1 for Atom-level masking (atomic motif scaffolding).

\subsubsection{Cropping Methods}

Training on complete biomolecular assemblies is computationally infeasible due to GPU memory limitations. Following the approach established in AlphaFold3~\cite{Abramson2024_Nature_AF3}, we implement stochastic cropping of biomolecular entities during training to ensure computational feasibility. The cropping methodology encompasses three distinct strategies: {Contiguous Cropping}, {Spatial Cropping}, and {Spatial Interface Cropping}, which are randomly sampled during training with weights [0.2, 0.4, 0.4], respectively.

\textbfr{Contiguous Cropping} performs sequential truncation along the chain direction. The protocol proceeds as follows:  
(1) all chains are randomly permuted to shuffle their order within the complex;  
(2) chains are iteratively processed, with the cropping range for each chain dynamically determined based on the number of tokens already selected ($N_{\text{added}}$) and the total number of tokens remaining in unprocessed chains ($N_{\text{remaining}}$), ensuring the total crop size does not exceed the threshold;  
(3) within the determined cropping length range, a random starting position is selected, from which the specified number of tokens are consecutively selected.  
Notably, metal ions are excluded from this strategy, as sequentially adjacent metal ions and tokens may be spatially distant and lack meaningful interactions, rendering their coordinate prediction challenging for the model.

\textbfr{Spatial Cropping} selects tokens based on spatial distance, thereby preserving spatially contiguous molecular regions that exhibit tight intermolecular interactions. The protocol comprises:  
(1) random selection of a resolved reference token within a specified chain as the spatial center;  
(2) computation of distances between the central atom of each token in the complex and the spatial center, constructing a distance matrix with minimal stochastic noise to break ties;  
(3) ranking tokens by ascending distance and selecting the $K$ nearest tokens;  
(4) post-selection validation to ensure ligand integrity—if partial tokens of a ligand or non-standard residue are selected, all tokens of that molecular entity are removed.  
This strategy leverages \texttt{ref\_space\_uid} to identify all tokens belonging to the same molecular entity and filters out entities spanning the selected/unselected boundary.

\textbfr{Spatial Interface Cropping} represents an enhanced variant of spatial cropping, specifically optimized for intermolecular interaction interfaces. The workflow includes:  
(1) identification of interface tokens within specified chains that contact other chains, defined as tokens whose central atom lies within 15\AA~of any central atom in other chains;  
(2) random selection of one interface token as the spatial cropping reference center, ensuring the cropped region focuses on intermolecular interaction zones;  
(3) execution of the standard spatial cropping procedure using the selected reference token as the spatial center.

\textbfr{Ligand-specific cropping strategy} is implemented as a specialized cropping strategy for small-molecule design tasks, which includes:  
(1) when a training sample involves a small molecule, the cropping method is mandatorily set to Spatial Interface Cropping to ensure the cropped region encompasses protein–ligand or nucleic acid–ligand interaction interfaces;  
(2) if multiple ligands are selected following Spatial Interface Cropping, only the target ligand is retained while other ligand molecules are removed, thereby preventing multi-ligand interference during training.

Following token-level cropping, the MSA and template features are synchronously processed. Through these cropping strategies, ODesign achieves efficient training under GPU memory constraints while maintaining molecular structural integrity and accurate representation of interaction interfaces, thereby providing high-quality training samples for downstream structure prediction tasks.

\subsection{Architecture}

ODesign employs a two-stage design paradigm that first generates the molecular backbone followed by sequence design. The framework comprises five core modules: (1) an \texttt{Embedding Module} for initial element representation; (2) a \texttt{Conditional Module} encoding the target's three-dimensional structure; (3) a \texttt{Pairformer Module} for modeling intermolecular interactions; (4) a \texttt{Conditional Diffusion Module} for decoding all-atom coordinates; and (5) an \texttt{OInvFold Module} for unified sequence generation. The main inference loop is illustrated in Algorithm~\ref{alg:main_inference_loop}:

\subsubsection{Unified Generative Token Mechanism}

To achieve unified representation across multiple molecular modalities, ODesign consolidated the minimal chemical unit of each modality into a \textbfr{unified generative token}, denoted as \texttt{"-"}, with modality-specific suffixes controlling the design modality. Table~\ref{tab:chemcomp} summarizes the generative tokens, their chemical composition, and corresponding input features for each modality.

\subsubsection{Condition Module with 3D Target Structure}

During molecular design, researchers typically possess reference conformations of druggable targets, which may be derived from X-ray crystallography experiments or molecular dynamics simulations. To effectively integrate structural priors of target molecules into the model, the conditional control module implements a dual control mechanism. {The first mechanism} incorporates the initial coordinates of the conditional molecule's three-dimensional structure as input features within the element embedding layer, serving as an "initial guess." {The second mechanism} introduces distance information into the Pairformer module to capture interactions between target and designed structures. Specifically, ODesign integrates a distance constraint module between the embedding layer and the Pairformer. This module takes the distance matrix between representative atoms of minimal chemical units as input, transforms the distance matrix into a distance histogram (distogram) according to predefined distance thresholds, extracts high-dimensional feature representations, and ultimately fuses them with the pairwise information output from the embedding module. The pseudocode for the distance constraint module is described as Algorithm~\ref{alg:constraint_template_embedder}.

During training, the model randomly selects one or both control mechanisms for conditional guidance (with equal probability of 1/3 for first control only, second control only, or dual control), thereby enhancing model robustness. Moreover, benefiting from the dual control mechanism of the conditional module, ODesign supports two distinct design modes: {rigid-target design} mode and {flexible-target design} mode, which enable generation with fixed and flexible target conformations, respectively.

\textbfr{Flexible-Target Design} is suited for targets with expansive design space and substantial conformational flexibility at the interaction interface. During generation, ODesign bypasses the first control mechanism and employs only the second mechanism for conditional control. In this configuration, all conditional information originates from internal distances within the target's 3D structure, thereby affording greater conformational flexibility to the target and enabling co-generation of both target and binder structures.

\textbfr{Rigid-Target Design} design is applicable to targets featuring conserved interaction interfaces and conformations. During generation, ODesign simultaneously employs both the first and second control mechanisms. In this mode, beyond the soft constraints imposed by distance information from the target's 3D structure, the model enforces hard constraints by fixing the true coordinates of the target structure.

\textbfr{Epitope-Specific Design} is introduced as a {hotspot residue-based epitope guidance strategy} during ODesign training, to enable epitope-directed molecular design on specified target epitopes. This strategy identifies and annotates key residues at the target–binder (designed molecule) interaction interface, thereby guiding the model to generate molecular structures capable of specifically binding to the target epitope. Hotspot residue identification is based on dual constraints of spatial distance and entity IDs. Specifically, the system first computes the minimum interatomic distances between all token pairs (by calculating pairwise distances among all atoms within tokens and selecting the minimum value), then applies the following filtering criteria: (1) the minimum distance between token pairs must be below a threshold (default 8\AA), ensuring capture of only genuine interaction interfaces; (2) token pairs must originate from the masked entity and the target entity, respectively, with distinct entity IDs, ensuring hotspot residues are located at intermolecular interfaces rather than within a single molecule. The mask constructed through these criteria (i.e., the \texttt{is\_hotspot} feature in Table~\ref{tab:input_features}) precisely localizes the target interaction epitope. The hotspot residue feature is input to the model as a binary vector, where 1 indicates a hotspot residue and 0 indicates a non-hotspot or masked token. This feature is concatenated with other token-level features (such as residue type) during the input embedding stage, with specific integration occurring within the initial Embedding Module.

To enhance model generalization and robustness, ODesign implements a stochastic sampling strategy for hotspot residue features. Specifically, we first compute the minimum distance from each hotspot residue to all other tokens and rank hotspot residues in ascending order by this distance, prioritizing retention of residues closest to the interface core. Subsequently, we sample the number of retained hotspots using an exponential decay distribution (with peak position at 4 residues and decay rate of 0.3), with the sampling range spanning 5–20 residues. This distribution design biases the model toward retaining moderate numbers of hotspots (near the peak) while maintaining adaptability to very few or many hotspots. For hotspot residues not retained during sampling, their \texttt{is\_hotspot} features are reset to 0. This sampling mechanism ensures that the model learns to perform inference from varying numbers and positions of hotspot residues during training, preventing over-reliance on complete hotspot information while aligning with the typical number of epitope residues provided in practical applications.

\subsubsection{Pairformer Module}

The Pairformer module (Algorithm~\ref{alg:pairformer_module}) is designed to model complex intermolecular interaction patterns by iteratively updating single-token and pairwise representations to capture spatial geometric constraints and chemical interaction information both within and between molecules. This module constitutes the core component of the ODesign architecture, responsible for transforming initial embedding features into high-dimensional representations enriched with interaction information. We posit that the weight parameters of a converged AlphaFold3-like Pairformer encode rich knowledge of intermolecular interactions across diverse molecular modalities. Accordingly, ODesign inherits these pre-trained weights and performs fine-tuning upon this foundation.

The Pairformer module employs 48 stacked Pairformer blocks, incorporating core architectural components including triangular multiplicative updates and triangular attention mechanisms. Additionally, ODesign implements a recycling mechanism to progressively refine structural predictions. During each recycling iteration, the Pairformer output is fed back into the module for iterative refinement. In the training phase, the number of recycling iterations is randomly sampled from the range [1, $N_{\text{cycle}}$] (default $N_{\text{cycle}} = 10$), encouraging the model to generate reasonable predictions at varying recycling depths and thereby enhancing robustness. During inference, a fixed maximum number of recycling iterations is employed to achieve optimal performance.

\subsubsection{Conditional Diffusion Module}

The conditional diffusion module (Algorithm~\ref{alg:conditional_diffusion_module}) serves as the core decoding component of the ODesign architecture, designed to transform high-dimensional token representations into precise all-atom 3D Cartesian coordinates, conditioned on the intermolecular interaction information learned by the Pairformer. This module adopts a Denoising Diffusion Probabilistic Model (DDPM)~\cite{ho2020denoising_ddpm} framework with an EDM~\cite{karras2022elucidating_edm} scheduling and preconditioning strategy, specifically optimized for biomolecular structure prediction tasks. Through an iterative denoising process, the module generates 3D molecular structures from random noise.

During training, the module first applies random 3D rigid-body transformations (rotation and translation) to the ground-truth structural coordinates. This data augmentation strategy reinforces the model's SE(3) equivariance, ensuring robust handling of input structures in arbitrary spatial orientations and positions while preventing overfitting to specific reference frames. Subsequently, gaussian noise is added to the augmented coordinates according to a sampled noise level, generating noised coordinates. The diffusion network receives four types of input: (1) noised atomic coordinates, (2) noise level parameters, (3) single-token features from the Pairformer, and (4) pairwise conditional features from the Pairformer. The network then outputs denoised predicted coordinates.

During both training and inference, the first control mechanism of the target 3D structure conditioning module is implemented through the following operations:  
(1) prior to noise addition, noise vectors for conditional atom positions are set to zero, thereby fixing their coordinates;  
(2) following denoising, conditional atom coordinates are replaced with their ground-truth values;  
(3) during data augmentation, the translation vector and rotation matrix are recorded and subsequently applied as inverse transformations to the denoised structure, ensuring restoration to the original coordinate.  
This mechanism ensures that the target structure remains stable throughout the generation process while permitting the designed molecule to evolve freely.

\subsubsection{OInvFold Module}

Following backbone structure generation via the diffusion module, ODesign assigns specific chemical sequence information to the generated scaffold through the \texttt{OInvFold} module. To accommodate the distinct structural characteristics of different molecular modalities, we designed a unified multimodal inverse folding architecture that dynamically switches internal geometric encoding and sequence decoding mechanisms based on the input molecule's modality. \texttt{OInvFold} transforms global Cartesian coordinates into local relative coordinates for processing by a shared SE(3)-equivariant graph neural network that learns geometric features. The definition of the basic unit for local representation is determined by the model input:

\textbfr{Sequence Design for Proteins and Nucleic Acids.}
Considering the structural similarity of proteins, DNA, and RNA as linear polymeric chains, the model identifies each residue (amino acid or nucleotide) as the fundamental unit. A local coordinate frame is constructed for each residue to capture its spatial geometric information, which then serves as the basis for autoregressive sequence decoding.

To establish a unified geometric representation across modalities, we selected representative backbone atoms for constructing local coordinate frames in each modality. For proteins, we chose the four core backbone atoms: N, C$\alpha$, C, and O. For RNA and DNA, we selected key atoms constituting the phosphate-ribose backbone, including P, O5', C5', C4', C3', O3', and the base-connecting N1 or N9 atom. Using these selected atoms, the model constructs an independent local coordinate frame for each residue. This step is crucial, as it transforms global Cartesian coordinates into local relative coordinates that are invariant to overall molecular rotation and translation, enabling subsequent networks to focus on learning intrinsic geometric relationships. For proteins, the C$\alpha$ atom serves as the local coordinate frame origin, with three orthogonal axes determined by the spatial positions of C$\alpha$, N, and C atoms. For RNA, the O5' atom serves as the origin, with the coordinate system defined by P, O5', and C5' atoms. For DNA, accounting for subtle backbone differences, we designate C5' as the origin, with the frame constructed using O5', C5', and C4' atoms. Since local relative poses alone may provide insufficient long-range coherence, we perform principal component analysis (PCA) on each sample to obtain a stable set of global principal axes, constructing $k$ virtual global frames with corresponding edges and their local coordinate systems. These ``global virtual nodes/edges'' are concatenated with the actual local graph and fed into the encoder, preserving a lightweight global reference even in non-grid connectivity.

Once local coordinate frames are established for all residues, coordinates of neighboring atoms are transformed into these local reference frames. Geometric information is then fed into a unified SE(3)-equivariant graph neural network encoder. When processing linear polymers, the model activates its autoregressive decoding pathway. The causal decoder receives the encoded features and sequentially predicts residue types (specific amino acids or nucleotides) along the chain direction, analogous to text generation. This sequential generation naturally aligns with the chain architecture and sequence dependency of proteins and nucleic acids.

\textbfr{Atom Type Design for Small Molecule Ligands.}
In contrast to linear biopolymers, small molecules do not follow sequential patterns; their atoms are connected through complex chemical bond networks forming graph-like topologies. When the input is a ligand, the model switches to finer granularity, treating atoms as the basic units. Since ligand lack repetitive backbone structures, the model constructs an independent local coordinate frame for each individual atom. This coordinate frame is defined by the central atom and its two spatially nearest neighbors, precisely capturing the unique chemical microenvironment surrounding each atom. Similar to the protein and nucleic acid processing pathway, an SE(3)-equivariant graph neural network encoder first extracts geometric features from each atom's local environment. The key difference lies in the decoding stage: the model employs a non-autoregressive decoder that directly and simultaneously predicts elemental types and chemical properties for all atoms in parallel, rather than generating them sequentially. This design better accommodates the non-sequential topological structure of small molecules and enables efficient chemical information assignment across the entire molecule.

\subsection{Loss}

\subsubsection{Composition}

The core objective (Equation~\ref{eq:1}) of ODesign is to predict the 3D all-atom or backbone structures of biomolecules, which is fundamentally identical to the task addressed by AlphaFold3. Accordingly, ODesign adopts the validated structure-related loss functions established in AlphaFold3. Specifically, the loss function comprises four components: $L_{\text{distogram}}$ quantifies the deviation between the token-wise distance distribution predicted by the Pairformer module and the ground-truth values; $L_{\text{MSE}}$ measures the mean squared error (MSE) between predicted and true conformational coordinates in three-dimensional space; $L_{\text{bond}}$ enforces accuracy of covalent bond lengths both within and between molecules in the predicted conformation, calculated as the root mean squared error (RMSE) between predicted and true ligand bond lengths; $L_{\text{smooth\_lddt}}$ evaluates the local distance difference test (LDDT) for each atom within a defined radius in the predicted conformation; $L_{\text{reconstruction}}$ assesses the model's ability to reconstruct ligand chemical bonds, computed as the cross-entropy loss between predicted and true bond connections in ligands. Notably, because predicted conformations lack a fixed coordinate frame, rigid-body alignment between predicted and true conformations is required prior to loss calculation.

\begin{equation}
\label{eq:1}
\begin{aligned}
L & = \alpha_{\text{diffusion}} \cdot\left(\left(\frac{t^{2}+\sigma_{\text{data}}^{2}}{\left(t+\sigma_{\text{data}}\right)^{2}}\right) \cdot\left(L_{\text{MSE}}+\alpha_{\text{bond}} \cdot L_{\text{bond}}\right)+L_{\text{smooth\_lddt}}\right) \\
& \quad +\alpha_{\text{distogram}} \cdot L_{\text{distogram}}+\alpha_{\text{bond}} \cdot L_{\text{reconstruction}}
\end{aligned}
\end{equation}

\subsubsection{Global Align and Partial Align}
In AlphaFold3, predicted and true conformations are aligned via the Kabsch algorithm prior to error computation. However, ODesign's input incorporates true coordinates for conditional atoms, implying that the coordinate frame for atoms to be predicted is known. Based on this observation, we formulated a hypothesis: rather than performing global alignment of predicted and true conformations, could the coordinate frame of conditional atoms be leveraged to constrain the coordinate frame of predicted atoms, thereby providing additional prior information to enhance model performance?

In implementation, the original AlphaFold3 applies random translation and rotation to initial coordinates of samples within each batch before input to the diffusion module, serving as data augmentation to enhance the model's equivariance. When conditional atoms are present, we record the coordinate transformation operations during each data augmentation step and apply the inverse transformation after the model predicts coordinates, ensuring that the transformed conditional atom coordinates remain identical to their initial values—effectively restoring the coordinate frame to the reference system constructed from the conditional atoms. We term this alignment strategy \textbfr{Partial Align}, in contrast to \textbfr{Global Align}, which denotes AlphaFold3's approach of aligning the entire complex.

\begin{figure}
    \centering
    \includegraphics[width=0.7\linewidth]{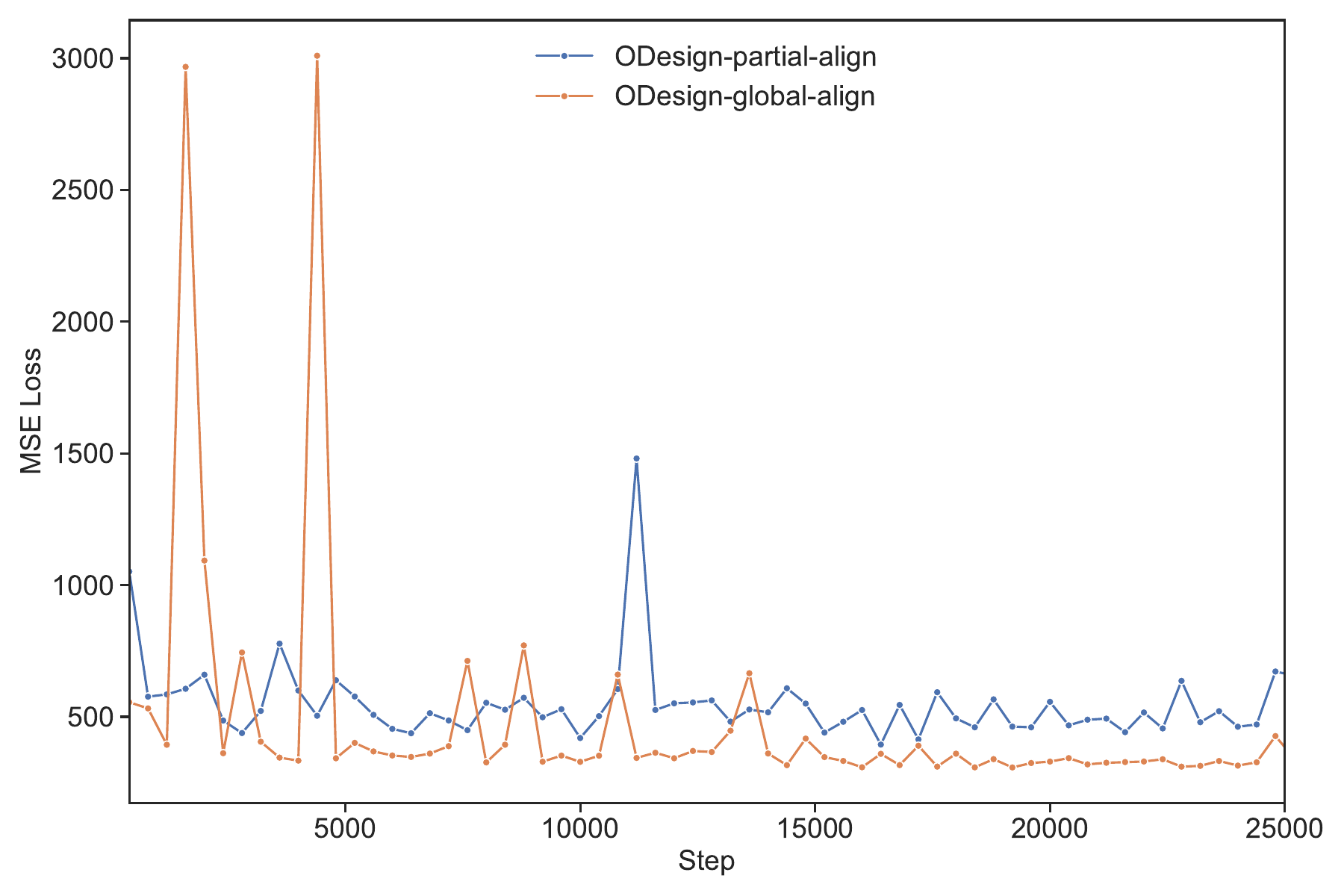}
    \caption{Global Align vs. Partial Align}
    \label{fig:s1}
\end{figure}

However, Boltz-1~\cite{wohlwend2025boltz}, an open-source implementation of AlphaFold3, proposed a perspective diametrically opposed to the above hypothesis: allowing the model to freely explore the coordinate frame distribution of specific biological systems (i.e., Global Align) rather than fixing the coordinate frame, is more beneficial for model performance. Consequently, we systematically investigated the impact of global versus partial alignment strategies on ODesign performance. As illustrated in Figure~\ref{fig:s1}, when training from a pre-trained checkpoint, the model employing the global alignment strategy exhibited significantly lower validation loss compared to partial alignment. This validates that the conclusion drawn by Boltz-1 applies equally to our task: even in the presence of conditional atom constraints, allowing the model to autonomously explore the coordinate frame remains superior to using a fixed coordinate frame constructed from conditional atoms.

\subsection{Modalities}
\begin{figure}
    \centering
    \includegraphics[width=0.7\linewidth]{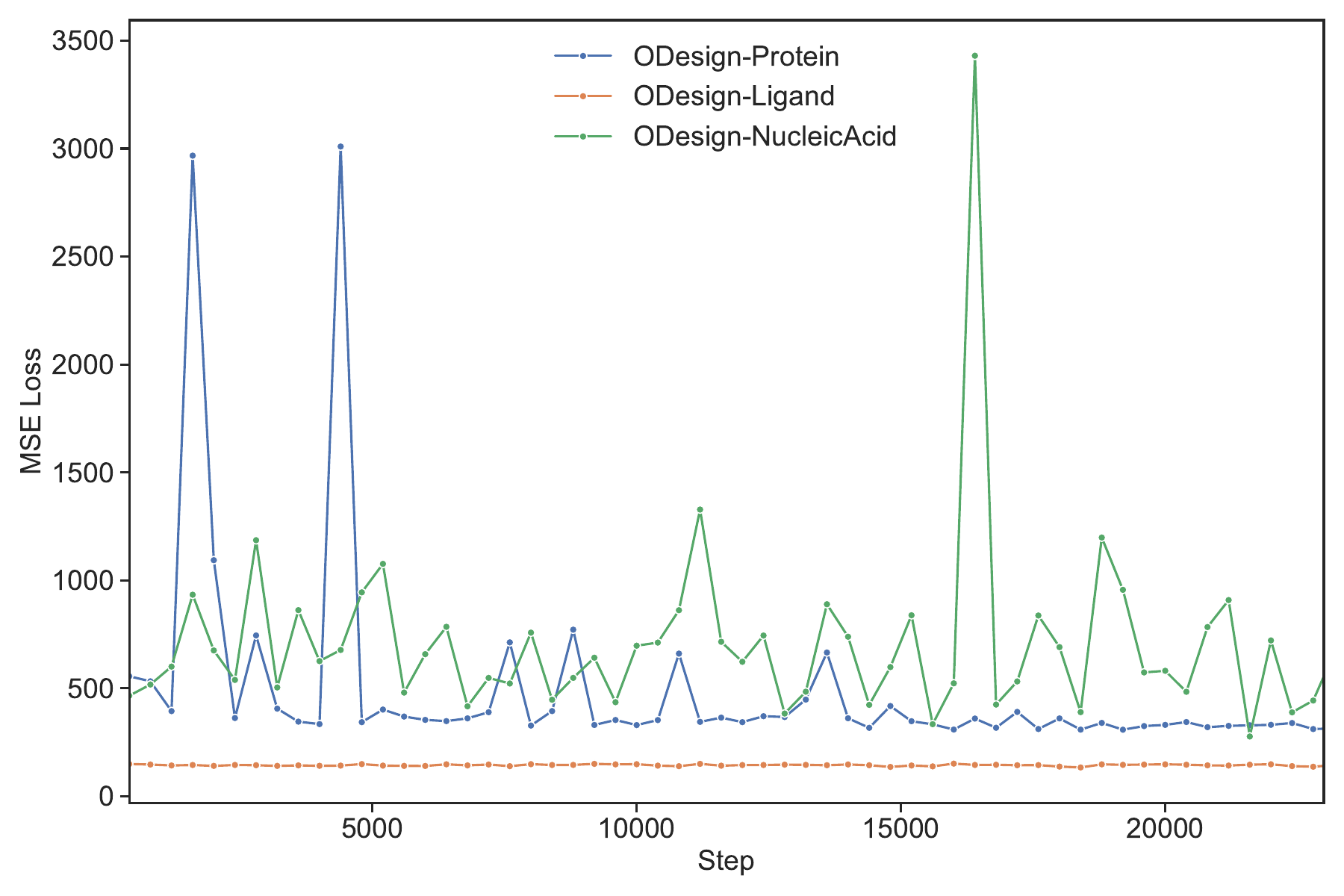}
    \caption{Mean square error loss across different modalities}
    \label{fig:s2}
\end{figure}

ODesign is capable of designing both monomeric structures and complexes across multiple molecular modalities. However, owing to disparities in dataset scale and quality across different modalities, model performance varies substantially among modalities. As shown in Figure~\ref{fig:s2}, ODesign achieves the lowest loss on the ligand modality, exhibits higher loss on the protein modality, and demonstrates the highest loss on the nucleic acid modality (particularly for RNA). This outcome aligns with expectations based on the training data characteristics.

The training dataset contains the largest volume and highest quality of protein and ligand data, whereas nucleic acid data are comparatively scarce. During ligand design, targets remain fixed, resulting in a stable coordinate frame that limits additional loss from weighted alignment. The relatively small number of atoms in ligands compared to proteins further contributes to the lower observed loss. In contrast, protein design involves molecules with substantially more atoms, and the additional loss from weighted alignment is more pronounced, resulting in higher overall loss values.

For nucleic acids, particularly RNA, the increased complexity arises from multiple factors. These molecules possess a greater number of atoms and exhibit higher conformational flexibility compared to proteins and ligands, rendering nucleic acid structure prediction inherently more challenging and contributing to the elevated loss values observed in this modality.

\section{ODesign Inference}

\subsection{Input JSON Format}

During inference, ODesign supports JSON-formatted inputs, with the structure illustrated in JSON 1-7. The top-level structure of the JSON file is a list, where each dictionary represents an individual sample. The total number of samples generated is determined by the number of random seeds and the sampling count per seed. The fields in the JSON file are defined as follows: the \texttt{name} field specifies the sample identifier to distinguish different samples; the \texttt{ref\_file} field specifies the reference file, supporting both \texttt{.pdb} and \texttt{.cif} formats---in conditional generation tasks, the conditioning information is derived from this reference file; the \texttt{chains} field provides detailed information about the chain composition of the sample, including both conditional and design components; the \texttt{condition\_atom} field is a dictionary enabling atomic-level control over conditional information; the \texttt{hotspot} field is a string specifying hotspot residues at the interface; the \texttt{center\_method} field is used to specify the noise center at the initial stage of diffusion; and the \texttt{motif\_scaffolding} field is a boolean parameter for motif scaffolding tasks. When set to \texttt{True}, it can tolerate UNK residues with atoms N, C$\alpha$, C, O as condition.

The \texttt{chains} field warrants further explanation. It is a list where each element is a dictionary representing information for a single chain. Each dictionary contains four fields: \texttt{chain\_type}, \texttt{sequence}, \texttt{length}, and \texttt{msa}, with the latter two being optional. The \texttt{chain\_type} field indicates the chain type, including \texttt{proteinChain}, \texttt{dnaChain}, \texttt{rnaChain}, and \texttt{ligand}. The \texttt{sequence} field can be segmented by commas into multiple fragments. If a fragment contains ``/'', it represents the conditional component derived from the corresponding chain and residue range in the \texttt{ref\_file}; otherwise, it represents the design component, where ``-'' separates the lower and upper bounds of the design length range. The \texttt{msa} field provides MSA information for the chain; if no MSA is available, this field remains empty. The \texttt{length} field specifies the total length of the chain to control the overall length of the design component; if not provided, no length constraint is imposed.

\subsubsection{Protein Design}
\textbfr{Protein-Binding Protein Design.}
JSON~\ref{json:proteinprotein} presents the JSON file used for designing protein binders for target IL7Ra. The \texttt{ref\_file} is obtained from the RCSB PDB database. The design comprises two protein chains: the first, ``B/65-257'' represents residues 65--257 from chain B in the complex, serving as the conditional component; the second chain is a protein of 65 amino acid residues, maintaining the same length as the native ligand. The hotspot residues are defined as residues 129, 130, 131, 132 from chain B in the \texttt{ref\_file}. In this example, hotspot residues are randomly selected from conditional residues within 8\AA of the native ligand. Since the conditional component is derived from a natural structure, its MSA information can be obtained via MMseqs2~\cite{steinegger2017mmseqs2} pre-search against the UniRef100~\cite{suzek2007uniref} database.

\textbfr{Ligand-Binding Protein Design.}
JSON~\ref{json:ligprotein} presents the JSON file used for designing FAD-binding proteins. The \texttt{ref\_file} is a CIF file containing the FAD–protein complex. The design consists of one ligand chain and one protein chain. The conditional component is the ligand from chain E in the \texttt{ref\_file}. The design component is a protein chain of 411 amino acid residues, with the length referencing the native protein binder.

\textbfr{Motif Scaffolding.}
JSON~\ref{json:motifscaffold} presents a JSON file for the motif scaffolding task. The \texttt{ref\_file} is obtained from MotifBench~\cite{Zheng2025_arXiv_MotifBench}. The design comprises a single protein chain consisting of alternating design and conditional segments. The \texttt{sequence} field is a string segmented by commas. For example, ``1-80'' indicates the first segment is a design component with length ranging from 1 to 80 residues; ``A/1-11'' indicates the second segment comprises conditional residues 1--11 from chain A in the \texttt{ref\_file}; ``25-35'' indicates the third segment is a design component with length ranging from 25 to 35 residues, and so forth. With the length constrained to 175, the total length across all segments sums to 175. All segments are concatenated to form a single protein chain containing both conditional and design components.

\textbfr{Atom Scaffolding.}
JSON~\ref{json:atomscaffold} presents a JSON file for the atomic scaffolding task. The \texttt{ref\_file} is obtained from the Atomic Motif Enzyme (AME) benchmark~\cite{Ahern2025_bioRxiv_RFDiffusion2}. The design consists of a protein chain comprising nine segments and two ligand chains. In the protein chain: the first segment is a design component with length ranging from 1 to 70 residues; the second segment is conditional residue 1 from chain A in the \texttt{ref\_file}, and so forth. The total length across all segments is constrained to 180 amino acid residues. Each ligand chain contains a different residue from chain Z of \texttt{ref\_file}. The \texttt{condition\_atom} field enables atomic-level control over the conditional component. For example, in the second segment (chain A, residue 1), only atoms with names NE2, CD2, and CE1 are retained as conditional information, while the remaining atoms must be redesigned by the model.

\subsubsection{Nucleic Acids Design}
\textbfr{RNA Backbone Design.}
JSON~\ref{json:rnafree} presents an example JSON file for RNA backbone design. Since this is unconditional generation, the \texttt{ref\_file} is empty. The design consists of a single RNA chain with a length of 50 ribonucleotides.

\textbfr{Protein-Binding RNA Design.}
JSON~\ref{json:proteinrna} presents an example JSON file for protein-binding RNA design. The \texttt{ref\_file} is obtained from the FoldBench~\cite{xu2025foldbench} test set. The design comprises one protein chain and one RNA chain. The conditional component is a protein chain consisting of residues 1--96 from chain C in the \texttt{ref\_file}. Since it is a structure from PDB, MSA information can be obtained via MMseqs2 pre-search against the UniRef100 database. The design component is an RNA chain of 61 ribonucleotides, with the length referencing the native RNA binder in the \texttt{ref\_file}. The hotspot residues comprise the four residues in chain C closest to the native RNA binder.

\subsubsection{Ligand Design}
JSON~\ref{json:liggen} shows an example JSON file for protein-binding ligand design. The \texttt{ref\_file} is the protein–ligand complex with PDB ID 7v11. The design consists of one protein chain and one ligand. The conditional component comprises all residues within a 10\AA{} radius of the native ligand, forming the binding pocket. The design component is a small molecule containing 29 atoms. All conditional residues are designated as hotspot residues.

\subsection{Diffusion Sample Initialization}
\begin{figure}
    \centering
    \includegraphics[width=0.9\linewidth, trim=190 100 170 60, clip]{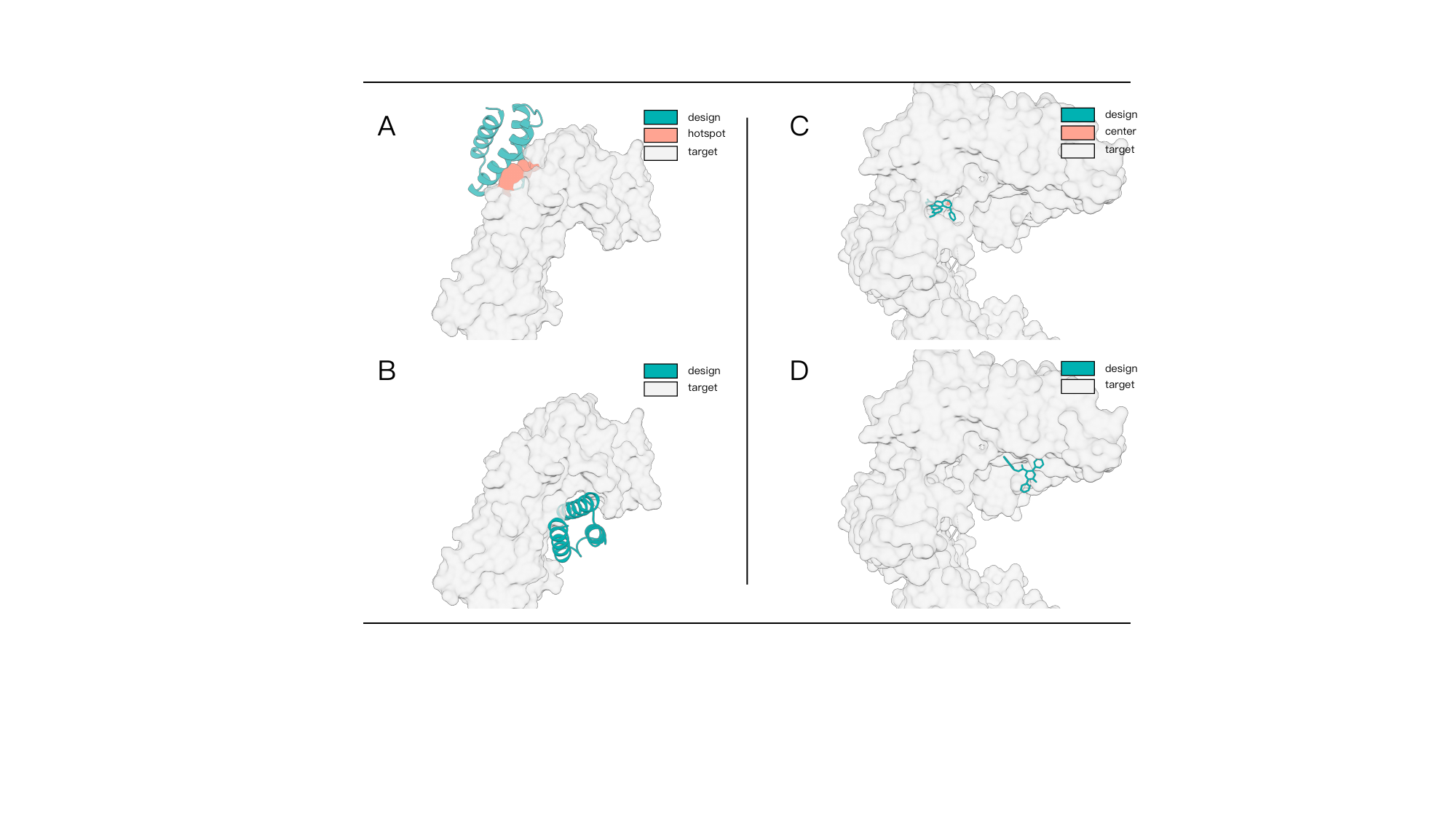}
    \caption{Effect of hotspot specification and centering methods on sampling outcomes}
    \label{fig:s3}
\end{figure}

To accommodate different design tasks, our inference pipeline supports multiple initialization (centering) methods, including \textbfr{hotspot center}, \textbfr{global center}, \textbfr{user-provided center}, and \textbfr{user-provided residue center}. The selected initialization method determines the central point of the noise distribution for the commencement of diffusion sampling. In cases where no initialization method is explicitly specified, the diffusion sampling process defaults to noise centered at the coordinate origin. The various initialization methodologies and their corresponding applicable tasks are detailed in the following sections, with Figure~\ref {fig:s3} showing the corresponding effect of hotspot specification and centering methods on sampling outcomes.

\subsubsection{Hotspot Center}
The hotspot center method can be activated only when hotspot residues are specified. It sets the noise center at the diffusion sampling starting point to the geometric center of the hotspot residues. This method is suitable for tasks emphasizing interface remodeling and interaction modeling, such as ligand-binding protein design, protein-binding protein design, and protein-binding ligand design.

\subsubsection{Global Center}
The global center method is meaningful only in conditional design scenarios. It uses the geometric center of the structure in the \texttt{ref\_file} as the noise center at the diffusion sampling starting point. This method is applicable to conditional generation tasks that consider global information, such as nucleic acid-binding ligand design and protein-binding ligand design.

\subsubsection{User-Provided Center}
The user-provided center method uses 3D coordinates specified by the user as the noise center at the diffusion sampling starting point. This method is suitable for design tasks with strong prior knowledge, such as protein-binding ligand design. This initialization approach helps guide the model toward the correct binding pocket, thereby improving generation quality.

\subsubsection{User-Provided Residue Center}
The user-provided residue center method uses the geometric center of user-specified conditional residues as the noise center at the diffusion sampling starting point. When users need to design a ligand that interacts with specific residues, this initialization method can be employed to position the ligand near the residues of interest. The prerequisite for using this method is that users possess strong design preferences or prior knowledge.

\subsection{Partial Diffusion}
In real-world application scenarios, beyond \textit{de novo} binder design from scratch, researchers often need to modify existing binding molecules. For example, an existing antibody or mini-protein binder may already exhibit certain affinity but still requires modifications to enhance stability, modulate specificity, or improve expressibility. For such ``suboptimal binders'', we can treat them as initial molecules and introduce controlled perturbations and corrections within the diffusion or inverse folding framework, thereby exploring the optimization space while preserving the binding mode. To achieve this, we employ the partial diffusion method, which operates as a back-and-forth strategy to explore the structural space in the vicinity of the reference binder.

Specifically, in the ``back'' stage, we gradually push the original structure toward an intermediate state of the diffusion process by applying noise dependent on the signal-to-noise ratio (SNR). Subsequently, in the ``forth'' stage, we utilize the model to perform multi-step denoising to optimize the structure. 

In the back stage, the SNR is defined as:
\begin{equation}
SNR(t)=\frac{\sigma_{\text{data}}^2}{\sigma^2(t)}
\end{equation}

Accordingly, following the EDM preconditioner used during model training, we add noise $\varepsilon \sim N(0, I)$ to the input structure as follows:
\begin{equation}
\begin{aligned}
X_t=\frac{1}{\sqrt{\sigma_{\text{data}}^2+\sigma^2(t)}}X_0+\frac{\sigma(t)}{\sqrt{\sigma_{\text{data}}^2+\sigma^2(t)}}\varepsilon
\end{aligned}
\end{equation}

Subsequently, in the ``forth'' stage, we input the noised $X_t$ into the model for multi-step denoising. Specifically, we select various $SNR(t)$ values, such as [0.1, 0.5, 1.0, ...], as starting points for the denoising generation process for different protein structures. It should be noted that when the SNR is smaller, the generated modified molecule exhibits greater divergence from the target molecule, resulting in increased diversity but making it more challenging to preserve the target molecule's function. Conversely, when the SNR is larger, diversity decreases, but the generatively modified molecule maintains higher similarity to the target.

\subsection{Inference Speed}
ODesign is a generative model built upon the AlphaFold3 framework. Each design iteration requires a single forward pass through a 48-layer Pairformer and one pass through a diffusion model solver with 200 Euler steps. Consequently, ODesign incurs only the computational cost equivalent to running the fold model once per design iteration. In contrast, RFdiffusion~\cite{Watson2023_Nature_RFDiffusion} and RFdiffusion-AA~\cite{Krishna2024_Science_RFAA} incorporate the fold model as their diffusion module, requiring computational costs equivalent to 200 fold model executions per iteration. Similarly, hallucination-based models rely on the fold model for design generation and necessitate multiple iterative fold model evaluations, resulting in substantially higher computational overhead compared to ODesign.

Furthermore, benefiting from I/O optimizations and triangle attention CUDA kernels, our sampling speed achieves significant acceleration, surpassing that of RFdiffusion and RFdiffusion-AA, and substantially exceeding hallucination-based models such as Boltz-1~\cite{wohlwend2025boltz} and BindCraft~\cite{Pacesa2025_Nature_BindCraft}. To demonstrate this computational advantage, we benchmarked the sampling times of ODesign, RFdiffusion-AA, and Boltz-1 on the ligand-binding protein design task using an 8$\times$H100 (80~GB) GPU cluster, calculating the time required to generate a single sample per target. Detailed results are presented in Table~\ref{tab:runtime}, confirming that ODesign achieves superior computational efficiency compared to existing methods, consistent with our theoretical analysis.

\section{ODesign Benchmarks}
\subsection{Protein}
\subsubsection{Free Generation}
Free backbone generation refers to the \textit{de novo} design of proteins without any prior conditioning, representing the model's explorable protein design space. 

We selected classical models including ESM3~\cite{hayes2025simulating_esm3}, Chroma~\cite{ingraham2023illuminating_chroma}, FrameFlow~\cite{yim2023fast_frameflow}, Proteina~\cite{Geffner2025_arXiv_Proteina}, FrameDiff~\cite{yim2023se_framediff}, Genie2~\cite{lin2024out_genie2}, and RFdiffusion as baselines. To comprehensively evaluate the models' capability to generate both short and long protein backbones, we fixed the test length range at 100--800 amino acid residues. Within this range, we sampled 100 protein backbones at 100-residue intervals using each model, designed 8 sequences per backbone using ProteinMPNN~\cite{dauparas2022robust_proteinmpnn}, and refolded the structure of each sequence using ESMFold~\cite{lin2023evolutionary_esmfold}. 

ProteinMPNN was configured with a temperature parameter of 0.1, while all other parameters retained their default settings. ESMFold v1 was employed for structure prediction. These configurations for ProteinMPNN and ESMFold were consistently applied throughout all subsequent analyses. 

In Proteina, a structure is deemed designable if one sequence achieves a backbone-aligned RMSD $<$ 2.0\AA{} between the designed and refolded structures. However, since the protein design space may contain numerous disordered structures, relying solely on RMSD to assess design success is insufficient. Given that pLDDT can serve as a filter for disordered structures, we defined designability as pLDDT $>$ 70 and backbone-aligned RMSD $<$ 2.0\AA.

\subsubsection{Protein-Binding Protein Design}
Protein-binding protein design involves generating protein binders against specified protein targets, with classical applications including antibody design. Eleven targets, including EGFR and CD3d, were selected from CaoData~\cite{Cao2022_Nature_TargetOnly}, a dataset derived from classical de novo protein design campaigns based on target structures. These targets primarily comprise two categories: human cell-surface or extracellular proteins involved in signal transduction, and pathogen surface proteins. 

Three models capable of protein-binding protein design (RFdiffusion, BoltzDesign1~\cite{Cho2025_bioRxiv_Boltzdesign1}, and BindCraft) were selected as baseline methods. For each target, the design length was defined to match that of the original experimentally validated target-binding protein from CaoData, ensuring fair comparison across methods. For ODesign and RFDiffusion, 100 protein backbones were generated for each target. For each backbone, eight protein sequences were designed using ProteinMPNN with fixed target sequences, yielding 800 candidate sequences, which were subsequently refolded using AlphaFold3. Since BoltzDesign1 and BindCraft are hallucination-based models requiring iterative structure prediction, their sampling times are substantially longer. To reduce computational cost, we generated only 10 target-binding proteins per target for these two hallucination models.

For RFdiffusion, we employed the ``base'' checkpoint with the ``base'' configuration file for inference. For BoltzDesign, MSA was performed, but LigandMPNN~\cite{dauparas2025atomic_ligandmpnn}, AlphaFold3, and RoseTTAFold were skipped; all other parameters were set to default configurations. For BindCraft, all parameters were set to default configurations. During structure prediction with AlphaFold3, only the target's MSA was provided, while templates and the binder's MSA were excluded.

Based on the AF2-IG-easy filter definition in PXdesign (ipAE<10.85, pLDDT>0.8, ipTM>0.5, binder bound/unbound RMSD < 3.5\AA), our filters were defined as ipAE<10.85, pLDDT>80, ipTM>0.5, complex RMSD < 2.5\AA. The pLDDT threshold was set to 80 rather than 0.8 because pLDDT values in AlphaFold3 are not normalized. Additionally, the RMSD metric was modified to complex RMSD < 2.5\AA, as used in AlphaProteo, to evaluate the designability of the overall complex structure. For ODesign and RFDiffusion, a backbone was classified as a filter-passing backbone if at least one sequence generated from that backbone satisfied the filter criteria. For each case, 100 backbones were sampled using these models (10 designs for BoltzDesign1 and BindCraft), and both sampling and refolding times were recorded on a single H100 GPU. Using Equation~\ref{eq:4}, the number of filter-passing backbones (designs) that could be generated in 24 hours on a single H100 GPU was calculated.

\begin{equation}
\label{eq:4}
N_{\text{1d}}=\frac{\text{24h}}{T_{\text{sample}}/N_{\text{sample}}+T_{\text{refold}}}\times\frac{N_{\text{filter-passing}}}{N_{\text{sample}}}
\end{equation}

\subsubsection{Ligand-Binding Protein Design}
Ligand-binding protein design involves generating protein binders against specified ligand targets, with applications in biosensor reagent development. The test dataset comprised four ligands from RFdiffusion-AA: IAI, OQO, FAD, and SAM. IAI and OQO are synthetic ligands with low training set similarity, making them suitable for evaluating model generalization capability. FAD is a natural redox cofactor representing the cofactor class in model training, while SAM is a common cosubstrate participating in methyl transfer reactions in biological systems.

Models capable of ligand-binding protein design (RFdiffusion-AA and BoltzDesign1) were selected as baseline methods. For each ligand, the design length was defined to match that of the original ligand-binding protein, ensuring fair comparison across methods. For ODesign and RFdiffusion-AA, we generated 100 protein backbones per ligand, designed 8 protein sequences per backbone using LigandMPNN, yielding 800 candidate sequences. Due to the computational expense of the hallucination-based BoltzDesign1, we generated only 10 samples per ligand. One structure was predicted for each sequence using Chai-1.

For RFdiffusion-AA, all parameters were set to default configurations. For LigandMPNN, the temperature parameter was set to 0.1, with all other parameters using default configurations; these LigandMPNN settings were maintained throughout all subsequent analyses. BoltzDesign1 hyperparameters were set to official default configurations. During structure prediction with Chai-1 (version 0.6.1 PyPI package), MSA and template searches were disabled, while all other parameters were set to default configurations; these Chai-1 settings were maintained throughout all subsequent analyses.

The same success criterion from BoltzDesign1 was adopted: refolded complex pLDDT > 0.7 and iPAE<10. For ODesign and RFdiffusion-AA, a backbone was classified as a filter-passing backbone if at least one sequence generated from that backbone satisfied the filter criteria. For each case, 100 backbones were sampled using ODesign and baseline models (10 designs for BoltzDesign1), and both sampling and refolding times were recorded on a single H100 GPU. Using Equation \ref{eq:4}, the number of filter-passing backbones (designs) that could be generated in 24 hours on a single H100 GPU was calculated.

\subsubsection{Motif Scaffolding}
Motif scaffolding requires designing diverse protein structures (scaffolds) to accommodate given motifs—geometric coordinates of residues associated with biochemical function—while preserving their spatial arrangements. Motif scaffolding has numerous applications in enzyme active site design, functional binding interface construction, and stabilization of specific structural or functional modules. MotifBench, a classical benchmark for quantifying model performance on motif scaffolding tasks, was selected. Specifically, all 30 targets from MotifBench were utilized as the test set, including orphan proteins, enzyme active sites, protein-binding interfaces, and challenging specific structures (e.g., fragment loops, parallel $\beta$-sheets).

RFdiffusion, RFdiffusion-AA, and Proteina were selected as baseline methods. Following MotifBench's definitions of motifs and design lengths, 100 protein backbones were designed for each target using different models, and 4 sequences were generated for each backbone using ProteinMPNN. According to MotifBench's definition, one structure was predicted for each sequence using ESMFold.

Notably, the official RFdiffusion-AA repository requires ligand input; to support MotifBench's ligand-free inputs, we modified the code to remove the mandatory ligand requirement. For all baselines, default configurations were used.

According to MotifBench, success was defined as N, C$\alpha$, C-aligned motif RMSD $< 1.0$\AA{}  and C$\alpha$-aligned C$\alpha$ RMSD < 2.0\AA. Following MotifBench's scoring methodology, diversity and novelty of successful samples were calculated using Foldseek~\cite{van2024fast_foldseek}, and the motif score was computed by integrating diversity, novelty, and success rate.

\subsubsection{Atomic Motif Scaffolding}
Atomic motif scaffolding requires designing scaffolds to accommodate given ligands and active sites while preserving their precise geometric coordinates, demanding exceptional fine-grained control capability. AME is a classical benchmark proposed in RFdiffusion2 for evaluating model performance on atomic motif enzyme tasks. All 41 systems from AME were selected as the test set, spanning EC classes 1–5 and covering active sites for diverse enzymatic catalytic reactions.

RFdiffusion-AA and RFdiffusion2 were selected as baselines. Following AME's definitions of motif atoms and design lengths, we generated 100 protein backbones per target using different models and designed 8 sequences per backbone using LigandMPNN. One structure was predicted for each sequence using Chai-1~\cite{Chai12024_bioRxiv}. RFdiffusion2 was employed with default configurations.

Following RFdiffusion2's methodology for assessing whether sequences fold to expected structures, success was defined as: when aligned on the backbone (N, C$\alpha$, C) of catalytic residues in at least one LigandMPNN sequence predicted by Chai-1, the RMSD of all heavy atoms in catalytic residues is less than 1.5\AA{} with no steric clashes. For each system, 100 backbones were sampled using ODesign and baseline models, and both sampling and refolding times were recorded on a single H100 GPU. Using Equation \ref{eq:4}, the number of successful backbones that could be generated in 24 hours on a single H100 GPU was calculated.

\subsubsection{Interface Design}
Interface design represents a constrained variant of ligand-binding protein design. Accordingly, test targets for interface design were selected from four cases in the ligand-binding protein design benchmark: IAI, OQO, FAD, and SAM. Following PocketGen's~\cite{Zhang2024_NMI_PocketGen} pocket definition methodology, residues located within 3.5\AA{} of the ligand were designated as the redesign component, averaging 14 redesign residues per target.

PocketGen, which employs ProteinMPNN and ESMFold/AlphaFold2 for sequence prediction and refolding, was selected as the baseline method. However, this pipeline omitted small molecules during prediction, which may introduce noise into the evaluation results for protein-ligand complexes. Therefore, LigandMPNN and Chai-1, which can account for ligands during prediction, were used for sequence generation and structure prediction, respectively. For each case, 100 protein backbones were generated using different models, and 8 sequences per backbone were designed using LigandMPNN. Each designed sequence was subsequently refolded using Chai-1 with the ligand included in the structure prediction.

The same metrics reported in PocketGen (i.e., scRMSD, pocket RMSD, and pLDDT) were evaluated for each sequence.

\subsection{Nucleic Acid}
\subsubsection{DNA Free Generation}
As ODesign represents the first deep generative model capable of DNA backbone design, no well-established benchmark exists for this task.

To systematically assess ODesign's performance in designing both short and long DNA backbones, ODesign was required to sample DNA structures at 10-nucleotide intervals, generating 50 DNA backbones per length within the 10–150 nucleotide range. For each backbone, 8 DNA sequences were designed using the \texttt{OInvFold} module. Following the validity criterion established in the RNA free generation benchmark, one structure for each DNA backbone was predicted using AlphaFold3, without MSA search or template information.

Following the definition of RNA backbone designability in RNAFrameFlow, DNA backbone designability was evaluated using two metrics: RMSD < 5\AA{} and TM-score > 0.45. A DNA backbone was classified as designable under the corresponding metric if at least one of its 8 designed sequences achieved C4' RMSD < 5\AA{} (or TM-score > 0.45) between the original and refolded structures.

\subsubsection{Protein-Binding DNA Design}
Protein-binding DNA design involves generating DNA aptamers against specified protein targets, with applications in nucleic acid drug discovery. Four protein–DNA complexes from the PDB released after January 13, 2023, were curated, ensuring independence from the training set to rigorously evaluate generalization capability. This task has potential applications in downstream DNA aptamer discovery.

For each target, ODesign generated 100 DNA aptamers matching the ground-truth length. Each DNA backbone was then assigned 4 sequences via our in-house inverse folding module, and 1 structure was predicted for each sequence using AlphaFold3. During structure prediction with AlphaFold3, the target protein's MSA was provided, while templates were excluded.

Building upon RFDiffusion3's~\cite{butcher2025novo_rfdiffusion3} filter definition for protein–nucleic acid interactions (DNA phosphate atoms-aligned C$\alpha$ RMSD < 5\AA{}), an analogous success criterion was adopted: C$\alpha$-aligned DNA phosphate atoms RMSD < 5\AA{}.

\subsubsection{RNA Free Generation}
Free backbone generation refers to the \textit{de novo} design of RNAs without any prior conditioning, representing the model's explorable RNA design space. For RNA free generation, RNAFrameFlow~\cite{anand2025rna_rnaframeflow} was selected as the baseline. To enable direct comparison with conditional generation tasks, RNAFrameFlow's design length was extended to range from 10 to 150 ribonucleotides, thereby evaluating comprehensive performance across both short and long RNA backbones. Within this length range, sampling at 10-nucleotide intervals, 50 RNA backbones were generated per length using each model, and 8 RNA sequences were designed per backbone using gRNAde~\cite{Joshi2024_Methods_gRNAde}. For each RNA sequence, one structure was predicted using AlphaFold3 without MSA search.

Following RNAFrameFlow's validity definition, structures were predicted using AlphaFold3, and sequences achieving RMSD < 5\AA{} or TM-score > 0.45 between designed and refolded conformations were considered valid. A backbone was classified as designable if at least one of its 8 sequences satisfied the filter criteria (C4' RMSD < 5\AA{} or TM-score > 0.45) between the designed and refolded structures.

\subsubsection{Protein-Binding RNA Design}
Protein-binding RNA design involves generating RNA aptamers against specified protein targets, with classical applications in nucleic acid drug discovery. ODesign represents the first deep generative model capable of protein-binding RNA aptamer design. In the absence of established benchmarks for this task, we constructed a test set comprising complexes excluded from the training data. Ten protein–RNA complexes released after January 13, 2023, were selected from PDB, ensuring dataset independence to assess generalization capability.

For each target, 100 RNA aptamer backbones were generated using ODesign, with the design length defined to match that of the ground-truth RNA structure, ensuring fair evaluation. Four sequences were designed per backbone using the \texttt{OInvFold} module with fixed target sequences, yielding 400 candidate sequences per target, followed by structure prediction using AlphaFold3. During structure prediction with AlphaFold3, only the target's MSA was provided, while templates and the RNA's MSA were excluded.

Mirroring the evaluation protocol established for protein-binding DNA design, we defined successful designs using the criterion of C$\alpha$-aligned RNA phosphate atoms RMSD < 5\AA{}.

\subsection{Ligand}

\subsubsection{Protein-Binding Ligand Design}

Protein-binding ligand design involves generating small molecules against specified protein targets, a critical task in drug discovery. Six protein targets were selected for benchmarking: two clinically relevant therapeutic targets (DRD2 and ADRB1), alongside four protein--ligand complexes from the ligand-binding protein design task.

Multiple approaches were employed for comparison, including representative diffusion-based methods (TargetDiff~\cite{guan20233d_targetdiff}, D3FG~\cite{Lin2024_NeurIPS_FGDiff}) and autoregressive methods (Pocket2Mol~\cite{Peng2022_ICML_Pocket2Mol}, SurfGen~\cite{Zhang2023_NatCompSci_TopSurf}, ResGen~\cite{Zhang2023_NMI_ResGen}). For ODesign, molecular size was first inferred from prior distributions conditioned on pocket size~\cite{guan20233d_targetdiff,Lin2022_arXiv_DiffBP}, structural backbones were generated, and atom types were subsequently assigned using the \texttt{OInvFold} module. Default configurations were utilized for inference across all baseline methods. Each model generated 100 molecules per target on a single H100 GPU, and both sampling and docking times were recorded.

Previous ligand design models targeting protein receptors have typically evaluated generation quality along two dimensions: chemical properties and molecular interactions. Following established evaluation protocols~\cite{guan20233d_targetdiff,qu2024molcraft}, we employed validity and success rate as metrics to jointly assess both generation quality and computational efficiency. Validity was defined as successful recognition by RDKit and docking by AutoDock Vina~\cite{eberhardt2021autodock_vina}. Success required satisfaction of both drug-likeness (QED~$>$~0.25) and binding affinity (Vina Dock score~$<$~-8.18) criteria. Using Equation~\ref{eq:4}, the extrapolated daily throughput of valid and successful molecules that could be generated in 24~hours on a single H100 GPU was calculated based on recorded sampling and docking times.

\subsubsection{DNA-Binding Ligand Design}

DNA-binding ligand design advances drug discovery by generating small molecules against specific DNA targets. Given the absence of established benchmarks for DNA-binding ligand design, we selected three DNA--ligand complexes from the Protein Data Bank (PDB) as test cases, specifically those released after January~13,~2023, to ensure they fall outside the training data cutoff dates of current structure prediction models.

For each target, hotspot nucleotides were defined as those positioned within 10~\AA{} of any reference ligand atom. The molecular size for generated ligands was randomly sampled within $\pm$5 atoms of the corresponding reference ligand to ensure comparability. For each DNA target, 100 candidate ligands were generated, with atom types subsequently assigned using the \texttt{OInvFold} model, followed by structure prediction using AlphaFold3.

Evaluation metrics included distributions of AlphaFold3-predicted minimum iPAE, iPTM, refolded ligand PTM, and DNA-C1$'$-aligned ligand RMSD before and after refolding. The first two metrics (iPAE and iPTM) assess binding interface confidence, indicating the plausibility of stable interaction modes between the ligand and DNA target. The latter two metrics (ligand PTM and RMSD) evaluate ligand structural confidence and conformational consistency, respectively, providing validation through established structure prediction methods.

\subsubsection{RNA-Binding Ligand Design}

RNA-binding ligand design involves generating small molecules against specified RNA targets. Following the evaluation framework established for DNA-binding ligand design, we assessed ODesign on three RNA--ligand complexes from the PDB released after January~13,~2023.

Analogous to the DNA protocol, hotspot nucleotides were identified within 10~\AA{} of reference ligand atoms, and molecular size was sampled within $\pm$5 atoms of the reference. One hundred ligands were generated per RNA target, with atom types assigned via the \texttt{OInvFold} module.

Evaluation metrics comprised AlphaFold3-predicted minimum iPAE, iPTM, refolded ligand PTM, and RNA-C1$'$-aligned ligand RMSD distributions before and after refolding. These metrics collectively assess binding interface confidence and ligand structural fidelity.

\newpage

\begin{table*}[htbp]
\centering
\renewcommand{\thetable}{S1}
\caption{{Input Features of ODesign}}
\label{tab:input_features}
\renewcommand{\arraystretch}{1.25}
\small                    
\begin{tabularx}{\textwidth}{p{3cm} p{3cm} X}

\toprule
\rowcolor[gray]{0.85} \textbf{Feature} & \textbf{Dimension} & \textbf{Description} \\
\midrule

\rowcolor[gray]{0.95} \multicolumn{3}{l}{\textbf{Token features}} \\
\midrule
residue\_index & [$N_\text{token}$] & Index of the residue within its chain. \\
 
token\_index & [$N_\text{token}$] & Index of the token within the entire complex. \\
 
asym\_id & [$N_\text{token}$] & Unique chain identifier, numbered sequentially from 0, with distinct values for each chain. \\
 
entity\_id & [$N_\text{token}$] & Unique biological entity identifier; each entity comprises one or more symmetry chains with identical sequences. \\
 
sym\_id & [$N_\text{token}$] & Unique symmetry chain identifier within the same biological entity. For example, if chains A, B, and C share a sequence but D does not, their sym\_ids, asym\_ids, and entity\_ids would be [0, 1, 2, 0], [0, 1, 2, 3], and [0, 0, 0, 1]. \\
 
res\_type & [$N_\text{token}$, 35] & One-hot encoding of residue types. The vocabulary encompasses 35 possible types: 20 standard amino acids + unknown (UNK); 4 RNA nucleotides + unknown (C); 4 DNA nucleotides + unknown (DC); GAP; and 3 unified generative tokens (-P, -N, -L). Ligands are marked as “UNK.” \\
 
is\_hotspot & [$N_\text{token}$] & Binary indicator denoting whether a specific token is located at the interaction interface (0 = False, 1 = True). \\
 
is\_masked & [$N_\text{token}$] & Binary mask indicating whether a token is marked as a unified generative token. \\
 
is\_protein / is\_rna / is\_dna / is\_ligand & [$N_\text{token}$, 4] & Binary masks indicating token modality types. \\
 
msa & [$N_\text{msa}$, $N_\text{token}$, 32] & One-hot encoding of the multiple sequence alignment (MSA). \\
 
has\_deletion & [$N_\text{msa}$, $N_\text{token}$] & Binary feature indicating if there is a deletion to the left of each position in the MSA. \\
 
deletion\_value & [$N_\text{msa}$, $N_\text{token}$] & Raw deletion counts (number of deletions to the left of each MSA position), transformed to the interval [0, 1] using the mapping function $2/\pi \cdot \arctan(d/3)$. \\
 
profile & [$N_\text{token}$, 32] & Residue type distribution derived from the MSA. \\
 
deletion\_mean & [$N_\text{token}$] & Mean deletion count across the MSA. \\
 
token\_bonds & [$N_\text{token}$, $N_\text{token}$] & Binary adjacency matrix encoding covalent bond connectivity; records only bonds within ligands and between ligands and polymers. \\
 
\midrule
\rowcolor[gray]{0.95} \multicolumn{3}{l}{\textbf{Atom features}} \\
 \midrule
ref\_element & [$N_\text{atom}$, 129] & One-hot encoding of atomic element types: 118 known elements from the periodic table, 10 unknown types, and the unified generative token “–”. \\
 
ref\_atom\_name\_chars & [$N_\text{atom}$, 4, 64] & One-hot encoding of atom names; names are padded to length 4 and each character encoded as ord(c) – 32. \\
 
ref\_charge & [$N_\text{atom}$] & Charge for each atom. \\
 
ref\_pos & [$N_\text{atom}$, 3] & Coordinates of the reference conformation after random translation and rotation; masked tokens are substituted with representative reference conformations from the same modality. \\
 
ref\_space\_uid & [$N_\text{atom}$] & Numerical encoding of the chain id and residue index associated with the reference conformation. Each unique (chain id, residue index) tuple is assigned a distinct integer upon first occurrence. \\
 \bottomrule

\end{tabularx}
\end{table*}

\begin{table*}[htbp]
\centering
\renewcommand{\thetable}{S2}
\caption{{Feature Masking Strategy}}
\label{tab:feature_masking}
\renewcommand{\arraystretch}{1.25}
\small
\begin{tabularx}{\textwidth}{p{2.5cm} p{3.5cm} X}
\toprule
\rowcolor[gray]{0.85} \textbf{Feature} & \textbf{Masked Value} & \textbf{Description} \\
\midrule

\rowcolor[gray]{0.95} \multicolumn{3}{l}{\textbf{Token features}} \\
\midrule
res\_type & Protein: -P; Nucleic acid: -N; Ligand: -L & Residue type; after masking, classified by modality: -P (protein), -N (nucleic acid), -L (small-molecule ligand). \\

msa & 0 & One-hot encoding of MSA. \\

has\_deletion & False & Binary feature indicating whether there is a deletion to the left of each position in the MSA. \\

deletion\_value & 0 & Raw deletion counts (number of deletions to the left of each MSA position), transformed to the interval [0, 1] using the mapping function $2/\pi \cdot \arctan(d/3)$. \\

profile & 0 & Residue type distribution derived from the MSA. \\

deletion\_mean & 0 & Mean deletion count across the MSA. \\

token\_bonds & 0 & Binary adjacency matrix encoding covalent bond connectivity; records only bonds within ligands and between ligands and polymers. \\

\midrule
\rowcolor[gray]{0.95} \multicolumn{3}{l}{\textbf{Atom features}} \\
\midrule
ref\_element & Ligand: – & Element type; modified only for masked ligands. \\

ref\_atom\_name\_chars & Ligand: – & Atom name; modified only for masked ligands. \\

ref\_charge & Ligand: 0 & Atomic charge; modified only for masked ligands. \\

ref\_pos & Protein: GLY; DNA: DC; RNA: C; Ligand: 0 & Coordinates of the reference conformation following random translation and rotation; masked tokens are substituted with reference conformations from modality-specific representative residues. \\
\bottomrule
\end{tabularx}
\end{table*}
\begin{table*}[htbp]
\centering
\renewcommand{\thetable}{S3}
\caption{{Masking Strategy Sampling Probabilities}}
\label{tab:masking_sampling}
\renewcommand{\arraystretch}{1.25}
\small
\begin{tabularx}{\textwidth}{p{3.5cm} p{3cm} X p{2cm}}
\toprule
\rowcolor[gray]{0.85}  \textbf{Modality (Weight)} & \textbf{Strategy (Weight)} & \textbf{Task} & \textbf{Probability} \\
\midrule
\multirow{4}{*}{Protein (0.4)} 
  & All (0.1)   & Free Backbone Generation   & 0.04 \\
  & Entity (0.1)& Binder Design              & 0.04 \\
  & Token (0.7) & Motif Scaffolding          & 0.28 \\
  & Atom (0.1)  & Atomic Motif Scaffolding   & 0.04 \\
\midrule
\multirow{4}{*}{Nucleic Acid (0.3)} 
  & All (0.1)   & Free Backbone Generation   & 0.03 \\
  & Entity (0.1)& Binder Design              & 0.03 \\
  & Token (0.7) & Motif Scaffolding          & 0.21 \\
  & Atom (0.1)  & Atomic Motif Scaffolding   & 0.03 \\

\multirow{4}{*}{Ligand (0.3)} 
  & All (0.1)   & Free Backbone Generation   & 0.03 \\
  & Entity (0.1)& Binder Design              & 0.03 \\
  & Token (0.7) & Motif Scaffolding          & 0.21 \\
  & Atom (0.1)  & Atomic Motif Scaffolding   & 0.03 \\
\bottomrule
\end{tabularx}
\end{table*}

\begin{table*}[htbp]
\centering
\renewcommand{\thetable}{S4}
\caption{{Chemical Composition and Input Features of Generative Tokens}}
\label{tab:chemcomp}
\renewcommand{\arraystretch}{1.25}
\small
\begin{tabularx}{\textwidth}{p{3cm} p{4cm} p{3cm} X}
\toprule
\rowcolor[gray]{0.85} 
\textbf{Modality} & \textbf{Chemical Composition} & \textbf{Feature} & \textbf{Value} \\
\midrule

\multirow{2}{*}{Protein} 
  & {N, CA, C, O} 
  & res\_type & -P \\
  &  & ref\_pos & Backbone atom coordinates from the CCD conformation of GLY \\

\midrule
\multirow{2}{*}{DNA – Nucleic Acid} 
  & {P, OP1, OP2, O5$'$, C5$'$, C4$'$, O4$'$, C3$'$, O3$'$, C1$'$, C2$'$, N1/N9} 
  & res\_type & -N \\
  &  & ref\_pos & Backbone atom coordinates from the CCD conformation of DC \\

\midrule
\multirow{2}{*}{RNA – Nucleic Acid} 
  & {P, OP1, OP2, O5$'$, C5$'$, C4$'$, O4$'$, C3$'$, O3$'$, C1$'$, C2$'$, O2$'$, N1/N9} 
  & res\_type & -N \\
  &  & ref\_pos & Backbone atom coordinates from the CCD conformation of C \\

\midrule
\multirow{4}{*}{Ligand} 
  & \multirow{4}{*}{--} 
  & res\_type & -L \\
  &  & ref\_pos & 0 \\
  &  & ref\_element & -- \\
  &  & ref\_atom\_name\_chars & -- \\
\bottomrule
\end{tabularx}
\end{table*}

\begin{table*}
\caption{ {Runtime of various design methods across different targets (unit: seconds).}}
\label{tab:runtime}
\centering
\begin{tabular}{lrrrr}
\toprule
\rowcolor[gray]{0.85} \textbf{Target} & \textbf{RFdiffusion} & \textbf{bindcraft} & \textbf{boltzdesign} & \textbf{ODesign} \\
\midrule
cd3d     & 31.9 & 4573.8 & 266.2 & 6.5 \\
egfr     & 31.9 & 3293.6 & 628.0 & 6.5 \\
fgfr2    & 31.9 & 3593.5 & 419.5 & 6.5 \\
h3       & 31.9 & 2053.5 & 639.4 & 6.5 \\
il7ra    & 31.9 & 19624.2 & 565.2 & 6.5 \\
insulinr & 31.9 & 782.6 & 570.3 & 6.5 \\
pd1      & 31.9 & 38951.8 & 759.3 & 6.5 \\
pdgfr    & 31.9 & 4708.6 & 501.7 & 6.5 \\
pdl1     & 31.9 & 1645.0 & 769.6 & 6.5 \\
tgfb     & 31.9 & 10227.5 & 359.9 & 6.5 \\
tie2     & 31.9 & 16948.2 & 402.7 & 6.5 \\
trka     & 31.9 & 2613.3 & 376.9 & 6.5 \\
virb8    & 31.9 & 1553.7 & 377.1 & 6.5 \\
\bottomrule
\end{tabular}
\end{table*}
\clearpage

\begin{figure}
    \centering
    \includegraphics[width=0.8\linewidth]{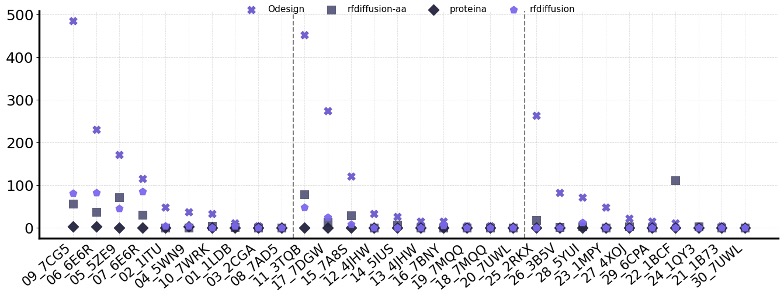}
    \caption{Case by Case Results on Motif Scaffolding}
    \label{fig:s4}
\end{figure}
\clearpage
\newpage

\begin{figure}
\centering
\definecolor{codeblue}{rgb}{0.25,0.5,0.5}
\definecolor{codekw}{rgb}{0.85, 0.18, 0.50}
\definecolor{codesign}{RGB}{0, 0, 255}
\definecolor{codefunc}{rgb}{0.85, 0.18, 0.50}
\definecolor{codegreen}{rgb}{0.0, 0.6, 0.4}

\lstdefinelanguage{PythonFuncColor}{
  language=Python,
  keywordstyle=\color{blue}\bfseries,
  commentstyle=\color{codeblue},  
  stringstyle=\color{orange},
  showstringspaces=false,
  basicstyle=\ttfamily\small,
  literate=
    {*}{{\color{codesign}* }}{1}
    {-}{{\color{codesign}- }}{1}
    {+}{{\color{codesign}+ }}{1}
    {Get_Standard_Token}{{\color{codefunc}get\_standard\_token}}{1}
    {return}{{\color{codekw}return }}{1}
    {in}{{\color{codekw}in }}{1}
    {True}{{\color{codegreen}True}}{1}
    {False}{{\color{codegreen}False}}{1}
    {protein}{{\color{black}protein}}{1}
}

\lstset{
  language=PythonFuncColor,
  backgroundcolor=\color{white},
  basicstyle=\fontsize{8.9pt}{9.9pt}\ttfamily\selectfont,
  columns=fullflexible,
  breaklines=true,
  captionpos=b,
}

\begin{minipage}{0.85\linewidth}
\begin{algorithm}[H]
\caption{{Get Standard Token}.}\label{alg:get_standard_token}\vspace{-0.5em}
\begin{lstlisting}
def Get_Standard_Token(token_array):
    is_standard_protein = token_array.res_type in [
        ALA, ARG, ASN, ASP, CYS, GLN, GLU, GLY, HIS, 
        ILE, LEU, LYS, MET, PHE, PRO, SER, TRP, TYR, VAL
    ]
    is_standard_nucleic_acid = token_array.res_type in [
        A, G, N, U, DA, DG, DN, DT
    ]
    is_standard_ligand = (
        (token_array.mol_type == "ligand")
        and (token_array.mol_atom_num > 1)
    )
    is_std_token = is_standard_protein or is_standard_nucleic_acid or \ 
                    is_standard_ligand
    return is_std_token
\end{lstlisting}\vspace{-0.5em}
\end{algorithm}
\end{minipage}
\vspace{-1em}
\end{figure}
\begin{figure}
\definecolor{codeblue}{rgb}{0.25,0.5,0.5}
\definecolor{codekw}{rgb}{0.85, 0.18, 0.50}
\definecolor{codesign}{RGB}{0, 0, 255}
\definecolor{codefunc}{rgb}{0.85, 0.18, 0.50}
\definecolor{codegreen}{rgb}{0.0, 0.6, 0.4}

\lstdefinelanguage{PythonFuncColor}{
  language=Python,
  keywordstyle=\color{blue}\bfseries,
  commentstyle=\color{codeblue},  
  stringstyle=\color{orange},
  showstringspaces=false,
  basicstyle=\ttfamily\small,
  literate=
    {*}{{\color{codesign}* }}{1}
    {-}{{\color{codesign}- }}{1}
    {+}{{\color{codesign}+ }}{1}
    {All_Mask_Strategy}{{\color{codefunc}all\_mask\_strategy}}{1}
    {Get_Standard_Token}{{\color{codefunc}get\_standard\_token}}{1}
    {Token_To_Atom}{{\color{codefunc}token\_to\_atom}}{1}
    {return}{{\color{codekw}return }}{1}
    {in}{{\color{codekw}in }}{1}
    {True}{{\color{codegreen}True}}{1}
    {False}{{\color{codegreen}False}}{1}
    {protein}{{protein}}{1}
    {chains}{{chains}}{1}
}

\lstset{
  language=PythonFuncColor,
  backgroundcolor=\color{white},
  basicstyle=\fontsize{8.9pt}{9.9pt}\ttfamily\selectfont,
  columns=fullflexible,
  breaklines=true,
  captionpos=b,
}
\centering
\begin{minipage}{0.85\linewidth}
\begin{algorithm}[H]
\caption{{All Mask Strategy (Free Backbone Generation)}.}\label{alg:all_mask_strategy}\vspace{-0.5em}
\begin{lstlisting}
def All_Mask_Strategy(token_array, atom_array):
    # get standard tokens
    is_std_token = Get_Standard_Token(token_array)

    # mask generation
    is_masked_token = is_std_token

    # postprocess: symmetrize, remove side chains, map to atoms
    is_masked_atom = Token_To_Atom(
        token_array, atom_array, is_masked_token
    )

    return is_masked_token, is_masked_atom
\end{lstlisting}\vspace{-0.5em}
\end{algorithm}
\end{minipage}
\end{figure}

\begin{figure}
\definecolor{codeblue}{rgb}{0.25,0.5,0.5}
\definecolor{codekw}{rgb}{0.85, 0.18, 0.50}
\definecolor{codesign}{RGB}{0, 0, 255}
\definecolor{codefunc}{rgb}{0.85, 0.18, 0.50}
\definecolor{codegreen}{rgb}{0.0, 0.6, 0.4}

\lstdefinelanguage{PythonFuncColor}{
  language=Python,
  keywordstyle=\color{blue}\bfseries,
  commentstyle=\color{codeblue},  
  stringstyle=\color{orange},
  showstringspaces=false,
  basicstyle=\ttfamily\small,
  literate=
    {*}{{\color{codesign}* }}{1}
    {-}{{\color{codesign}- }}{1}
    {+}{{\color{codesign}+ }}{1}
    {Entity_Mask_Strategy}{{\color{codefunc}entity\_mask\_strategy}}{1}
    {Get_Standard_Token}{{\color{codefunc}get\_standard\_token}}{1}
    {Get_Symmetric_Mask}{{\color{codefunc}get\_symmetric\_mask}}{1}
    {Token_To_Atom}{{\color{codefunc}token\_to\_atom}}{1}
    {Unique}{{\color{codefunc}unique}}{1}
    {Length}{{\color{codefunc}length}}{1}
    {Rand_Int}{{\color{codefunc}rand\_int}}{1}
    {Random_Choice}{{\color{codefunc}random\_choice}}{1}
    {return}{{\color{codekw}return }}{1}
    {in}{{\color{codekw}in }}{1}
    {True}{{\color{codegreen}True}}{1}
    {False}{{\color{codegreen}False}}{1}
    {protein}{{protein}}{1}
    {chains}{{chains}}{1}
}

\lstset{
  language=PythonFuncColor,
  backgroundcolor=\color{white},
  basicstyle=\fontsize{8.9pt}{9.9pt}\ttfamily\selectfont,
  columns=fullflexible,
  breaklines=true,
  captionpos=b,
}
\centering
\begin{minipage}{0.85\linewidth}
\begin{algorithm}[H]
\caption{{Entity Mask Strategy (Binder Design)}.}\label{alg:entity_mask_strategy}\vspace{-0.5em}
\begin{lstlisting}
def Entity_Mask_Strategy(token_array, atom_array):
    # get standard tokens
    is_std_token = Get_Standard_Token(token_array)

    # mask generation
    unique_entity_ids = Unique(token_array.entity_id[is_std_token])
    N = Length(unique_entity_ids)
    if N == 1:
        N_mask = 1
    else:
        N_mask = Rand_Int(1, N - 1)
    masked_entity_ids = Random_Choice(unique_entity_ids, N_mask)
    is_masked_token = (token_array.entity_id in masked_entity_ids) and is_std_token

    # postprocess: symmetrize, remove side chains, map to atoms
    is_masked_token = Get_Symmetric_Mask(
        token_array, atom_array, is_masked_token
    )
    is_masked_atom = Token_To_Atom(
        token_array, atom_array, is_masked_token
    )

    return is_masked_token, is_masked_atom
\end{lstlisting}\vspace{-0.5em}
\end{algorithm}
\end{minipage}
\end{figure}
\begin{figure}
\definecolor{codeblue}{rgb}{0.25,0.5,0.5}
\definecolor{codekw}{rgb}{0.85, 0.18, 0.50}
\definecolor{codesign}{RGB}{0, 0, 255}
\definecolor{codefunc}{rgb}{0.85, 0.18, 0.50}
\definecolor{codegreen}{rgb}{0.0, 0.6, 0.4}

\lstdefinelanguage{PythonFuncColor}{
  language=Python,
  keywordstyle=\color{blue}\bfseries,
  commentstyle=\color{codeblue},
  stringstyle=\color{orange},
  showstringspaces=false,
  basicstyle=\ttfamily\small,
  literate=
    {*}{{\color{codesign}* }}{1}
    {-}{{\color{codesign}- }}{1}
    {+}{{\color{codesign}+ }}{1}
    {get_standard_token}{{\color{codefunc}get\_standard\_token}}{1}
    {get_symmetric_mask}{{\color{codefunc}get\_symmetric\_mask}}{1}
    {token_to_atom}{{\color{codefunc}token\_to\_atom}}{1}
    {unique}{{\color{codefunc}unique}}{1}
    {length}{{\color{codefunc}length}}{1}
    {rand_int}{{\color{codefunc}rand\_int}}{1}
    {random_choice}{{\color{codefunc}random\_choice}}{1}
    {return}{{\color{codekw}return }}{1}
    {in}{{\color{codekw}in }}{1}
    {True}{{\color{codegreen}True}}{1}
    {False}{{\color{codegreen}False}}{1}
    {protein}{{protein}}{1}
    {chains}{{chains}}{1}
    {Min}{{\color{codefunc}min}}{1}
    {std_indices}{{std\_indices }}{1}
    {(std_indices)}{{(std\_indices)}}{1}
    {Where}{{\color{codefunc}where}}{1}
    {Length}{{\color{codefunc}length}}{1}
    {Zeros}{{\color{codefunc}zeros}}{1}
    {avg_length}{{avg\_length }}{1}
    {mask_length}{{mask\_length }}{1}
    {std_indices[tmp]}{{std_indices[tmp]}}{1}
}

\lstset{
  language=PythonFuncColor,
  backgroundcolor=\color{white},
  basicstyle=\fontsize{8.9pt}{9.9pt}\ttfamily\selectfont,
  columns=fullflexible,
  breaklines=true,
  captionpos=b,
}
\centering
\begin{minipage}{0.85\linewidth}
\begin{algorithm}[H]
\caption{{Token Mask Strategy (Motif Scaffolding)}.}\label{alg:token_mask_strategy}\vspace{-0.5em}
\begin{lstlisting}
def token_mask_strategy(token_array, atom_array):
    # get standard tokens
    is_std_token = get_standard_token(token_array)

    # mask generation
    std_indices = Where(is_std_token)
    L = Length(std_indices)
    M = rand_int(1, Min(5, L))
    is_masked_token = Zeros(Length(token_array))
    left_segments = M
    start_idx, end_idx = 0, 0

    for j in range(1, M + 1):
        avg_length = (L - end_idx) / left_segments
        if avg_length == 0:
            break
        mask_length = rand_int(1, avg_length + 1)
        tmp_start = rand_int(start_idx, start_idx + avg_length - mask_length + 1)
        end_idx = tmp_start + mask_length
        is_masked_token[std_indices[tmp_start:end_idx]] = True
        left_segments = left_segments - 1
        start_idx = end_idx

    # postprocess: symmetrize, remove side chains, map to atoms
    is_masked_token = get_symmetric_mask(
        token_array, atom_array, is_masked_token
    )
    is_masked_atom = token_to_atom(
        token_array, atom_array, is_masked_token
    )

    return is_masked_token, is_masked_atom
\end{lstlisting}\vspace{-0.5em}
\end{algorithm}
\end{minipage}
\end{figure}
\begin{figure}
\definecolor{codeblue}{rgb}{0.25,0.5,0.5}
\definecolor{codekw}{rgb}{0.85, 0.18, 0.50}
\definecolor{codesign}{RGB}{0, 0, 255}
\definecolor{codefunc}{rgb}{0.85, 0.18, 0.50}
\definecolor{codegreen}{rgb}{0.0, 0.6, 0.4}

\lstdefinelanguage{PythonFuncColor}{
  language=Python,
  keywordstyle=\color{blue}\bfseries,
  commentstyle=\color{codeblue},  
  stringstyle=\color{orange},
  showstringspaces=false,
  basicstyle=\ttfamily\small,
  literate=
    {*}{{\color{codesign}* }}{1}
    {-}{{\color{codesign}- }}{1}
    {+}{{\color{codesign}+ }}{1}
    {atom_mask_strategy}{{\color{codefunc}atom\_mask\_strategy}}{1}
    {get_standard_token}{{\color{codefunc}get\_standard\_token}}{1}
    {get_symmetric_mask}{{\color{codefunc}get\_symmetric\_mask}}{1}
    {token_to_atom}{{\color{codefunc}token\_to\_atom}}{1}
    {length}{{\color{codefunc}length}}{1}
    {rand}{{\color{codefunc}rand}}{1}
    {zeros}{{\color{codefunc}zeros}}{1}
    {return}{{\color{codekw}return }}{1}
    {in}{{\color{codekw}in }}{1}
    {True}{{\color{codegreen}True}}{1}
    {False}{{\color{codegreen}False}}{1}
    {protein}{{protein}}{1}
    {chains}{{chains}}{1}
    {mol_atom_index}{{mol\_atom\_index}}{1}    
    {append}{{\color{codefunc}append}}{1}
    {get}{{\color{codefunc}get}}{1}
    {get standard}{{get standard}}{1}
    {get_standard}{{get\_standard}}{1}
}

\lstset{
  language=PythonFuncColor,
  backgroundcolor=\color{white},
  basicstyle=\fontsize{8.9pt}{9.9pt}\ttfamily\selectfont,
  columns=fullflexible,
  breaklines=true,
  captionpos=b,
}
\centering
\begin{minipage}{0.85\linewidth}
\begin{algorithm}[H]
\caption{{Atom Mask Strategy (Atomic Motif Scaffolding)}.}\label{alg:atom_mask_strategy}\vspace{-0.5em}
\begin{lstlisting}
def atom_mask_strategy(token_array, atom_array):
    # get standard tokens
    is_std_token = get_standard_token(token_array)

    # mask generation
    is_masked_token = is_std_token
    is_masked_cond_atom = (rand(length(token_array)) < 0.2) and is_std_token

    # build symmetric atom UID dictionary
    sym_atom_uid = dict()
    for idx in range(0, length(atom_array)):
        atom = atom_array[idx]
        atom_uid = atom.entity_mol_id + "_" + atom.mol_atom_index
        uid_list = sym_atom_uid.get(atom_uid, [])
        uid_list.append(idx)
        sym_atom_uid[atom_uid] = uid_list

    # postprocess: symmetrize, remove side chains, map to atoms
    is_masked_token = get_symmetric_mask(
        token_array, atom_array, is_masked_token
    )
    is_masked_atom = token_to_atom(
        token_array, atom_array, is_masked_token,
        is_masked_cond_atom, sym_atom_uid
    )

    return is_masked_token, is_masked_atom
\end{lstlisting}\vspace{-0.5em}
\end{algorithm}
\end{minipage}
\end{figure}

\begin{figure}
\definecolor{codeblue}{rgb}{0.25,0.5,0.5}
\definecolor{codekw}{rgb}{0.85, 0.18, 0.50}
\definecolor{codesign}{RGB}{0, 0, 255}
\definecolor{codefunc}{rgb}{0.85, 0.18, 0.50}
\definecolor{codegreen}{rgb}{0.0, 0.6, 0.4}

\lstdefinelanguage{PythonFuncColor}{
  language=Python,
  keywordstyle=\color{blue}\bfseries,
  commentstyle=\color{codeblue},
  stringstyle=\color{orange},
  showstringspaces=false,
  basicstyle=\ttfamily\small,
  literate=
    {*}{{\color{codesign}* }}{1}
    {-}{{\color{codesign}- }}{1}
    {+}{{\color{codesign}+ }}{1}
    {length}{{\color{codefunc}length}}{1}
    {deepcopy}{{\color{codefunc}deepcopy}}{1}
    {any}{{\color{codefunc}any}}{1}
    {return}{{\color{codekw}return }}{1}
    {in}{{\color{codekw}in }}{1}
    {for}{{\color{codekw}for }}{1}
    {if}{{\color{codekw}if }}{1}
    {else}{{\color{codekw}else }}{1}
    {True}{{\color{codegreen}True}}{1}
    {False}{{\color{codegreen}False}}{1}
    {append}{{\color{codefunc}append}}{1}
    {get}{{\color{codefunc}get}}{1}
    {get_symmetric_mask}{{\color{codefunc}get\_symmetric\_mask}}{1}
    {mol_atom_index}{{mol\_atom\_index}}{1}    
    {center_atom_index}{{center\_atom\_index}}{1}    
    {atom_indices}{{atom\_indices}}{1}    
    {obtain}{{obtain }}{1}    
}

\lstset{
  language=PythonFuncColor,
  backgroundcolor=\color{white},
  basicstyle=\fontsize{8.9pt}{9.9pt}\ttfamily\selectfont,
  columns=fullflexible,
  breaklines=true,
  captionpos=b,
}
\centering
\begin{minipage}{0.85\linewidth}
\begin{algorithm}[H]
\caption{{Get Symmetric Token Mask}.}\label{alg:get_symmetric_mask}\vspace{-0.5em}
\begin{lstlisting}
def get_symmetric_mask(token_array, atom_array, asym_token_mask):
    # build symmetric atom UID dictionary
    sym_atom_uid = dict()
    center_atom_array = atom_array[token_array.center_atom_index]
    for idx in range(0, length(center_atom_array)):
        atom = center_atom_array[idx]
        atom_uid = atom.entity_mol_id + "_" + atom.mol_atom_index
        uid_list = sym_atom_uid.get(atom_uid, [])
        uid_list.append(idx)
        sym_atom_uid[atom_uid] = uid_list

    # obtain symmetric token mask
    sym_token_mask = deepcopy(asym_token_mask)
    for atom in center_atom_array[~asym_token_mask]:
        atom_uid = atom.entity_mol_id + "_" + atom.mol_atom_index
        atom_indices = sym_atom_uid.get(atom_uid, [])
        atom_mask = asym_token_mask[atom_indices]
        if any(atom_mask):
            sym_token_mask[atom_indices] = True

    return sym_token_mask
\end{lstlisting}\vspace{-0.5em}
\end{algorithm}
\end{minipage}
\end{figure}

\begin{figure}
\definecolor{codeblue}{rgb}{0.25,0.5,0.5}
\definecolor{codekw}{rgb}{0.85, 0.18, 0.50}
\definecolor{codesign}{RGB}{0, 0, 255}
\definecolor{codefunc}{rgb}{0.85, 0.18, 0.50}
\definecolor{codegreen}{rgb}{0.0, 0.6, 0.4}

\lstdefinelanguage{PythonFuncColor}{
  language=Python,
  keywordstyle=\color{blue}\bfseries,
  commentstyle=\color{codeblue},
  stringstyle=\color{orange},
  showstringspaces=false,
  basicstyle=\ttfamily\small,
  literate=
    {*}{{\color{codesign}* }}{1}
    {-}{{\color{codesign}- }}{1}
    {+}{{\color{codesign}+ }}{1}
    {token_to_atom}{{\color{codefunc}token\_to\_atom}}{1}
    {length}{{\color{codefunc}length}}{1}
    {zeros}{{\color{codefunc}zeros}}{1}
    {get_sym_condition_atoms}{{\color{codefunc}get\_sym\_condition\_atoms}}{1}
    {return}{{\color{codekw}return }}{1}
    {for}{{\color{codekw}for }}{1}
    {if}{{\color{codekw}if }}{1}
    {else}{{\color{codekw}else }}{1}
    {in}{{\color{codekw}in }}{1}
    {and}{{\color{codekw}and }}{1}
    {not}{{\color{codekw}not }}{1}
    {True}{{\color{codegreen}True }}{1}
    {False}{{\color{codegreen}False}}{1}
    {mask_atom_indices}{{mask\_atom\_indices }}{1}  
    {[mask_atom_indices]}{{[mask\_atom\_indices]}}{1}  
    {condition_atom_indices}{{condition\_atom\_indices }}{1} 
    {center_atom_index}{{center\_atom\_index}}{1} 
    {atom_indices}{{atom\_indices}}{1}  
    {protein}{{protein}}{1}    
    {ligand}{{ligand}}{1}    
}

\lstset{
  language=PythonFuncColor,
  backgroundcolor=\color{white},
  basicstyle=\fontsize{8.9pt}{9.9pt}\ttfamily\selectfont,
  columns=fullflexible,
  breaklines=true,
  captionpos=b,
}
\centering
\begin{minipage}{0.85\linewidth}
\begin{algorithm}[H]
\caption{{Map Token Mask To Atom Mask}.}\label{alg:token_to_atom}\vspace{-0.5em}
\begin{lstlisting}
def token_to_atom(token_array, atom_array, is_masked_token, 
                  is_masked_cond_atom=None, sym_atom_uid=None):
    # check condition atoms
    if is_masked_cond_atom is None:
        is_masked_cond_atom = zeros(length(token_array), bool)

    # map token to atom
    is_condition_atom = zeros(length(atom_array), bool)
    for idx in range(0, length(token_array)):
        token = token_array[idx]
        mask_atom_indices = token.atom_indices

        if is_masked_token[idx] == True:
            if token.mol_type != "ligand":
                res_name = atom_array.res_name[token.center_atom_index]
                backbone_mask = res_name in BACKBONE_ATOM_NAMES
                mask_atom_indices = token.atom_indices[backbone_mask]

                if is_masked_cond_atom[idx] == True and token.mol_type == "protein":
                    condition_atom_indices = get_sym_condition_atoms(
                        token, sym_atom_uid
                    )
                    mask_atom_indices = mask_atom_indices or condition_atom_indices

        is_condition_atom[mask_atom_indices] = True

    is_masked_atom = ~is_condition_atom
    return is_masked_atom
\end{lstlisting}\vspace{-0.5em}
\end{algorithm}
\end{minipage}
\end{figure}
\begin{figure}
\definecolor{codeblue}{rgb}{0.25,0.5,0.5}
\definecolor{codekw}{rgb}{0.85, 0.18, 0.50}
\definecolor{codesign}{RGB}{0, 0, 255}
\definecolor{codefunc}{rgb}{0.85, 0.18, 0.50}
\definecolor{codegreen}{rgb}{0.0, 0.6, 0.4}

\lstdefinelanguage{PythonFuncColor}{
  language=Python,
  keywordstyle=\color{blue}\bfseries,
  commentstyle=\color{codeblue},
  stringstyle=\color{orange},
  showstringspaces=false,
  basicstyle=\ttfamily\small,
  literate=
    {*}{{\color{codesign}* }}{1}
    {-}{{\color{codesign}- }}{1}
    {+}{{\color{codesign}+ }}{1}
    {get_sym_condition_atoms}{{\color{codefunc}get\_sym\_condition\_atoms}}{1}
    {construct_bond_graph}{{\color{codefunc}construct\_bond\_graph}}{1}
    {shortest_path_length}{{\color{codefunc}shortest\_path\_length}}{1}
    {argmax}{{\color{codefunc}argmax}}{1}
    {random_choice}{{\color{codefunc}random\_choice}}{1}
    {geometric}{{\color{codefunc}geometric}}{1}
    {get_keys}{{\color{codefunc}get\_keys}}{1}
    {range}{{\color{codefunc}range}}{1}
    {rand()}{{\color{codefunc}rand}}{1}  
    {return}{{\color{codekw}return }}{1}
    {for}{{\color{codekw}for }}{1}
    {if}{{\color{codekw}if }}{1}
    {else}{{\color{codekw}else }}{1}
    {in}{{\color{codekw}in }}{1}
    {and}{{\color{codekw}and }}{1}
    {or}{{\color{codekw}or }}{1}
    {not}{{\color{codekw}not }}{1}
    {True}{{\color{codegreen}True}}{1}
    {False}{{\color{codegreen}False}}{1}
    {geometric_distribution}{{geometric distribution}}{1}  
    {randomly}{{randomly}}{1}  
    {within}{{within}}{1}  
    {sym_condition_atom_indices}{{sym\_condition\_atom\_indices }}{1}  
    {condition_atom_indices}{{condition\_atom\_indices }}{1}  
    {[condition_atom_indices]}{{[condition\_atom\_indices]}}{1}  
    {atom_indices}{{atom\_indices}}{1}  
    {atom_indices=}{{atom\_indices = }}{1}  
    {mol_atom_index}{{mol\_atom\_index }}{1}  
    {initialization}{{initialization }}{1}  
}

\lstset{
  language=PythonFuncColor,
  backgroundcolor=\color{white},
  basicstyle=\fontsize{8.9pt}{9.9pt}\ttfamily\selectfont,
  columns=fullflexible,
  breaklines=true,
  captionpos=b,
}
\centering
\begin{minipage}{0.85\linewidth}
\begin{algorithm}[H]
\caption{{Condition Atom Selection}.}\label{alg:get_sym_condition_atoms}\vspace{-0.8em}
\begin{lstlisting}
def get_sym_condition_atoms(token, atom_array, sym_atom_uid):
    # initialization
    res_atoms = atom_array[token.atom_indices]
    O_INDEX = 3

    if rand() < 0.8:
        # select atom furthest from backbone oxygen
        bond_graph = construct_bond_graph(res_atoms)
        distances = shortest_path_length(bond_graph, O_INDEX)
        seed_atom_idx = argmax(distances)
    else:
        # randomly select seed atom
        seed_atom_idx = random_choice(token.atom_indices)

    # sample number of chemical bonds from geometric_distribution
    n = geometric(0.5) - 1

    # get all atoms within n bonds from seed atom
    bond_graph = construct_bond_graph(res_atoms)
    paths = shortest_path_length(bond_graph, seed_atom_idx, cutoff=n)
    condition_atom_indices = get_keys(paths)

    # get symmetric atoms
    sym_condition_atom_indices = []
    for atom in atom_array[condition_atom_indices]:
        atom_uid = atom.entity_mol_id + "_" + atom.mol_atom_index
        atom_indices= sym_atom_uid.get(atom_uid, [])
        sym_condition_atom_indices = (sym_condition_atom_indicesoratom_indices)

    return sym_condition_atom_indices
\end{lstlisting}\vspace{-0.5em}
\end{algorithm}
\end{minipage}
\end{figure}

\begin{figure}
\definecolor{codeblue}{rgb}{0.25,0.5,0.5}
\definecolor{codekw}{rgb}{0.85, 0.18, 0.50}
\definecolor{codesign}{RGB}{0, 0, 255}
\definecolor{codefunc}{rgb}{0.85, 0.18, 0.50}
\definecolor{codegreen}{rgb}{0.0, 0.6, 0.4}

\lstdefinelanguage{PythonFuncColor}{
  language=Python,
  keywordstyle=\color{blue}\bfseries,
  commentstyle=\color{codeblue},
  stringstyle=\color{orange},
  showstringspaces=false,
  basicstyle=\ttfamily\small,
  literate=
    {*}{{\color{codesign}* }}{1}
    {-}{{\color{codesign}- }}{1}
    {+}{{\color{codesign}+ }}{1}
    {main_inference_loop}{{\color{codefunc}main\_inference\_loop}}{1}
    {embedder_module}{{\color{codefunc}embedder\_module}}{1}
    {pairwise_distance_with_mask}{{\color{codefunc}pairwise\_distance\_with\_mask}}{1}
    {constraint_template_embedder}{{\color{codefunc}constraint\_template\_embedder}}{1}
    {pairformer_module}{{\color{codefunc}pairformer\_module}}{1}
    {conditional_diffusion_module}{{\color{codefunc}conditional\_diffusion\_module}}{1}
    {distogram_head}{{\color{codefunc}distogram\_head}}{1}
    {bond_reconstruction_head}{{\color{codefunc}bond\_reconstruction\_head}}{1}
    {oinvfold_module}{{\color{codefunc}oinvfold\_module}}{1}
    {return}{{\color{codekw}return }}{1}
    {for}{{\color{codekw}for }}{1}
    {if}{{\color{codekw}if }}{1}
    {in}{{\color{codekw}in }}{1}
    {else}{{\color{codekw}else }}{1}
    {True}{{\color{codegreen}True}}{1}
    {False}{{\color{codegreen}False}}{1}
    {chains}{{chains}}{1}
    {protein}{{protein}}{1}
    {bins}{{bins}}{1}
    {min\_bin}{{min\_bin}}{1}
    {max\_bin}{{max\_bin}}{1}
    {no\_bins}{{no\_bins}}{1}
    {Pairformer}{{Pairformer}}{1}
    {Diffusion}{{Diffusion}}{1}
    {OinvFold}{{OinvFold}}{1}
    {input}{{input}}{1}
    {init}{{init}}{1}
}

\lstset{
  language=PythonFuncColor,
  backgroundcolor=\color{white},
  basicstyle=\fontsize{8.9pt}{9.9pt}\ttfamily\selectfont,
  columns=fullflexible,
  breaklines=true,
  captionpos=b,
}
\centering
\begin{minipage}{0.85\linewidth}
\begin{algorithm}[H]
\caption{{Main Inference Loop}.}\label{alg:main_inference_loop}\vspace{-0.5em}
\begin{lstlisting}
def main_inference_loop(f_star, x_true, mask_generative, N_cycle):
    # Embedder Module
    s_input, s_i_init, z_ij_init = embedder_module(f_star)

    # Conditional Module Part 1
    z_ij_distance = pairwise_distance_with_mask(x_true, mask_generative)
    z_ij_distogram = constraint_template_embedder(z_ij_distance, z_ij_init)

    # Pairformer Module
    s_i, z_ij = pairformer_module(f_star, s_input, s_i_init, z_ij_init, z_ij_distogram, N_cycle)

    # Conditional Diffusion Module & Conditional Module Part 2
    x_l_pred = conditional_diffusion_module(
        f_star, s_input, s_i, z_ij, x_true, mask_generative
    )
    P_ij_distogram = distogram_head(z_ij)
    P_ij_bond = bond_reconstruction_head(z_ij)

    # OinvFold Module
    sequence = oinvfold_module(x_l_pred)

    return x_l_pred, sequence
\end{lstlisting}\vspace{-0.5em}
\end{algorithm}
\end{minipage}
\end{figure}

\begin{figure}
\definecolor{codeblue}{rgb}{0.25,0.5,0.5}
\definecolor{codekw}{rgb}{0.85, 0.18, 0.50}
\definecolor{codesign}{RGB}{0, 0, 255}
\definecolor{codefunc}{rgb}{0.85, 0.18, 0.50}
\definecolor{codegreen}{rgb}{0.0, 0.6, 0.4}

\lstdefinelanguage{PythonFuncColor}{
  language=Python,
  keywordstyle=\color{blue}\bfseries,
  commentstyle=\color{codeblue},
  stringstyle=\color{orange},
  showstringspaces=false,
  basicstyle=\ttfamily\small,
  literate=
    {*}{{\color{codesign} * }}{1}
    {-}{{\color{codesign} -}}{1}
    {+}{{\color{codesign}+ }}{1}
    {constraint_template_embedder}{{\color{codefunc}constraint\_template\_embedder}}{1}
    {linspace}{{\color{codefunc}linspace}}{1}
    {one\_hot}{{\color{codefunc}one\_hot}}{1}
    {layernorm}{{\color{codefunc}layernorm}}{1}
    {linear\_no\_bias}{{\color{codefunc}linear\_no\_bias}}{1}
    {relu}{{\color{codefunc}relu}}{1}
    {pairformer\_stack}{{\color{codefunc}pairformer\_stack}}{1}
    {return}{{\color{codekw}return }}{1}
    {in}{{\color{codekw}in }}{1}
    {for}{{\color{codekw}for }}{1}
    {if}{{\color{codekw}if }}{1}
    {else}{{\color{codekw}else }}{1}
    {True}{{\color{codegreen}True}}{1}
    {False}{{\color{codegreen}False}}{1}
    {bins}{{bins}}{1}
    {min\_bin}{{min\_bin}}{1}
    {max\_bin}{{max\_bin}}{1}
    {no\_bins}{{no\_bins}}{1}
    {input}{{input}}{1}
    {init}{{init}}{1}
    {transform}{{transform}}{1}
    {processing}{{processing}}{1}
}

\lstset{
  language=PythonFuncColor,
  backgroundcolor=\color{white},
  basicstyle=\fontsize{8.9pt}{9.9pt}\ttfamily\selectfont,
  columns=fullflexible,
  breaklines=true,
  captionpos=b,
}
\centering
\begin{minipage}{0.87\linewidth}
\begin{algorithm}[H]
\caption{{ConstraintTemplateEmbedder}.}\label{alg:constraint_template_embedder}\vspace{-0.8em}
\begin{lstlisting}
def constraint_template_embedder(z_ij_distance, z_ij_init, min_bin, max_bin, no_bins):
    # get boundaries
    boundaries = linspace(min_bin, max_bin, no_bins - 1)

    # transform distance matrix to distogram
    bins = sum(z_ij_distance > boundaries, dim=-1)
    distogram = one_hot(bins, no_bins)

    # processing
    vij = linear_no_bias(layernorm(z_ij_init) + linear_no_bias(distogram))
    _, vij = pairformer_stack(s=None, z=vij)
    uij = layernorm(vij)
    uij = linear_no_bias(relu(vij))

    return uij
\end{lstlisting}\vspace{-0.5em}
\end{algorithm}
\end{minipage}
\end{figure}

\begin{figure}
\definecolor{codeblue}{rgb}{0.25,0.5,0.5}
\definecolor{codekw}{rgb}{0.85, 0.18, 0.50}
\definecolor{codesign}{RGB}{0, 0, 255}
\definecolor{codefunc}{rgb}{0.85, 0.18, 0.50}
\definecolor{codegreen}{rgb}{0.0, 0.6, 0.4}

\lstdefinelanguage{PythonFuncColor}{
  language=Python,
  keywordstyle=\color{blue}\bfseries,
  commentstyle=\color{codeblue},
  stringstyle=\color{orange},
  showstringspaces=false,
  basicstyle=\ttfamily\small,
  literate=
    {*}{{\color{codesign}* }}{1}
    {-}{{\color{codesign}- }}{1}
    {+}{{\color{codesign}+ }}{1}
    {pairformer_module}{{\color{codefunc}pairformer\_module}}{1}
    {msa_module}{{\color{codefunc}msa\_module}}{1}
    {pairformer\_stack}{{\color{codefunc}pairformer\_stack}}{1}
    {layernorm}{{\color{codefunc}layernorm}}{1}
    {linear\_no\_bias}{{\color{codefunc}linear\_no\_bias}}{1}
    {return}{{\color{codekw}return }}{1}
    {for}{{\color{codekw}for }}{1}
    {in}{{\color{codekw}in }}{1}
    {if}{{\color{codekw}if }}{1}
    {else}{{\color{codekw}else }}{1}
    {True}{{\color{codegreen}True}}{1}
    {False}{{\color{codegreen}False}}{1}
        {input}{{input}}{1}
    {init}{{init}}{1}
    {transform}{{transform}}{1}
    {processing}{{processing}}{1}
}

\lstset{
  language=PythonFuncColor,
  backgroundcolor=\color{white},
  basicstyle=\fontsize{8.9pt}{9.9pt}\ttfamily\selectfont,
  columns=fullflexible,
  breaklines=true,
  captionpos=b,
}
\centering
\begin{minipage}{0.87\linewidth}
\begin{algorithm}[H]
\caption{{Pairformer Module}.}\label{alg:pairformer_module}\vspace{-0.8em}
\begin{lstlisting}
def pairformer_module(f_star, s_input, s_i_init, z_ij_init, z_ij_distogram, N_cycle):
    s_hat_i, z_hat_ij = 0, 0
    for _ in range(N_cycle):
        z_ij = z_ij_init + linear_no_bias(layernorm(z_hat_ij))
        # distogram condition rather than template embedder
        z_ij = z_ij + z_ij_distogram  
        z_ij = z_ij + msa_module(f_star, z_ij, s_input)

        s_i = s_i_init + linear_no_bias(layernorm(s_hat_i))
        s_i, z_ij = pairformer_stack(s_i, z_ij)

        s_hat_i, z_hat_ij = s_i, z_ij

    return s_i, z_ij
\end{lstlisting}\vspace{-0.5em}
\end{algorithm}
\end{minipage}
\end{figure}

\begin{figure}
\definecolor{codeblue}{rgb}{0.25,0.5,0.5}
\definecolor{codekw}{rgb}{0.85, 0.18, 0.50}
\definecolor{codesign}{RGB}{0, 0, 255}
\definecolor{codefunc}{rgb}{0.85, 0.18, 0.50}
\definecolor{codegreen}{rgb}{0.0, 0.6, 0.4}

\lstdefinelanguage{PythonFuncColor}{
  language=Python,
  keywordstyle=\color{blue}\bfseries,
  commentstyle=\color{codeblue},
  stringstyle=\color{orange},
  showstringspaces=false,
  basicstyle=\ttfamily\small,
  literate=
    {*}{{\color{codesign}* }}{1}
    {-}{{\color{codesign}- }}{1}
    {+}{{\color{codesign}+ }}{1}
    {conditional_diffusion_module}{{\color{codefunc}conditional\_diffusion\_module}}{1}
    {centre_random_augmentation}{{\color{codefunc}centre\_random\_augmentation}}{1}
    {diffusion}{{\color{codefunc}diffusion}}{1}
    {reverse_transformation}{{\color{codefunc}reverse\_transformation}}{1}
    {length}{{\color{codefunc}length}}{1}
    {rand}{{\color{codefunc}rand}}{1}
    {randn}{{\color{codefunc}randn}}{1}
    {zeros}{{\color{codefunc}zeros}}{1}
    {edm_precond}{{\color{codefunc}edm\_precond}}{1}
    {diffusion_denoising}{{\color{codefunc}diffusion\_denoising}}{1}
    {return}{{\color{codekw}return }}{1}
    {for}{{\color{codekw}for }}{1}
    {in}{{\color{codekw}in }}{1}
    {if}{{\color{codekw}if }}{1}
    {else}{{\color{codekw}else }}{1}
    {True}{{\color{codegreen}True}}{1}
    {False}{{\color{codegreen}False}}{1}
    {input}{{input}}{1}
    {coordinates}{{coordinates}}{1}
    {init}{{init }}{1}
    {transform}{{transform}}{1}
    {diffusion_loop}{{diffusion loop}}{1}
    {gamma_min}{{gamma\_min }}{1}
    {gamma_min=}{{gamma\_min=}}{1}
    {gamma_min,}{{gamma\_min,}}{1}
}

\lstset{
  language=PythonFuncColor,
  backgroundcolor=\color{white},
  basicstyle=\fontsize{8.9pt}{9.9pt}\ttfamily\selectfont,
  columns=fullflexible,
  breaklines=true,
  captionpos=b,
}
\centering
\begin{minipage}{0.88\linewidth}
\begin{algorithm}[H]
\caption{{Conditional Diffusion Module}.}\label{alg:conditional_diffusion_module}\vspace{-0.8em}
\begin{lstlisting}
def conditional_diffusion_module(
    f_star, s_input, s_i_trunk, z_ij_trunk,
    x_true, mask_generative, gamma_0, gamma_min,
    noise_schedule, noise_scale, step_scale
):
    # init sample
    if noise_schedule is None:
        noise_schedule = []
    x_l = noise_scale * randn(length(x_true), 3)  # c0 * N(0, I_3)
    x_l[mask_generative] = x_true[mask_generative]  # fix coordinates

    # diffusion_loop
    for c_tau_idx in range(len(noise_schedule)):
        c_tau = noise_schedule[c_tau_idx]

        x_l, dx_reverse, R_reverse = centre_random_augmentation(x_l)
        gamma = gamma_0 if c_tau > gamma_min else 0.0

        t_hat = noise_schedule[max(c_tau_idx - 1, 0)] * (gamma + 1.0)
        eps = noise_scale * ((t_hat**2 - noise_schedule[max(c_tau_idx - 1, 0)]**2) ** 0.5) * randn(length(x_l), 3)
        eps[mask_generative] = 0.0  # fix coordinates

        x_l_noisy = edm_precond(x_l, eps)
        x_l_denoised = diffusion_denoising(
            x_l_noisy, t_hat, f_star, s_input, s_i_trunk, z_ij_trunk
        )

        delta_l = (x_l - x_l_denoised) / max(t_hat, 1e-8)
        dt = c_tau - t_hat
        x_l = x_l_noisy + step_scale * dt * delta_l

        x_l = reverse_transformation(x_l, dx_reverse, R_reverse)  # restore coords

    return x_l
\end{lstlisting}\vspace{-0.5em}
\end{algorithm}
\end{minipage}
\end{figure}

\clearpage

\newpage
\renewcommand{\thelstlisting}{S\arabic{lstlisting}}
\lstset{
  captionpos=b,
  frame=lines,
  framerule=0.8pt,
  basicstyle=\ttfamily\small,
  numbers=none
}

\renewcommand{\lstlistingname}{JSON}
\begin{multicols}{2}
\setlength{\columnsep}{1em}
\begin{minipage}{1\linewidth}
\begin{lstlisting}[caption={Configuration for protein\_binding\_protein},label={json:proteinprotein},frame=lines, framerule=0.8pt]
{
  "name": "IL7Ra",
  "ref_file": ".../IL7Ra/IL7Ra_complex.pdb",
  "chains": [
    {
      "chain_type": "proteinChain",
      "sequence": "B/65-257",
      "msa": {
        "precomputed_msa_dir": ".../IL7Ra/1",
        "pairing_db": "uniref100"
      }
    },
    {
      "chain_type": "proteinChain",
      "sequence": "65-65"
    }
  ],
  "hotspot": "B/129,B/130,B/131,B/132",
  "center_method": "hotspot_center"
}
\end{lstlisting}
\end{minipage}
\begin{minipage}{1\linewidth}
\begin{lstlisting}[caption={Configuration for ligand\_binding\_protein}, label={json:ligprotein},frame=lines, framerule=0.8pt]
{
  "name": "7bkc",
  "ref_file": ".../FAD/7bkc_FAD.cif",
  "chains": [
    {
      "chain_type": "proteinChain",
      "sequence": "411-411"
    },
    {
      "chain_type": "ligand",
      "sequence": "E/505-505"
    }
  ],
  "hotspot": "E/505",
  "center_method": "hotspot_center"
}
\end{lstlisting}
\end{minipage}
\begin{minipage}{1\linewidth}
\begin{lstlisting}[caption={Configuration for motif\_scaffolding},label={json:motifscaffold},frame=lines, framerule=0.8pt]
{
  "name": "30_7UWL",
  "ref_file": ".../motifbench_motif_pdbs/30_7UWL.pdb",
  "motif_scaffolding": true,
  "chains": [
    {
      "chain_type": "proteinChain",
      "sequence": "1-80,A/1-11,25-35...",
      "length": 175
    }
  ]
}
\end{lstlisting}
\end{minipage}
\begin{minipage}{1\linewidth}
\begin{lstlisting}[caption={Configuration for atom\_scaffolding},label={json:atomscaffold},frame=lines, framerule=0.8pt]
{
  "name": "M0129_1os7",
  "ref_file": ".../atomscaffold_motif_pdbs/M0129_1os7.pdb",
  "chains": [
    {
      "chain_type": "proteinChain",
      "sequence": "1-70,A/1-1,1-70,...",
      "length": 180
    },
    { "chain_type": "ligand", "sequence": "Z/1-1" },
    { "chain_type": "ligand", "sequence": "Z/2-2" },
    { "chain_type": "ligand", "sequence": "Z/3-3" }
  ],
  "condition_atom": {
    "A/1": ["NE2","CD2","CE1"],
    "B/1": ["OD1","CG","CB","OD2"],
    "C/1": ["NE2","CD2","CE1","CG","ND1"],
    "D/1": ["NH1","CZ"]
  },
  "hotspot": "A/1,B/1,C/1,D/1,Z/1,Z/2,Z/3",
  "center_method": "hotspot_center"
}
\end{lstlisting}
\end{minipage}
\begin{minipage}{1\linewidth}
\begin{lstlisting}[caption={Configuration for rna\_free\_gen},label={json:rnafree},frame=lines, framerule=0.8pt]
{
    "name": "rna_50",
    "ref_file": "",
    "chains": [
        {
            "chain_type": "rnaChain",
            "sequence": "50-50"
        }
    ]
}
\end{lstlisting}
\end{minipage}
\begin{minipage}{1\linewidth}
\begin{lstlisting}[caption={Configuration for protein\_binding\_rna},label={json:proteinrna},frame=lines, framerule=0.8pt]
{
  "name": "8gxb-assembly1",
  "ref_file": ".../ref_pdb/8gxb-assembly1.cif",
  "chains": [
    {
      "chain_type": "proteinChain",
      "sequence": "C/1-96",
      "msa": {
        "precomputed_msa_dir": ".../msa/0",
        "pairing_db": "uniref100"
      }
    },
    {
      "chain_type": "rnaChain",
      "sequence": "61-61"
    }
  ],
  "center_method": "hotspot_center",
  "hotspot": "C/18,C/90,C/79,C/49"
}
\end{lstlisting}
\end{minipage}
\begin{minipage}{1\linewidth}
\begin{lstlisting}[caption={Configuration for ligand\_generation},label={json:liggen},frame=lines, framerule=0.8pt]
{
    "name": "rna_50",
    "ref_file": "",
    "chains": [
        {
            "chain_type": "rnaChain",
            "sequence": "50-50"
        }
    ]
}
\end{lstlisting}
\end{minipage}
\end{multicols}

\end{document}